\documentclass{article}
\usepackage[letterpaper,top=2cm,bottom=2cm,left=2.5cm,right=2.5cm,marginparwidth=1.75cm]{geometry}
\usepackage{amsfonts}      
\usepackage{amsthm}
\usepackage{amsmath}
\usepackage{amssymb}
\usepackage{hyperref}
\usepackage{cleveref}
\usepackage[lofdepth,lotdepth]{subfig}
\usepackage{graphicx}
\usepackage{xcolor}
\usepackage{bbm}
\usepackage[shortlabels]{enumitem}
\usepackage{multicol}
\usepackage{array} 
\usepackage{stackrel}
\usepackage{setspace}

\newtheorem{theorem}{Theorem}
\numberwithin{theorem}{section}
\newtheorem{lemma}[theorem]{Lemma}
\newtheorem{corollary}[theorem]{Corollary}
\newtheorem{example}{Example}
\numberwithin{example}{section}
\newtheorem{remark}{Remark}
\numberwithin{remark}{section}
\newtheorem{definition}{Definition}
\numberwithin{definition}{section}
\newtheorem{proposition}[theorem]{Proposition}
\newcommand{\E}{\mathbb{E}}

\DeclareMathOperator*{\esssup}{ess\,sup}
\DeclareMathOperator*{\essinf}{ess\,inf}
\DeclareMathOperator{\sign}{sign}

\newcommand{\hiddenfootnote}[1]{%
    \begingroup%
    \let\thefootnote\relax\footnotetext{#1}%
    \endgroup%
}

\def\Pr{{\rm \mathbf {Pr}}}

\newcommand{\Pm}{\mathcal{P}}

\newcommand{\X}{\mathcal{X}}
\newcommand{\Y}{\mathcal{Y}}
\newcommand{\Z}{\mathcal{Z}}
\newcommand{\ml}[2]{\mathcal{L}\left(#1  \!\!  \to  \!\!   #2\right)} 
\newcommand{\eps}{\varepsilon}

\newcommand{\cP}{\mathcal{P}}
\newcommand{\cJ}{\mathcal{J}}

\newcommand{\ds}{\displaystyle}
\newcommand{\argmin}{\operatornamewithlimits{argmin}}
\newcommand{\argmax}{\operatornamewithlimits{argmax}}

\newcommand{\range}{(0,1) \cup (1,\infty)}

\newcommand{\revised}[1]{{\color{black} #1}}

\hypersetup{
    colorlinks,
    linkcolor={blue!50!black},
    citecolor={blue!50!black},
    urlcolor={blue!50!black}
}

\begin{document}

\title{Sibson $\alpha$-Mutual Information and its Variational Representations}

\author{Amedeo Roberto Esposito, Michael Gastpar, Ibrahim Issa}
\hiddenfootnote{A.R. Esposito is with the Okinawa Institute of Science and Technology (OIST), Onna-son, Japan.}\hiddenfootnote{M. Gastpar is with the School of Computer and Communication Sciences, {\'E}cole Polytechnique F{\'e}d{\'e}rale de Lausanne (EPFL), Lausanne, Switzerland.} \hiddenfootnote{I. Issa is with the American University of Beirut.}

\hiddenfootnote{This paper was presented in part at the {\it 2024 IEEE International Symposium on Information Theory,} Athens, Greece.}

\maketitle

\begin{abstract}
Information measures can be constructed from R\'enyi divergences much like mutual information from Kullback-Leibler divergence.
One such information measure is known as Sibson $\alpha$-mutual information and has received renewed attention recently in several contexts: concentration of measure under dependence, statistical learning, hypothesis testing, and estimation theory.
In this paper, we survey and extend the state of the art. In particular, we introduce variational representations for Sibson $\alpha$-mutual information and employ them in each described context to derive novel results. Namely, we produce generalized Transportation-Cost inequalities and Fano-type inequalities.  We also present an overview of known applications, spanning from learning theory and Bayesian risk to universal prediction. 
\end{abstract}

\tableofcontents

\begin{table*}[tbp]
\small
  \setlength{\abovedisplayskip}{1.5ex plus0pt minus1pt}
  \setlength{\belowdisplayskip}{\abovedisplayskip}
  \fbox{%
  \begin{minipage}[t][60\baselineskip][t]{\textwidth}
   
    \centerline{{\large\textbf{Summary}}}
    \vspace{0.85\baselineskip}

    \begin{multicols}{2}
      \textbf{Definition:} If
      $\alpha \in (0,1) \cup\, (1,\infty),$
      \begin{align*}
I_\alpha(X,Y) &= D_\alpha(P_{XY}\|P_XQ^\star_Y) \\
 &=\frac{\alpha}{\alpha-1} \log \E_{P_Y} \left[\left(\E_{P_X}\left[\left(\frac{dP_{XY}}{dP_XP_Y}\right)^\alpha\right]\right)^\frac{1}{\alpha}\right],
      \end{align*}
      where
         $Q^\star_Y(y)  \propto \left( \E_{P_X}\left[ P^\alpha_{Y|X}(y|X) \right] \right)^{\frac{1}{\alpha}}.$
      %
      \begin{align*}
        I_1(X,Y) &= I(X;Y) = \text{ mutual information}\\
        I_\infty(X,Y) &= \log \E_{P_Y}\left[\esssup_{P_X} \frac{dP_{XY}}{dP_XP_Y}\right]
        = \text{maximal leakage.}
      \end{align*}
      When $X$ and $Y$ are discrete random variables,
\begin{align*} 
    &I_\alpha(X,Y) & = \frac{\alpha}{\alpha-1} \log \sum_{y \in \Y}  \left( \sum_{x \in \X} P_X(x) P_{Y|X}(y|x)^{\alpha} \right)^{1/\alpha}.
\end{align*}      %
      \textbf{Equivalent definition via norms
      }(\Cref{corollary:definitionvianorms}):
      \begin{equation*}
 I_\alpha(X,Y) = \log \left\|  \left\|  \frac{dP_{Y|X}}{dP_{Y}}  \right\|_{L^{\alpha-1} \left(P_{X|Y}\right)}  \right \|_{L^{\frac{\alpha-1}{\alpha}}(P_Y)}.
      \end{equation*}
      
    \begin{samepage}
      \textbf{Elementary properties
      }(\Cref{thm:properties}): For $\alpha \in
      (0,\infty]$,
      \begin{enumerate}
        \item Asymmetry: $I_\alpha(X,Y)\neq I_\alpha(Y,X)$\label{prop:Tableasymm};
        \item Continuity in $\alpha$ of the mapping $\alpha\to I_\alpha(X,Y)$;
        \item Non-decreasability in $\alpha$: if $\alpha_1\leq \alpha_2$ then $I_{\alpha_1}(X,Y)\leq I_{\alpha_2}(X,Y)\leq \ml{X}{Y}$\label{prop:TablenonDecr};
        \item Non-negativity: $I_\alpha(X,Y)\geq 0$ with $I_\alpha(X,Y)=0$ if and only if $X$ and $Y$ are independent\label{prop:TablenonNeg};
        \item Additivity: if $\{(X_i,Y_i)\}_{i=1}^n$ is a set of independent pairs of random variables then $I_\alpha(X^n,Y^n)=\sum_{i=1}^n I_\alpha(X_i,Y_i)$\label{prop:Tableadd};  
        \item Data-Processing Inequality:  if $X-Y-Z$ forms a Markov Chain then, $I_\alpha(X,Z)\leq \min \{I_\alpha(X,Y),I_\alpha(Y,Z)\}$;
        \item Invariance:  If $f(\cdot)$ and $g(\cdot)$ are injective, then $I_\alpha(f(X),g(Y)) = I_\alpha(X,Y)$.
      \end{enumerate}
    \end{samepage}


    \begin{samepage}
      \textbf{Bounds via R\'enyi entropy }(\Cref{thm:boundsviaRenyi}): If $P_{XY}$ admits a pmf,
\begin{equation*}
    I_\alpha(X,Y) \leq \min \{ H_{\frac1\alpha}(X), H_{\frac1\alpha}(Y) \}.
\end{equation*}
    \end{samepage}


    \begin{samepage}
      \textbf{Convexity} (\Cref{thm:convexityproperties}): $I_\alpha(X,Y)$ is
      \begin{itemize}
      \item convex in $P_{Y|X}$ for $\alpha <1$ (for fixed $P_X$),
      \item concave in $P_X$ for $\alpha \geq 1$ (for fixed $P_{Y|X}$),
      \end{itemize}
     \end{samepage}

    \begin{samepage}
    \hspace{1em} while $\exp\left(\frac{\alpha-1}{\alpha}I_\alpha(X,Y)\right)$ is
      \begin{itemize}
      \item for $\alpha<1 $: concave in $P_{Y|X}$ (for fixed $P_X$),
      \item for $\alpha \in (0,1):$  convex in $P_X$ (for fixed $P_{Y|X}$),
      \item for $\alpha \in (1,\infty):$
convex in $P_{Y|X}$ (for fixed $P_X$),
concave in $P_X$ (for fixed $P_{Y|X}$).
      \end{itemize}
    \end{samepage}

      \textbf{Convexity
     } (\Cref{Lemma-SibsonViaKL}): For $\alpha \in (0,1) \cup\, (1,\infty)$,\\ $(1-\alpha)I_\alpha(X,Y)$ is concave in $\alpha.$
      \vspace{\belowdisplayskip}
      
      \textbf{Tensorization bound for conditionally independent observations }(\Cref{thm:tensorization}): For $\alpha \in (0,1) \cup\, (1,\infty)$, $\beta_i>0$ and $\sum_i 1/\beta_i=1,$
    \begin{align}
        (\alpha-1) I_\alpha(X, Y^n) \leq  \sum_{i=1}^n \left(\alpha - \frac{1}{\beta_i} \right) I_{\alpha \beta_i} (X,Y_i). \nonumber       
    \end{align}

      \textbf{Connection to Kullback-Leibler divergence }(\Cref{Lemma-SibsonViaKL}): For $\alpha \in (0,1) \cup\, (1,\infty)$,
\begin{align*}
   &(1-\alpha)I_\alpha(X,Y)\\ &= \min_{R_{XY}} \left\{ \alpha D(R_{XY}\|P_{XY}) + (1-\alpha) D(R_{XY}\|P_{X}R_Y) \right\}
\end{align*}
      \vspace{\belowdisplayskip}

\vspace{-1em}
\textbf{Variational Representation 1} (\Cref{thm-Sibson-var-alternative}): For $\alpha \in (0,1) \cup (1,\infty),$
 \begin{align*}
    I_\alpha(X,Y) &= \sup_{f:\X\times\mathcal{Y}\to\mathbb{R}} \frac{\alpha}{\alpha-1} \log \E_{P_{XY}}[\exp\left((\alpha-1)f(X,Y)\right)] \\
    &\hspace{6em} - \E_{R_Y^*} [\log \E_{P_X}[\exp\left(\alpha f(X,Y)\right)]] ,
    \end{align*}
where for $\alpha>1,$
$\frac{dR_Y^*}{dP_Y}(y) \propto \E_{P_{X|Y=y}} \left[\exp\left((\alpha-1)f(X,y)\right) \right]$
and for $0<\alpha<1,$ $R_Y^{*}(y) = Q_Y^\star(y).$
\vspace{\belowdisplayskip}

\textbf{Variational Representation 2} (\Cref{thm:varReprIalpha}): For $\alpha \in (1,\infty)$,
\begin{align*}
\exp\left(\frac{\alpha-1}{\alpha}I_\alpha(X,Y)\right)\! =\!\!\!\! \sup_{g:\X\times\mathcal{Y}\to\mathbb{R}^+}\!\!\ \frac{\E_{P_{XY}}[g(X,Y)]}{\max_y \left(\E_{P_X}[g^\beta(X,y)]\right)^\frac1\beta },
\end{align*}if $\alpha\in (0,1)$ replace $\sup$ with $\inf$, and $\max$ with $\min$.
\vspace{\belowdisplayskip}

      \textbf{Dependence vs. Independence }(\Cref{thm:probBoundIalpha}): Let $E$ be a measurable event and $E_y=\{ x: (x,y)\in E\}$. Then, for $\alpha>1,$
\begin{align*}    P_{XY}(E) \leq \max_y P_X(E_y)^\frac1\beta\exp\left(\frac{\alpha-1}{\alpha}I_\alpha(X,Y)\right),
\end{align*}
where $1/\alpha + 1/\beta =1.$ See~\Cref{comparison} for more details.\vspace{\belowdisplayskip}

\textbf{Transportation-Cost Inequality }(\Cref{thm:TpcIneqIalpha}):
Let $0<\alpha<1$, $\varphi(x)=\frac{1}{\alpha-1}\log(x)$, and $f:\X\times\Y\to\mathbb{R}$. If for all $y\in\mathbb{R}$, there exist $\kappa,c \in\mathbb{R}$ such that:
    \begin{equation*}
        \log\E_{P_X}[\exp(\kappa f(\cdot, y))] \leq \frac{\kappa^2 c}{2} - \text{Ent}^{P_{XY}}_\varphi(\exp((\alpha-1)\kappa f))
    \end{equation*}
    then
    \begin{equation*}
         \E_{P_{XY}}[f]-\E_{P_XP_Y}[f] \leq \sqrt{\frac{2cI_\alpha(X,Y)}{\alpha}}.
    \end{equation*}

\textbf{Fano-type inequality} (\Cref{thm:fano-like}): Let $X$ be a discrete random variable, $p^\star = \max_x P_X(x)$ and $\hat{X}_{\mathrm{MAP}}$ be the optimal estimator of $X$ given $Y$, and $\hat{p}=\Pr(X = \hat{X}_{\mathrm{MAP}})$. Then, for all $\alpha >1$ and $\gamma >0$,
    \begin{equation*}
        \begin{split}
             \hat{p} \leq  \frac{\left(p^\star(\gamma+1)^{\frac{\alpha}{\alpha-1}}+ \!\! 1-p^\star\right)^{\frac{\alpha-1}{\alpha}} \!\! \exp \! \left \lbrace \frac{\alpha-1}{\alpha} I_\alpha(X,Y) \right\rbrace -1 }{\gamma}
        \end{split}
    \end{equation*} 
    
    \textbf{Bayesian Risk Lower Bound} 
 (\Cref{thm:sibsMIResultBayesRisk}):
For every $\alpha>1$ and $\rho>0,$ the Bayesian risk in estimating $W$ from $X$ must satisfy
	\begin{align*}
	R_B\geq \rho\left(1- \exp\left(\frac{\alpha-1}{\alpha}\left(I_\alpha(W,X) + \log(L_W(\rho))\right) \right)\right),
    \end{align*}
    where $L_W(\rho)$ denotes the small-ball probability of $W$. 
\vspace{\belowdisplayskip}
\vspace{\belowdisplayskip}

    \end{multicols}
  \end{minipage}
   }
\end{table*}

\section{Introduction}

Shannon entropy, Kullback-Leibler (KL) divergence, and mutual information play a central role in information theory and appear in a wide array of applications including (but not limited to) hypothesis testing~\cite{chernoff:52}, gambling~\cite{kelly:56,latane:67}, probability distribution estimation~\cite{perez:2008}, and machine learning~\cite{tishby:2000}, among others~\cite{dembo:2009,cover:2006}. The R\'enyi generalization of Shannon entropy and KL divergence~\cite{renyi:61} have subsequently extended the corresponding results~\cite{renyiDiv}, and found new applications in guessing problems~\cite{arikan:96}, cut-off rates in block coding~\cite{csiszar:95}, nonparametric density estimation~\cite{vandegeer:93}, image ranking~\cite{hero:2002}, etc. As such, despite the existence of other generalizations, the R\'enyi generalizations are well established and well documented in the information theory literature.

By contrast, there are multiple proposals for (R\'enyi type) generalizations of mutual information~\cite{Sibson:69,arimoto:77,csiszar:95,verdu:2015,lapidothP:2019}, with no consensus in the literature on the ``correct'' one. We are interested in Sibson $\alpha$-mutual information~\cite{Sibson:69}, which is given by 
\begin{align}
    \label{eq:Sibson-Intro}
    I_\alpha(X,Y) =  \frac{\alpha}{\alpha-1} \log \E_{P_Y} \left[\left(\E_{P_X}\left[\left(\frac{dP_{XY}}{dP_XP_Y}\right)^\alpha\right]\right)^\frac{1}{\alpha}\right],
\end{align}
for $\alpha \in \range$. $I_\alpha(X,Y)$ considers an $\alpha$-th moment of the Radon-Nikodym derivative $ \frac{dP_{XY}}{dP_XP_Y}$ \revised{with respect to $P_X$ (and then averaged over $P_Y$)}. Equivalently, it is related to the cumulant generating function of $ \log \frac{dP_{XY}}{dP_XP_Y}$, the expectation of which yields (Shannon) mutual information; in this sense, it is a natural quantity to study.

This paper is intended to both make a case for the adoption of Sibson $\alpha$-mutual information~\cite{Sibson:69} as the ``standard'' generalization, and provide a reference document for it.  As such, we review (and in some cases strengthen) existing results concerning Sibson $\alpha$-mutual information, as well as provide new results.  

We give special attention to the notion of variational representations --- a powerful tool for the study of information measures, allowing us to employ them in many different settings and fields, including: 
functional analysis~\cite{DonskerVaradhan:75,Varadhan:84,Liu:18,LiuCCPV:18,Raginsky:16,anantharam:18,LiuCCV:20,Esposito:22,KurriKS:22,EspositoG:22}, probability theory~\cite{sanov:58,Varadhan:84,Marton:96,Marton:96b,Marton:98,RaginskyS:14,EspositoGI:2021,EspositoWG:21, EspositoM:23,RaginskyS:14}, geometry~\cite{amari:2000}, statistics~\cite{GuoSV:05,XuR:17trans,XuR:22,EspositoVG:22,EspositoVG:23}, etc. 
\revised{A well-known variational characterization} is the Donsker-Varadhan representation of the Kullback-Leibler divergence, originally stated in~\cite{DonskerVaradhan:75} (see also~\cite[Section 10]{Varadhan:84}). This simple but fundamental equality has been instrumental in providing groundbreaking results in probability theory and concentration of measure via the connection (through functional analytical arguments) to Wasserstein and total variation  distances~\cite{Bobkov:99,Marton:86,Marton:96,Varadhan:84,Vershynin:18,RaginskyS:14,RassoulS:15}. Moreover, it has recently been used to provide bounds on the exploration bias~\cite{RussoZ:16} and on the expected generalization error of learning algorithms~\cite{XuR:17}. Variational representations have also been found for R\'enyi $\alpha$-divergences~\cite{atar:15,anantharam:18} and $f$-divergences~\cite{BroniatowskiK:10,NguyenWJ:08}. Leveraging said representations allowed for the connection of most divergences to both transportation-cost inequalities and, consequently, the expected generalisation error of learning algorithms~\cite{EspositoG:22, Esposito:22}. 
Although Sibson $\alpha$-mutual information stems from R\'enyi $\alpha$-divergence, it does not currently possess a variational representation mimicking the existing ones. In this work, among the other contributions, we will provide various novel variational characterizations for Sibson $\alpha$-mutual information: 
\begin{itemize}
    \item one similar to the one provided for the Kullback-Leibler divergence~\cite{renyiDiv}, see~\Cref{Lemma-SibsonViaKL};
    \item one similar to the one advanced for R\'enyi divergences~\cite{atar:15,anantharam:18}, see~\Cref{thm-Sibson-var-alternative,thm:varReprIalpha}.
\end{itemize}

Moreover, we also attempted to provide a representation similar to the one\footnote{Similar representations, stemming from the one proposed for Maximal Leakage, have also been advanced in~\cite{LiaoKSC:19,KurriKS:22,KurriSK:22}. However, these representations typically involve other generalized information measures (such as Arimoto's mutual information) or entirely novel objects.  } used to define Maximal Leakage~\cite{MaximalLeakage:20} which resulted in~\Cref{thm:markovChainRepresentation}. However, for this representation, we cannot actually prove equality but only an inequality.

The outline of the paper is as follows. 
\Cref{sec:prelim} provides a brief review of basic properties of the R\'enyi entropy and divergence.
\Cref{sec:definition} reviews the definition of Sibson $\alpha$-mutual information from several perspectives, including a definition via norms. It also provides a number of basic examples.
\Cref{sec:propertiesknownresults} reviews the basic properties of Sibson $\alpha$-mutual information, thereby adding a number of novel properties to the state of the art. The properties show that Sibson $\alpha$-mutual information is consistent with an axiomatic view of a generalized mutual information measure (e.g., data processing inequality, zero if and only if $X$ and $Y$ are independent, boundedness for finite $X$ or $Y$, etc.).
\Cref{sec:variational} contains new contributions in the shape of several variational representations of Sibson $\alpha$-mutual information.
Applications are presented in~\Cref{sec:concentration} and~\Cref{sec:estimationTheory}. In~\Cref{sec:concentration}, applications to concentration of measure are introduced. In~\Cref{sec:estimationTheory}, applications to estimation theory are presented, including Fano-type inequalities and lower bounds on the Bayesian risk.

Extensions are studied in~\Cref{sec:conditional} and~\Cref{sec:discussion}.
Specifically, \Cref{sec:conditional} considers the problem of defining a conditional version of Sibson $\alpha$-mutual Information. Several definitions can be advanced and in~\Cref{sec:conditional}, rather than advocating for a specific choice, we propose a principled method to endow every possible definition with an operational meaning in a corresponding hypothesis testing problem. To conclude,~\Cref{sec:discussion} considers the problem of extending Sibson $\alpha$-mutual information to negative values of $\alpha$ and takes the first steps towards proposing yet another variational representation that is closer in spirit to the one used to define Maximal Leakage.

Of the results presented, the following are, to the best of our knowledge, entirely novel:~\Cref{thm:boundsviaRenyi},~\Cref{thm:tensorization}, the results in~\Cref{sec:variational},~\Cref{thm:TpcIneqIalpha},~\Cref{thm:expGenErrSibson},~\Cref{thm:fano-like},~\Cref{corr:fano-like-gamma-inf},~\Cref{thm:fano-alpha-Arimoto},~\Cref{thm:genFano} and~\Cref{thm:markovChainRepresentation}.




\section{Preliminaries}\label{sec:prelim}

We briefly review R\'enyi entropy, conditional R\'enyi entropy, and R\'enyi divergence, as they will be useful in the remainder of the paper. 

\subsection{R\'enyi Entropy} \label{sec:renyi-entropy}

R\'enyi entropy is a generalization of Shannon entropy; in fact it is the unique family of entropy measures that satisfy the additivity property (under some mild assumption on the form of the measure; for more details see~\cite{renyi:61} and~\cite{daroczy:64}).

\begin{definition}[R\'enyi Entropy] \label{def:renyi-entropy}
    Given a discrete random variable $X \in \X$ and $\alpha \in \range$, the R\'enyi entropy of order $\alpha$ is given by
    \begin{align}
        \label{eq:def-renyi-entropy}
        H_\alpha(X) = \frac{1}{1-\alpha} \log \sum_{x \in \X} P_X(x)^\alpha.
    \end{align}
\end{definition}
It is easy to check that R\'enyi entropy could be rewritten in terms of $L_p$ norms:
\begin{align}
    \label{eq:renyi-entropy-norms}
    H_\alpha (X) = \frac{\alpha}{1-\alpha} \log \left\| P_X \right\|_\alpha.
  \end{align}  
For $\alpha \geq 1$, the quantity inside the $\log$ is a norm; but it is only a quasi norm for \revised{$\alpha \in (0,1)$} as it does not satisfy the triangle inequality. This form inspired Arimoto's defintion of conditional R\'enyi entropy~\cite{arimoto:77}:

\begin{definition}[Conditional R\'enyi Entropy] \label{def:cond-renyi}
    Given a pair of discrete random variables $(X,Y) \in \X \times \Y$ and $\alpha \in \range$, the conditional R\'enyi entropy of $X$ given $Y$ of order $\alpha$ is given by 
    \begin{align}
        \label{eq:def-cond-renyi}
        H_\alpha(X | Y) & = \frac{\alpha}{1-\alpha} \log \E_{P_Y} \left[  \left\| P_{X|Y}(.|Y) \right\|_\alpha \right] \\
        & = \frac{\alpha}{1-\alpha} \log \sum_{y \in \Y} \left( \sum_{x \in \X} P_{XY}(x,y)^\alpha \right)^{\frac{1}{\alpha}}.
    \end{align}
\end{definition}

\subsection{R\'enyi Divergences} \label{sec:renyi-divergence}

R\'enyi divergences~\cite{renyi:61} generalize KL divergence and are central to the study of information measures. They also play a pivotal role in generalizing mutual information, like in the case of Sibson $\alpha$-mutual information. We follow the conventions discussed by van Erven and Harremo\"es~\cite{renyiDiv} (in their review of R\'enyi and Kullback-Leibler divergences).

\begin{definition} \label{def:renyidiv}
    Given two distributions $P$ and $Q$ on a common alphabet $\X$, and $\alpha \in (0,1) \cup (1, \infty)$, 
    the R\'enyi divergence of order $\alpha$ between $P$ and $Q$ is defined as follows:
    \begin{align} \label{eq:def-renyi}
        D_\alpha (P \| Q) = \frac{1}{\alpha-1} \log \int_{\X} p^{\alpha} q^{1-\alpha} d\mu,
    \end{align}
    where $\mu$ is any measure satisfying $P \ll \mu$ and $Q \ll \mu$,  and $\ds p = \frac{dP}{d \mu}$ and $\ds q = \frac{dQ}{d \mu}$ are the Radon-Nikodym derivatives with respect to $\mu$.  
    For $\alpha \in \{0,1,\infty\}$, we define $D_\alpha$ by continuous extensions:
    \begin{align}
        D_0 (P\| Q) & = \lim_{\alpha\rightarrow 0 } D_\alpha( P \| Q),  \\
         D_1(P\| Q) & = \lim_{\alpha\uparrow 1 } D_\alpha( P \| Q),  \\
         D_\infty( P \| Q) & = \lim_{\alpha \rightarrow \infty }D_\alpha( P \| Q). \label{eq:def-renyidiv-01inf}
    \end{align}
    \end{definition}

A few remarks about the definition are in order. First, it is easy to check that~\Cref{eq:def-renyi} does not depend on the choice of $\mu$ (as long as it dominates $P$ and $Q$). Second, it turns out that
\begin{align} \label{eq:renyidiv-1=KL}
    D_1(P\|Q) & = D(P\|Q),
\end{align}
where $D(\cdot \| \cdot)$ is the Kullback-Leibler divergence, so that R\'enyi divergence generalizes KL divergence. 
Furthermore, we may rewrite~\Cref{eq:def-renyi} by integrating with respect to $P$: 
\begin{align} \label{eq:renyidiv-wrtP}
    D_\alpha(P\|Q) &= \frac{1}{\alpha-1} \log \int_{\X} \left(\frac{p}{q} \right)^{\alpha-1} dP \\&= \frac{1}{\alpha-1} \log \E_P \left[ \left(\frac{p}{q} \right)^{\alpha-1} \right];
\end{align}
or with respect to $Q$ when $\alpha \in (0,1)$:
\begin{align}
    \label{eq:renyidiv-wrtQ}
     D_\alpha(P\|Q) &= \frac{1}{\alpha-1} \log \int_{\X} \left(\frac{p}{q} \right)^{\alpha} dQ\\ &= \frac{1}{\alpha-1} \log \E_Q \left[ \left(\frac{p}{q} \right)^{\alpha} \right];
\end{align}
or whenever $\ds \frac{dP}{dQ}$ exists, we can integrate with respect to either $P$ or $Q$:
\begin{align} \label{eq:renyidiv-PllQ}
    D_\alpha(P\|Q) &= \frac{1}{\alpha-1} \log \E_P \left[ \left( \frac{dP}{dQ} \right)^{\alpha-1} \right]\\ &= \frac{1}{\alpha-1} \log \E_Q \left[ \left( \frac{dP}{dQ} \right)^\alpha \right]. 
\end{align}
Finally, if $\X$ is a finite or countably infinite alphabet, the definition reduces to:
\begin{align}
    \label{eq:renyidiv-pmfs}
    D_\alpha(P\|Q) = \frac{1}{\alpha-1} \log \sum_{x \in \X} P(x)^\alpha Q(x)^{1-\alpha}.
\end{align}

As for the limiting cases, we have~\cite[Theorem 4, Theorem 6]{renyiDiv}
\begin{align}
    \label{eq:renyi-zero-inf}
    D_\infty(P||Q) & =  \log \left( \esssup_{P} \frac{p}{q} \right), \\
    D_0 (P||Q) & = - \log Q \left( p >0 \right). 
\end{align}

\subsection{Variational Representations of KL and R\'enyi divergence}

\revised{The Donsker-Varadhan variational representation of KL divergence~\cite[Section 10]{Varadhan:84}  can be stated as follows: 
let $P$ and $Q$ be two probability measures such that $P$ is absolutely continuous with respect to $Q$ (\textit{i.e.}, such that if for some event $E$, $Q(E)=0$ then $P(E)=0$ as well), one can show that~\cite[Theorem 5.6]{RassoulS:15}}
 \begin{equation}
     D(P\|Q) = \sup_{f\in B(\X)} \E_{P} \left[f \right] - \log \E_{Q} \left[ e^f \right],
 \end{equation}
 where $B(\X)$ is the space of bounded functions acting on $\X$.

One can derive analogous results for R\'enyi divergences.  First, one can connect R\'enyi divergence to KL divergence via the following lemma.

\begin{lemma}\label{prop:RenyiViaKL}
    For all $\alpha \in (0,1)\cup (1,\infty),$
\begin{align}
   (1-\alpha)D_\alpha (P\|Q) = \inf_R \{\alpha D(R\|P) + (1-\alpha) D(R\|Q) \}.
\end{align}
Furthermore, the objective function in braces is a convex and continuous function of $R.$
\end{lemma}

In this lemma, continuity should be understood in the sense of the discussion in~\cite[Section III.D]{renyiDiv}.
This lemma is well known, see e.g.~\cite[Theorem 30]{renyiDiv} or~\cite{shayevitz:2011}. For completeness, a proof is included in~\Cref{app:prop:RenyiViaKL}. Moreover, R\'enyi divergence admits a representation akin to the Donsker-Varadhan representation of KL divergence, first observed by~\cite{atar:15}. Indeed, as done in~\cite{anantharam:18}, one can utilize~\Cref{prop:RenyiViaKL} to derive:

\begin{lemma}[Variational Representation of R\'enyi divergence]\label{lemma-variationalRenyiDiv}
    For all $\alpha \in (0,1)\cup (1,\infty) $
\begin{equation}
    \begin{split}
        \frac{1}{\alpha} D_\alpha (P||Q)= \sup_f \frac{1}{\alpha-1} \log \E_P \left[ e^{(\alpha-1)f} \right]- \frac{1}{\alpha} \log \E_Q \left[e^{\alpha f} \right].
    \end{split}
\end{equation}
    
\end{lemma}

A simplified proof of this lemma is included in~\Cref{app:lemma-variationalRenyiDivProof}.

\section{Definition}\label{sec:definition}
\revised{We begin by stating the definition of Sibson’s $\alpha$-mutual information.}

\begin{definition}\label{def-SibsonalphafromDiv}
Let $P_{XY}$ be a joint distribution over $\X\times\Y$ and denote with $P_XP_Y$ the corresponding product of the marginals. Let $\alpha \in [0,\infty] $, then the Sibson $\alpha$-mutual information (sometimes also referred to as Sibson mutual information of order $\alpha$) between $X$ and $Y$ is defined as follows:
\begin{align}
    I_\alpha (X,Y) = \min_{Q_Y \in P(\Y)} D_\alpha (P_{XY} \| P_X Q_Y), \label{eq:defIalphaMin}
\end{align}
where $P(\Y)$ denotes the set of probability distributions over $\Y$. 
\end{definition}
The definition is a generalization of Shannon's mutual information, as $I_1(X,Y) = I(X;Y)$.
\revised{
\begin{remark}
It is possible to see that~\Cref{def-SibsonalphafromDiv} represents a generalization of the notion of Information Radius that Sibson built starting from R\'enyi divergence~\cite[Definition 2.1]{Sibson:69}:

    Let $(\mu_1,\ldots,\mu_n)$ be a family of probability measures and $(w_1,\ldots, w_n)$ be a set of weights satisfying $w_i\geq 0$ for $i=1,\ldots,n$ and such that $\sum_{i=1}^n w_i>0$. For $\alpha\geq1$, the information radius of order $\alpha$ is defined as:
    \begin{align*}
       \frac{1}{\alpha-1}\min_{\nu\ll\sum_iw_i\mu_i}\log\left(\sum_i w_i\exp((\alpha-1)D_\alpha(\mu_i\|\nu))\right)  \label{infoRadius}.
    \end{align*}
In particular, for illustrative purposes, consider a discrete setting. Assume that one has two random variables $X,Y$ jointly distributed according to $P_{XY}$, one can think of $\mu_i = P_{Y|X=i}$ and $w_i = P_X(i)$. Leveraging then the geometric averaging present in R\'enyi’s information measures \textit{i.e.},
\begin{align}
    &\hspace{-1em}D_\alpha(P_X P_{Y|X}\|P_X Q_Y) = \frac{1}{\alpha-1}\log \mathbb{E}_{P_X}\left[\exp\left((\alpha-1)D_\alpha(P_{Y|X}\|Q_Y \right)\right],
\end{align}  
one can see that the information radius is a special case of~\Cref{eq:defIalphaMin} which we consider as our de-facto definition of Sibson $\alpha$-mutual information.
\end{remark}}
Note that $D_\alpha (P_{XY}|| P_XQ_Y)$ is convex in $Q_Y$~\cite[Theorem 12]{renyiDiv}. Indeed, one can derive a closed-form expression for the minimizer $Q^\star_Y$ and, consequently, a simple closed-form expression for $I_\alpha(X,Y)$:
\begin{theorem} \label{thm:Ialpha} Consider $\alpha \in \range$. Then
    \begin{align}
        I_\alpha(X,Y) &= D_\alpha(P_{XY}\|P_XQ^\star_Y)\label{thm:Ialpha-Eqn-defwithQstar}\\
        &=  \frac{\alpha}{\alpha-1} \log \E_{P_Y} \left[\left(\E_{P_X}\left[\left(\frac{dP_{XY}}{dP_XP_Y}\right)^\alpha\right]\right)^\frac{1}{\alpha}\right],\label{thm:Ialpha-Eqn-explicitdef}
    \end{align}
    where \begin{align} \label{eq:qstarIalpha}
        \frac{d Q^\star_Y}{d P_Y} = \frac{\E_{P_X} \left[   \left(  \frac{d P_{XY}}{ d P_{X} P_Y} \right)^\alpha \right]^{\frac{1}{\alpha}} }{ \E_{ P_Y} \left[ \E_{P_X} \left[   \left(  \frac{d P_{XY}}{ d P_{X} P_Y} \right)^\alpha \right]^{\frac{1}{\alpha}} \right] }.
    \end{align}
    If $\alpha \geq 1$ and $P_{XY}$ is not absolutely continuous with respect to $P_X P_Y$, then $I_\alpha(X,Y) = + \infty$.
   Moreover,
   \begin{align}
       I_0(X,Y) & =  - \esssup_{P_Y} \log P_X \left( \frac{dP_{XY}}{dP_XP_Y}(X,Y) > 0  \right), \\
       \text{and } I_\infty(X,Y) & =  \lim_{\alpha \to \infty} I_\alpha(X,Y)\nonumber \\  &= \log \E_{P_Y} \left[  \esssup_{P_X}  \frac{dP_{XY}}{dP_XP_Y} \right].
   \end{align}
\end{theorem}
The proof of this theorem is given in~\Cref{app:proof:thm:Ialpha}.
\begin{remark}
    A more ``rigorous'' way of writing~\Cref{thm:Ialpha-Eqn-explicitdef} would be to consider a dominating measure $\mu$ and then write the Radon-Nikodym derivatives with respect to this measure. However, we opt for the above form as it is more illustrative, while measure-theoretic issues may hinder readability. 
\end{remark}
\begin{remark}
    \Cref{eq:qstarIalpha} may not be well-defined if the denominator is infinite, in which case indeed $I_\alpha(X,Y) = \infty$.
\end{remark}

If $X$ and $Y$ are discrete random variables, then the expression simplifies to
\begin{align} 
   I_\alpha(X,Y) & = \frac{\alpha}{\alpha-1} \log \sum_{y \in \Y} P_Y(y) \left( \sum_{x \in \X} P_X(x) \left( \frac{P_{Y|X}(y|x)}{P_Y(y)} \right)^{\alpha} \right)^{1/\alpha} \nonumber \\
    & = \frac{\alpha}{\alpha-1} \log \sum_{y \in \Y}  \left( \sum_{x \in \X} P_X(x) P_{Y|X}(y|x)^{\alpha} \right)^{1/\alpha},\label{eq:Ialpha-discrete-formula}
\end{align}
where, without loss of generality, we assumed that $P_Y$ has full support over $\Y$.
Moreover, the minimizing probability mass function in~\Cref{thm:Ialpha-Eqn-defwithQstar} can be expressed as
\begin{align}\label{eq:qstarIalpha:discrete}
    Q_Y^\star(y) &=  \frac{\left(\sum_{x \in \X} P_X(x) P_{Y|X}(y|x)^{\alpha}\right)^{1/\alpha}}{\sum_{\tilde{y} \in \Y}\left(\sum_{x \in \X} P_X(x) P_{Y|X}(\tilde{y}|x)^{\alpha}\right)^{1/\alpha}}.
\end{align}

Similarly, if $X$ and $Y$ are jointly continuous random variables then we can replace sums with integrals and PMFs with pdfs in~\Cref{eq:Ialpha-discrete-formula}.
The closed-form expression can be extended to $\alpha=\infty$:
\begin{equation}
    I_{\infty}(X,Y) = \log \sum_{y\in \Y} \max_{\substack{x:\\ P_X(x)>0}} P_{Y|X=x}(y) = \ml{X}{Y}, \label{eq-def-maxleakage}
\end{equation}
where $\ml{X}{Y}$ denotes the maximal leakage between $X$ and $Y$~\cite{MaximalLeakage:20}. A more extensive treatise on the connection between Maximal Leakage and $I_\alpha$ is deferred to~\Cref{sec:max-leakage}.
\begin{remark}
    One could also attempt to extend the definition of Sibson $\alpha$-mutual information for negative values of $\alpha$. An informal discussion of this extension and its potential applications is provided in~\Cref{app:negativeAlpha}.  Indeed, assuming a closed-form expression analogous to the positive case, such an object could naturally arise in contexts like those discussed in~\Cref{sec:depVsIndep,sec:bayesianRisk}. However, a rigorous formulation would involve additional technical challenges beyond the scope of this work. We therefore refrain from a formal treatment and simply refer the reader to instances where such an extension has been used, \textit{e.g.},~\cite[Appendix D.2]{EspositoVG:23}.
\end{remark}

\subsection{Examples}

The closed-form expression for Sibson $\alpha$-mutual information makes it relatively simple to compute. We provide a few examples.

\begin{example}[Binary Symmetric Channel]
Assume that $X=\text{Ber}(1/2)$, and that $P_{Y|X=x}(y)= \epsilon\mathbbm{1}_{x\neq y}+(1-\epsilon)\mathbbm{1}_{x=y}$, \textit{i.e.} one has a Binary Symmetric Channel (BSC) with error probability $\epsilon$. If $Y$ is the outcome of $X$ after being passed through the BSC, one has that $Y=\text{Ber}(1/2)$. Then for $\alpha\in (0,+\infty) $
\begin{align}
    I_\alpha(X,Y)=I_\alpha(Y,X) & = \frac{1}{\alpha-1}\log\left(\frac{1}{2^{1-\alpha}}\left(\epsilon^\alpha+(1-\epsilon)^\alpha\right)\right)\nonumber\\
    &= \log 2 + \frac{1}{\alpha-1}\log\left(\epsilon^\alpha+(1-\epsilon)^\alpha\right) \\
  &= \log 2 - h_{2,\alpha}(\epsilon)\nonumber,
\end{align}
\revised{where $h_{2,\alpha}(\epsilon)$ denotes the binary R\'enyi Entropy of the distribution $(\epsilon,1-\epsilon)$}.

For the limiting case $\alpha=0,$ we find
\begin{align}    I_0(X,Y) = I_0(Y,X) = 0;
\end{align}
\revised{
for the case $\alpha=1$, \textit{i.e.}, mutual information, we find
\begin{align}
            I_1(X,Y)=I_1(Y,X)  &=I(X;Y)
             \nonumber \\ &= \log2+ (1-\epsilon)\log(1-\epsilon)+\epsilon\log(\epsilon)  \nonumber \\
            &= \log2 - h_2(\epsilon),
\end{align}
}
and for the limiting case $\alpha=\infty,$ we find
\begin{align}
    I_\infty(X,Y) = I_\infty(Y,X) &= \log \left( 2 \max\{ \epsilon, 1-\epsilon \}  \right)  \nonumber \\
    &= \log2 - h_{2,\infty}(\epsilon).
\end{align}
\revised{~\Cref{fig:BSCIalpha} depicts the behavior of $I_\alpha(X,Y)$ of a Binary Symmetric Channel with $\epsilon=1/4$ as a function of $\alpha$.}
\begin{figure}
    \centering
    \includegraphics[scale = 0.1]{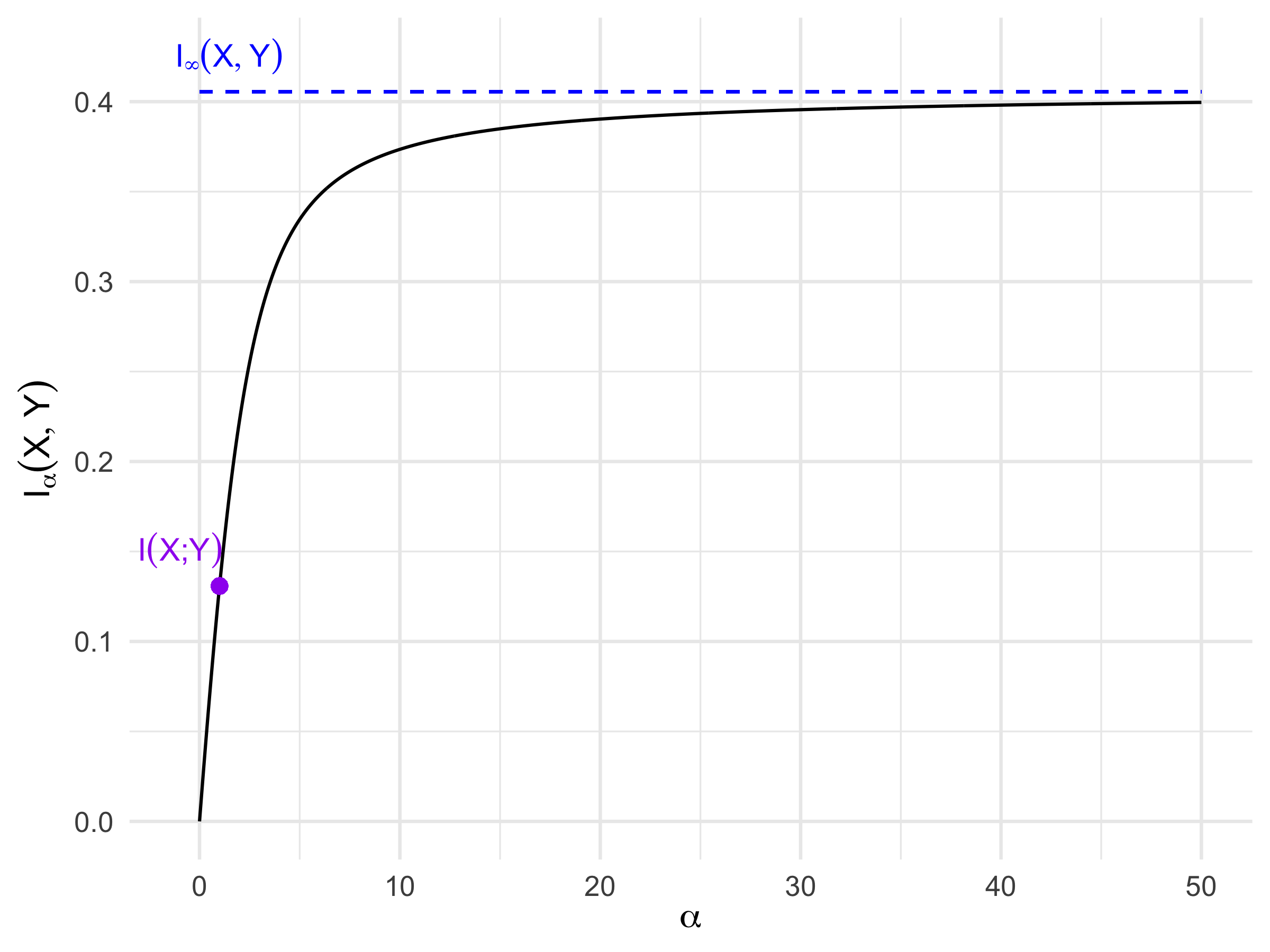}
    \caption{Behaviour of $I_\alpha(X,Y)$ as a function of $\alpha$ for the BSC($1/4$) with $X=\text{Ber}(1/2)$. The $\log$ is taken with base $2$. }
    \label{fig:BSCIalpha}
\end{figure}
\end{example}
The symmetry in the previous example is due to the choice of the joint distribution (a doubly symmetric binary source). However, $I_\alpha(X,Y)$ is not necessarily symmetric as the following example shows:
\begin{example}[Binary Erasure Channel] \label{ex:bec}
Assume that $X=\text{Ber}(1/2)$ and that $P_{Y|X=x}(y)=(1-\delta)\mathbbm{1}_{y=x}+\delta\mathbbm{1}_{y=e}$, \textit{i.e.}, one has a Binary Erasure Channel (BEC) with erasure probability $\delta$. Assume $Y$ is the outcome of $X$ after being passed through the BEC. Then, if $\alpha\in (0, +\infty)$
\begin{align}
    I_\alpha(X,Y) &= \frac{\alpha}{\alpha-1} \log(\delta +(1-\delta)2^{\frac{\alpha-1}{\alpha}})\\
    I_\alpha(Y,X) &= \frac{\alpha}{\alpha-1} \log\left(\left(\delta +(1-\delta)2^{\alpha-1}\right)^\frac1\alpha\right).
\end{align}
For the limiting case $\alpha=0,$ we find
\begin{align}    I_0(X,Y) = 0 \text{ and } I_0(Y,X) =\revised{\log\left(\frac{2}{1+\delta}\right)} ,
\end{align}
\revised{
for the case $\alpha=1$, \textit{i.e.}, mutual information, we find
\begin{align}
            I_1(X,Y)=I_1(Y,X)=I(X;Y) &= (1-\delta)\log2,
\end{align}
}
whereas
\begin{align}
    I_\infty(X,Y) = \log(2- \delta) \text{ and } I_\infty(Y,X) = \log 2.
\end{align}
Hence, this example illustrates the asymmetry of Sibson $\alpha$-mutual information.
\revised{~\Cref{fig:BECIalpha1} and ~\Cref{fig:BECIalpha2} depict the behavior of, respectively, $I_\alpha(X,Y)$ and $I_\alpha(Y,X)$ of a Binary Erasure Channel with $\delta=1/4$ as a function of $\alpha$.}
\begin{figure}
    \centering
    
    \begin{minipage}{0.47\textwidth}
        \centering
        \includegraphics[width=0.85\textwidth]{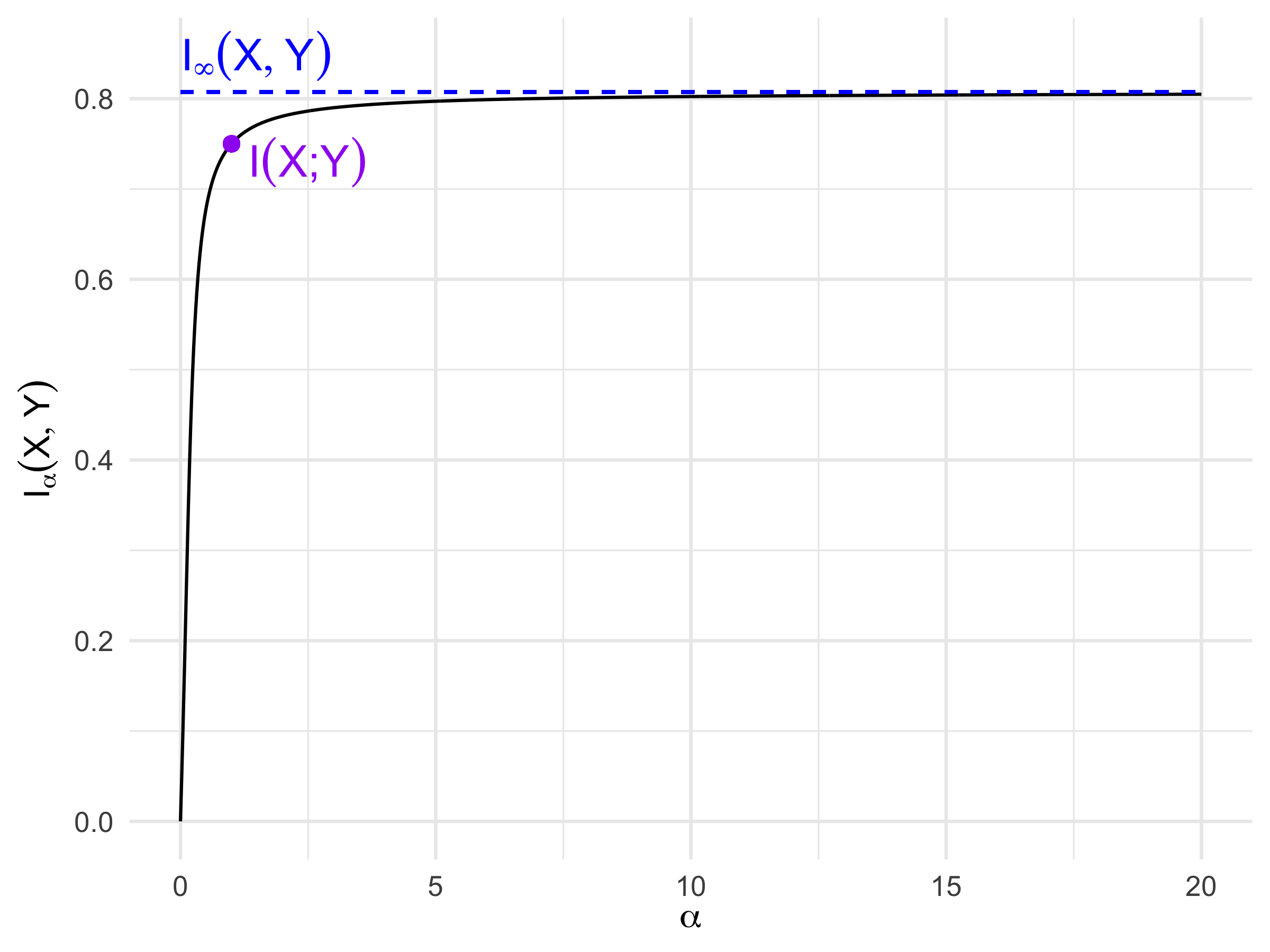}
        \caption{Behaviour of $I_\alpha(X,Y)$ as a function of $\alpha$ for the BEC($1/4$) with $X=\text{Ber}(1/2)$. The $\log$ is taken with base $2$.
        }
        \label{fig:BECIalpha1}
    \end{minipage}\hfill
    \begin{minipage}{0.47\textwidth} 
        \centering
        \includegraphics[width=0.78\textwidth]{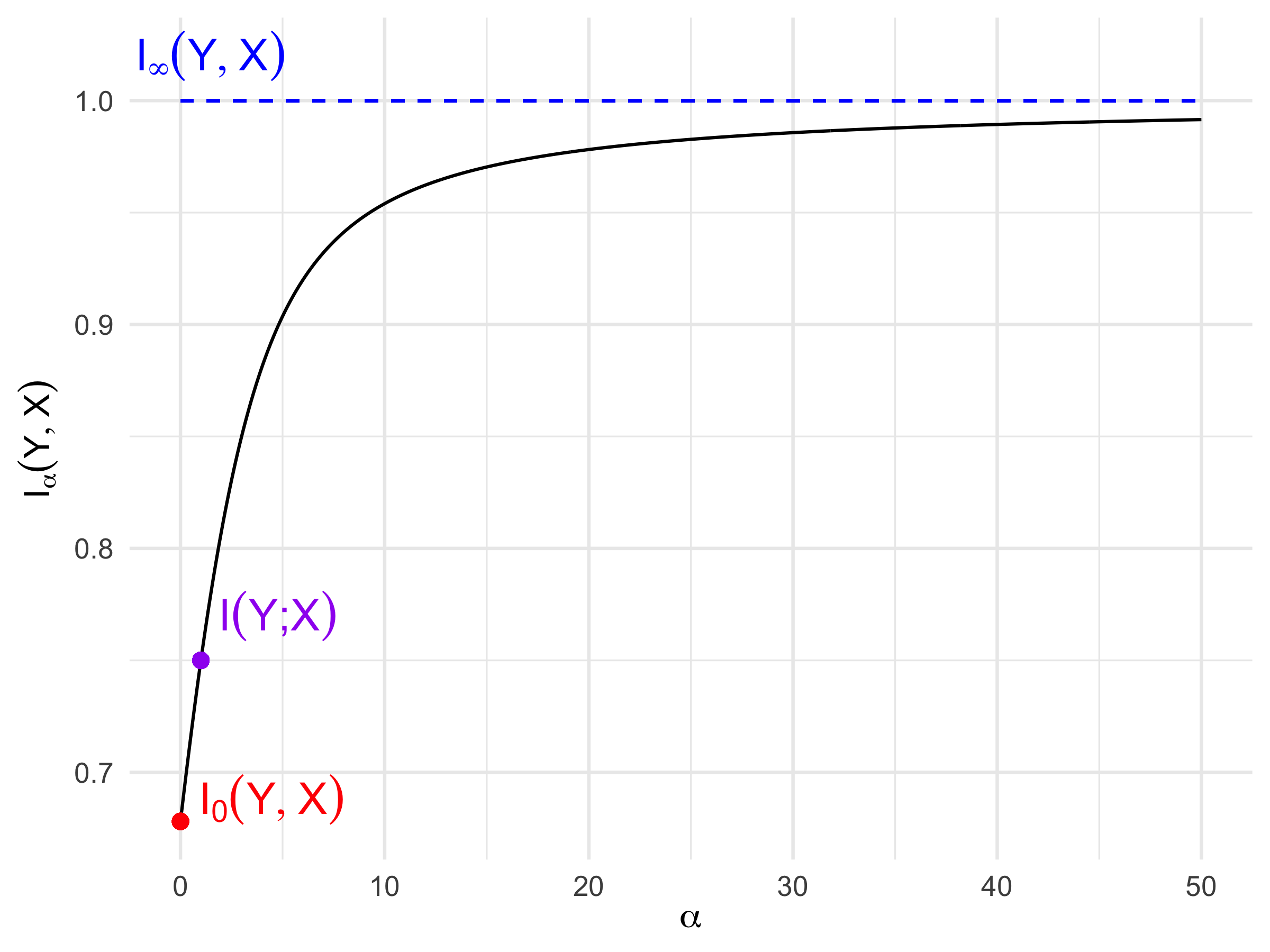}
      
        \caption{ Behaviour of $I_\alpha(Y,X)$ as a function of $\alpha$ for the BEC($1/4$) with $X=\text{Ber}(1/2)$. The $\log$ is taken with base $2$.
        }
        \label{fig:BECIalpha2}
    \end{minipage}
    \vspace{-1em}
\end{figure}
\end{example}
One can also compute the Sibson $\alpha$-mutual information for continuous random variables.
\begin{example}[Gaussian channel with Gaussian inputs]
Assume $X\sim \mathcal{N}(0,\sigma^2_X)$ \textit{i.e.}, $X$ is a Gaussian random variable with mean $0$ and variance $\sigma^2_X$. If $Y\sim \mathcal{N}(0,\sigma^2_Y)$ and $\alpha \in (0,+\infty)$
\begin{equation}
    I_\alpha(X,X+Y) = \frac{1}{2}\log\left( 1+\alpha \frac{\sigma^2_X}{\sigma^2_Y}\right).
\end{equation}
Clearly, for $\alpha=1,$ this recovers the classic and well known mutual information expression for jointly Gaussian random variables. Moreover, as $\alpha$ increases, $I_\alpha(X,X+Y)$ increases without bound irrespective of the values of $\sigma_X^2>0$ and $\sigma_Y^2.$
\end{example}

\revised{
  As a last example, let us consider a setting in which $I(X;Y)=+\infty$ but $I_\alpha(X,Y)<+\infty$ for $0<\alpha<1$.
  
\begin{example}
   Let $Y$ be a discrete random variable with support on $[2,+\infty)\cap \mathbb{N}$ and such that $P(y)= \frac{1}{C} \frac{1}{y(\log^2(y))}$ for $y\geq 2$ and with $C= 2.10974$. Let now $X=g(Y)$ with $g$ deterministic and bijective. One has that $I(X;Y)=H(Y)=\frac{1}{C}\sum_{y=2}^\infty \frac{1}{y\log^2(y)}\log\left(C y\log^2(y)\right)=+\infty$ (by the integral test).\\
However, \begin{align}
    I_\alpha(X,Y) = H_{1/\alpha}(Y)  
    &= \frac{\alpha}{\alpha-1} \log \sum_y P_Y(y)^\frac{1}{\alpha}  \nonumber \\
    &= \frac{\alpha}{\alpha-1} \log \sum_y \frac{1}{C^{\hat{\alpha}} y^{\hat{\alpha}} \log^{2\hat{\alpha}}(y)} < + \infty, \label{eq:sum}
\end{align}
where $\hat{\alpha}=1/\alpha$. Since $0<\alpha<1$ one has that $\hat{\alpha}>1$ and the sum in~\Cref{eq:sum} converges for every $0<\alpha<1$.
\end{example}
}
\subsection{Definitions Via (Quasi-)Norms}\label{sec:normPerspective}

Another insightful perspective on R\'enyi $\alpha$-divergence and Sibson $\alpha$-mutual information is via norms. To that end, given a random variable $X$ and a function $f$, the $L_p$-norm with respect to $P_X$ is defined as follows, 
\begin{align} \label{def:norm-distribution}
    \| f(X) \|_{L^p(P_X)} &= \left( \E_{P_X} [ |f(X)|^p ] \right)^{1/p},
\end{align}
where $p \geq 1$. For $p \in (0,1)$, the above definition yields a quasi-norm. Now, Sibson $\alpha$-mutual information can be written as:
\revised{\begin{corollary}\label{corollary:definitionvianorms}
    For all $\alpha \in (0,1) \cup (1,\infty)$, we can express $I_\alpha(X,Y)$ as
    \begin{align}
         \frac{(\alpha-1) }{\alpha}I_\alpha(X,Y)  = \log \left\|    \left\|   \frac{dP_{XY}}{dP_XP_{Y}}  \right\|_{L^{\alpha}\left(P_X\right)}\right\|_{L^1(P_Y)}.  \label{eq:normRepresentationEveryAlpha}
         \end{align}
    For $\alpha >1$, the following alternative representation holds:
    \begin{align}
         I_\alpha(X,Y) = \log \left\|    \left\|   \frac{dP_{Y|X}}{dP_{Y}}  \right\|_{L^{\alpha-1}\left(P_{X|Y}\right)}  \right\|_{L^{\frac{\alpha-1}{\alpha}}(P_Y)}.\label{eq:normRepresentationAlphaLargerThanOne}
    \end{align}
\end{corollary}}
The following is well known (a proof is included for completeness in~\Cref{app:l0normProof}):

\begin{lemma}\label{lem:pnorm-nondecreasing}
For all $0 \le p\le q,$ we have
    $\| f \|_{L^p({P_X})} \le \| f \|_{L^q({P_X})}.$
\end{lemma}

When we combine the expressions given in~\Cref{corollary:definitionvianorms} with~\Cref{lem:pnorm-nondecreasing}, we can conclude that $I_\alpha(X,Y)$ is a non-decreasing function of $\alpha.$
Another feature of the expressions given in~\Cref{corollary:definitionvianorms} is a natural connection to Shannon information measures.
To this end, we leverage the following lemma (proved, for completeness, in~\Cref{app:l0normProof}).

\begin{lemma}\label{lem:l0norm}
    \begin{align}
    \lim_{p \rightarrow 0+} \| X \|_{L^p(P_X)} &= \exp \E_{P_X} \left[ \log |X| \right].
    \end{align}
\end{lemma}

\begin{corollary}
    If $I_\alpha(X,Y) < \infty$ for some $\alpha >1$, then
\begin{align}
    \lim_{\alpha \to 1^+} I_\alpha (X,Y) \nonumber &= \log \exp \E_{P_Y} \left[ \log \exp \E_{P_{X|Y}} \left[ \log \frac{dP_{Y|X}}{dP_Y}  \right]       \right]  \nonumber  \\ &= \E_{P_{XY}} \left[ \log \frac{dP_{Y|X}}{dP_Y} \right] = I(X;Y).
\end{align}
\end{corollary}

\section{Properties and Known Results}\label{sec:propertiesknownresults}

We now present properties and well-known results about Sibson $\alpha$-mutual information. We start by showing that it satisfies ``axiomatic'' properties of an information measure (e.g., non-negativity, data processing inequality, etc.).  We further consider convexity properties, behaviour of $I_\alpha$ in terms of $\alpha$, as well as tensorization results.

In~\Cref{sec:capacityAndCoding}, we characterise the so-called Sibson capacity and relate Sibson $\alpha$-MI to the zero-error capacity and the random coding exponent. We will conclude by presenting its connection to information leakage measures (\Cref{sec:leakage}) and other generalizations of Shannon's mutual information that stem from R\'enyi divergences (Aritmoto, Csisz\'ar, Lapidoth-Pfister,~\Cref{sec:otherRenyiInformations}). Some of the proofs for this Section can be found in~\Cref{app:proof:sec:propertiesknownresults}. Other proofs will be omitted to avoid meaningful overlap with the literature. The material contained in this Section can also be found in~\cite{csiszar:95,Sibson:69,verdu:2015,KurriSK:22,MaximalLeakage:20,shannon:56,SiuWaiV:15,LapidothP:18,aishwarya:19,wuWIS:20}.

\subsection{Basic properties}\label{sec:basicProperties}

\begin{theorem}[Axiomatic properties]\label{thm:properties}
    Given two jointly distributed random variables $X,Y$, and $\alpha\in (0,+\infty)$, Sibson $\alpha$-mutual information $I_\alpha(X,Y)$ satisfies the following set of properties:
    \begin{enumerate}[i)]
        \item Non-negativity: $I_\alpha(X,Y)\geq 0$ with $I_\alpha(X,Y)=0$ if and only if $X$ and $Y$ are independent\label{prop:nonNeg};
        \item Data processing inequality:  Let $X,Y,Z$ be three random variables jointly distributed according to $P_{XYZ}$. If $X-Y-Z$ forms a Markov Chain then, 
        \begin{equation} \label{thm:dpiIalpha}
        I_\alpha(X,Z)\leq \min \{I_\alpha(X,Y),I_\alpha(Y,Z)\};
    \end{equation}
        \item Invariance under injective transformations: Let $f:\X\to\tilde{\X}$ and $g:\Y\to\tilde{\Y}$ be two injective mappings, then \begin{equation} \label{thm:invarianceinjectivetransform}
    I_\alpha(f(X),g(Y))=I_\alpha(X,Y)
\end{equation}
     \item Additivity: if $\{(X_i,Y_i)\}_{i=1}^n$ is a set of independent pairs of random variables then $I_\alpha(X^n,Y^n)=\sum_{i=1}^n I_\alpha(X_i,Y_i)$\label{prop:add}.
    \end{enumerate}
\end{theorem}

One may argue that the above properties should be satisfied by any ``reasonable'' information measure, and they are indeed satisfied by Sibson's mutual information. They essentially follow from the properties of R\'enyi divergence. For completeness, a proof of this theorem is included in~\Cref{app:proof:thm:properties}. 

Moreover, intuitively, a finite random variable cannot have infinite mutual information with another variable. This holds true for Sibson $\alpha$-mutual information as well:

\begin{theorem}[Boundedness]\label{thm:boundsviaRenyi}
Let $X,Y$ be two random variables jointly distributed according to $P_{XY}$ and assume that $P_{XY}$ admits a pmf. If $\alpha\in (0,+\infty)$, then
\begin{equation}
    I_\alpha(X,Y) \leq \min \{ H_{\frac1\alpha}(X), H_{\frac{1}{\alpha}} (Y)\} ,\label{thm:boundsviaRenyi:boundX}
\end{equation}
where equality to $H_{\frac{1}{\alpha}} (Y)$ holds if and only if $Y$ is a deterministic function of $X$.
\end{theorem}

We note that, while the upper bounds appear in the literature, to the best of our knowledge, the condition after~\Cref{thm:boundsviaRenyi:boundX} has not been stated before. A proof is included in~\Cref{app:proof:boundsviaRenyi}.

In the following, we provide analytical properties of Sibson $\alpha$-mutual information, in terms of the parameter $\alpha$, as well as in terms of $P_X$ and $P_{Y|X}$.

\begin{theorem}[Elementary properties]\label{thm:properties-alpha}
    Given two jointly distributed random variables $X,Y$, and $\alpha\in (0,+\infty)$, Sibson $\alpha$-mutual information $I_\alpha(X,Y)$ satisfies the following set of properties:
    \begin{enumerate}[i)]
    \item Asymmetry: $I_\alpha(X,Y)\neq I_\alpha(Y,X)$\label{prop:asymm};
        \item Continuity in $\alpha$: the mapping $\alpha\to I_\alpha(X,Y)$ is continuous\label{prop:continuity};
        \item Non-decreasability in $\alpha$: if $\alpha_1\leq \alpha_2$ then $I_{\alpha_1}(X,Y)\leq I_{\alpha_2}(X,Y)\leq \ml{X}{Y}.$\label{prop:nonDecr}
    \end{enumerate}
\end{theorem}

For completeness, a proof of this theorem is included in~\Cref{app:proof:thm:properties-alpha}. It is particularly important to understand convexity properties for studying optimization problems involving Sibson $\alpha$-mutual information:

\begin{theorem}[Convexity properties]\label{thm:convexityproperties}
Given two random variables $X,Y$ jointly distributed according to $P_{XY}$, $I_\alpha(X,Y)$ satisfies the following convexity/concavity properties:
\begin{enumerate}[i)]

    \item if $\alpha\leq 1$ then, given $P_X$, $I_\alpha(X,Y)$ is convex in $P_{Y|X}$\label{prop:convexConditional};
    \item if $\alpha \geq 1$ then, given $P_{Y|X}$, $I_\alpha(X,Y)$ is concave in $P_X$\label{prop:concaveMarginal};
    \label{convexity} 
    \item $\exp\left(\frac{\alpha-1}{\alpha}I_\alpha(X,Y)\right)$ is \label{prop:convexExp}
    \begin{enumerate}
        \item  convex in $P_{Y|X}$ for a given $P_X$ if $\alpha\in (1,+\infty)$\label{prop:convexExpConditionalAlphaLarger1};
        \item  concave in $P_{Y|X}$ for a given $P_X$ if $\alpha\in (0,1)$\label{prop:convexExpConditionalAlphaSmaller1};
        \item concave in $P_X$ for a given $P_{Y|X}$ if $\alpha\in (1,+\infty)$\label{prop:concaveExpConditional};
        \item convex in $P_X$ for a given $P_{Y|X}$ if $\alpha\in (0,1)$\label{prop:convexExpConditional}.
    \end{enumerate}
\end{enumerate}
\end{theorem}
For completeness, a proof of this theorem is included in~\Cref{proof:convexityproperties}.

\revised{A statement resembling a data-processing inequality also holds for Sibson $\alpha$-mutual information, namely, for the special case of two independent random variables $X$ and $Z,$ and an arbitrarily dependent third random variable $Y.$}

\begin{theorem}[{\cite[Lemma 2]{10161608}}]\label{thm:Riouletal}
If $X$ and $Z$ are independent, then
\begin{align}
    I_\alpha(X, (Y,Z)) & \le \revised{I_\alpha((X,Z) , Y)}.
\end{align}
\end{theorem}
For completeness, a proof is included in~\Cref{app:proof:thm:Riouletal}.

Finally, we provide a (seemingly new) tensorization result \revised{for an ensemble of random variables satisfying a conditional independence property, linking Sibson $\alpha$-mutual information terms of various orders.}

\begin{theorem}[Tensorization bound in the conditionally independent case] \label{thm:tensorization}
    Let $Y_1, Y_2, \ldots, Y_n$ be conditionally independent given $X$. For $\alpha \in (0,1) \cup\, (1,\infty)$,
    \begin{align}
        (\alpha-1) I_\alpha(X, Y^n) \leq  \sum_{i=1}^n \left(\alpha - \frac{1}{\beta_i} \right) I_{\alpha \beta_i} (X,Y_i),        
    \end{align}
    where $\beta_i > 0$ for all $i$, and $\ds \sum_{i=1}^n \frac{1}{\beta_i} =1$.
\end{theorem}

A proof is included in~\Cref{app:proof:thm:tensorization}. \revised{The above bound can be seen as a composition result for Sibson $\alpha$-mutual information. Similar composition results for mutual information measures (including Sibson's) have been studied in~\cite{wuWIS:20}, where the limit and exponential rate of convergence are derived. As done here, these results also assume conditionally independent observations. By contrast, the above result differs by deriving finite $n$ bounds (as opposed to asymptotic bounds) and using different Sibson orders $I_{\alpha \beta_i}(X,Y_i)$. Considering the case of $n=2$ and taking the limit as $\alpha$ goes to infinity recovers the composition result of Maximal Leakage~\cite[Lemma 6]{MaximalLeakage:20}. Finite $n$ composition bounds have been derived for a variant of Maximal Leakage called Binary Maximal Leakage~\cite{cung2024binary}.

Moreover, composition results are particularly interesting in the context of differential privacy~\cite{dwork2010boosting,kairouz2015composition}. In that setup however, \emph{adaptive} composition is allowed, i.e., the observations are not conditionally independent (with constraints imposed on $P_{Y_i|X,Y_1 = y_1, \ldots, Y_{i-1}=y_{i-1}}$). Adaptive composition bounds for Sibson $\alpha$-mutual information are not discussed in the literature, except for Maximal Leakage with binary inputs~\cite{issa2022adaptive}.}
    

\subsection{Capacity and Coding Theorems}\label{sec:capacityAndCoding}

\revised{In this section, we briefly discuss applications of Sibson $\alpha$-mutual information in classical coding theorems.}

\subsubsection{Maximizing over Input Distributions}

For a fixed channel $P_{Y|X}$, the Shannon capacity is given by
\begin{align} \label{eq:max_MI}
    \sup_{P_X} I(X;Y) &= \sup_{P_X} \inf_{Q_Y} D(P_{XY} || P_X Q_Y) \\ &= \inf_{Q_Y} \max_{x \in \X} D(P_{Y|X=x} || Q_Y),
\end{align}
where in the last equality, we assumed $\X$ is finite. Maximizing Sibson $\alpha$-mutual information, which we may term the Sibson capacity, yields an analogous result:
\begin{theorem}[{\cite[Proposition 1]{csiszar:95}}]   \label{thm:sibson-capacity} Suppose $\X$ is finite and fix $P_{Y|X}$. Then for all $\alpha >0$,
\begin{align} \label{eq:thm-sibson-capacity}
    \sup_{P_X} I_\alpha (X,Y) = \inf_{Q_Y} \sup_{x \in \X} D_\alpha (P_{Y|X=x} || Q_Y).
\end{align}    
\end{theorem}
For $\alpha=1$, this is simply~\Cref{eq:max_MI}. As observed by Csisz\'ar~\cite{csiszar:95}, this endows Sibson capacity with a geometrical interpretation analogous to that of Shannon capacity: it is the smallest radius that includes all $P_{Y|X=x}$ ``points'', where ``distance'' is measured according to $D_\alpha$. Csisz\'ar's proof can be found in~\Cref{app:sibson-capacity-proof}.

For the special class of (weakly) symmetric channels, it is well known that mutual information is maximized by uniform inputs. This result carries over to Sibson $\alpha$-mutual information with a modification of the definition of symmetric channels. In particular, we introduce $\alpha$-weakly symmetric channels:

\begin{definition}[$\alpha$-weakly symmetric channels] A discrete input-output channel $P_{Y|X}$ is $\alpha$-weakly symmetric if all the rows $P_{Y|X}(.|x)$ are permutations of each other, and all the column $\alpha$-power sums $\sum_{x} P_{Y|X}(y|x)^\alpha$ are equal.
\end{definition}

Now, we get:

\begin{theorem}[Capacity of $\alpha$-weakly symmetric channels]\label{thm-symmetric-channel-cap}
    For $X,Y$ discrete and finite random variables and for any $\alpha \in (0,1)\cup (1,\infty),$
    if $P_{Y|X}$ is symmetric in the sense that for any $x_1$ and $x_2,$ $p(y|x_1)$ is a permutation of $p(y|x_2),$ then $I_\alpha(X,Y)$ is maximized by uniform inputs.
\end{theorem}

The proof of this theorem is given in~\Cref{app:proof:thm:symmetric-channel-cap}. Note that this recovers the result for the Shannon capacity for symmetric channels when $\alpha$ is set to 1~\cite{cover:2006}. Nevertheless, the Sibson capacity does not have an operational interpretation akin to Shannon capacity. However, for $\alpha = 0$, $\sup_{P_X} I_0(X,Y)$ yields the zero-error capacity with feedback. Moreover, for $\alpha \in (0,1)$, the Sibson capacity is related to the sphere-packing and random coding exponents. For $\alpha \in (1,\infty)$, it may be given an operational interpretation in the context of information leakage measures, as discussed in~\Cref{sec:leakage}.

\subsubsection{Zero-Error Capacity with Feedback}

It is well known that feedback does not increase the capacity of discrete memoryless channels~\cite{shannon:56}. However, the zero-error capacity with feedback can indeed be larger than the zero-error capacity (without feedback). Indeed, the former is given by a maximization of Sibson $\alpha$-mutual information of order zero. In this setting, feedback is available at the encoder instantaneously, so that the choice of the next symbol can depend on previously received symbols. More precisely, for a channel with input alphabet $\X$ and output alphabet $\Y$, a (feedback-)code of blocklength $n$ and rate $R$ consists of $n$ functions $\{f_i\}_{i=1}^n$ where $f_i: \{1,2,\ldots,2^{nR}\} \times \Y^{i-1} \to \X$.
\begin{theorem}[{\cite[Theorem 7]{shannon:56}}]
    \label{thm:zero-error-feedback}
    Consider a discrete memoryless channel $P_{Y|X}$. Then the zero-error capacity with feedback of $P_{Y|X}$ is given by
    \begin{align}
        C_{0f} \left( P_{Y|X} \right) &= \sup_{P_X} I_0(X,Y) \nonumber\\ &= \sup_{P_X} \min_{y \in \Y} \log \! \left(  \sum_{x: P_{Y|X}(y|x)>0} \!\! P_X(x) \right)^{-1} \!.
        \label{eq:thm-zero-error-feedback}
    \end{align}
\end{theorem}

\subsubsection{Sphere-Packing and Random Coding Exponents}

The sphere-packing exponent is defined for a rate $R$, prior $P_X$ and conditional distribution $P_{Y|X}$ by
\begin{align}
    \label{eq:def-sphere-packing}
    E_{sp} (R,P_X, P_{Y|X}) = \!\!\! \min_{\substack{Q_{Y|X}: \\ I(P_X,Q_{Y|X}) \leq R}} \!\!\!\!\! D(Q_{Y|X} || P_{Y|X} | P_X).
\end{align}
The $\sup_{P_X} E_{sp} (R,P_X,P_{Y|X})$ yields an upper bound on the optimal exponent of the probability of error in channel coding, and it can be obtained through the following optimization involving Sibson capacities of order $\alpha \in (0,1]$:
\begin{theorem}[{\cite[Theorem 8]{verdu:2015}\cite{haroutunian1968estimates}}] Given a discrete memoryless channel $P_{Y|X}$ and $R \geq 0$,
    \label{thm:sphere-packing}
    \begin{align}
        \label{eq:thm-sphere-packing}
        \sup_{P_X} E_{sp} (R,P_X,P_{Y|X}) = \sup_{\rho \geq 0} \left\lbrace \rho \sup_{P_X} I_{\frac{1}{1+\rho}}(X,Y) -\rho R \right\rbrace.
    \end{align}
\end{theorem}
Restricting the range of the optimization to consider $I_\alpha$ for $\alpha \in [1/2,1]$ yields the random-coding exponent, which lower-bounds the optimal exponent of the probability of error in channel coding and is defined as
\begin{align}
    \label{eq:def-random-coding}
    E_{r} (R,P_X,P_{Y|X}) &= \min_{Q_{Y|X}} D(Q_{Y|X} || P_{Y|X} | P_X)+ [I(P_X,Q_{Y|X})-R]^+.
\end{align}
Then,
\begin{theorem}[{\cite[Theorem 8]{verdu:2015}\cite[Corollary 3]{Blahut:74}}] Given a discrete memoryless channel $P_{Y|X}$ and $R \geq 0$,
    \label{thm:random-coding}
    \begin{align}
        \label{eq:thm-random-coding}
        \sup_{P_X} E_{r} (R,P_X,P_{Y|X}) = \sup_{\rho \in [0,1]} \left\lbrace \rho \sup_{P_X} I_{\frac{1}{1+\rho}}(X,Y) -\rho R \right\rbrace.
    \end{align}
\end{theorem}
A more comprehensive treatment of the connection between Rényi information measures and error exponents is given by Verdú~\cite{verdu:2021error}. For a fixed input distribution, the sphere-packing and random-coding exponent functions are represented in terms of the Augustin--Csiszár mutual information (denoted by \(I_\alpha^C\) herein), with the optimizations ranging over \(\rho\geq 0\) and \(\rho\in[0,1]\), respectively. Maximizing over the input distribution and using the equality between the Augustin--Csiszár and Sibson capacities recovers the expressions in~\Cref{eq:thm-sphere-packing,eq:thm-random-coding}. Verdú further treats general alphabets and average input-cost constraints, thereby connecting Gallager's modified \(E_0\) formulation, the large-deviation formulation in terms of conditional relative entropy, and the formulation based on Rényi information measures. In particular, a method is given for computing the cost-constrained Augustin--Csiszár capacity through an optimization involving Sibson \(\alpha\)-mutual information, with applications to Gaussian and additive exponential-noise channels.

\subsection{Information Leakage Measures} \label{sec:leakage}

Sibson $\alpha$-mutual information of order $\infty$ and Sibson capacities of order $\alpha \in (1,\infty)$ have been endowed with operational characterizations in the context of information leakage measures. In particular, Issa \emph{et al.}~\cite{MaximalLeakage:20} define an information leakage measure in a ``guessing'' framework which turns out to equal Sibson $\alpha$-mutual information of order $\infty$. This framework was extended by Liao \emph{et al.}~\cite{LiaoKSC:19} who defined a family of information leakage measures that turn out to correspond to Sibson capacities of order $\alpha \in (1,\infty)$.

\subsubsection{Maximal Leakage} \label{sec:max-leakage}
For a given pair of random variables $(X,Y)$, Issa \emph{et al.}~\cite{MaximalLeakage:20} define ``maximal leakage'' in the terms of the multiplicative advantage of guessing any function of $X$ after observing $Y$:

\begin{definition}[Maximal Leakage~{\cite[Definition 1] {MaximalLeakage:20}}] \label{def:maxLeakage} \revised{Given a joint distribution $P_{XY}$ on alphabets $\X$ and $\Y$, the Maximal Leakage from $X$ to $Y$ is defined as:}
\begin{align}
    \label{def:max-leakage}
    \mathcal{L}(X \!\! \to \!\! Y) = \sup_{U: U-X-Y} \log \frac{ \sup_{P_{\hat{U}|Y}} \Pr(U = \hat{U}) }{ \max_u P_U(u)},
\end{align}
where the supremum is over all $U$ taking values in a finite, but arbitrary, alphabet.
\end{definition}

\revised{Maximal Leakage is motivated by the setting where $X$ represents private/secret information (e.g., password, secret key, etc.), and $Y$ represents information available to an adversary (typically through a ``side-channel''). The adversary attempts to guess $X$, or some function of $X$ which is not necessarily known to the system designer. As such, Maximal Leakage adopts a (conservative) worst-case approach by considering the maximum multiplicative guessing advantage, over all such (randomized) functions~\cite{MaximalLeakage:20}.}  It turns out that:

\begin{theorem}[{\cite[Theorem 1] {MaximalLeakage:20}}]  \label{thm:max-leakage}
Given a joint distribution $P_{XY}$, the Maximal Leakage from $X$ to $Y$ is given by:
    \begin{align}
        \label{eq:thm-max-leakage}
        \mathcal{L}(X \!\! \to \!\! Y) = I_\infty(X,Y).
    \end{align}
\end{theorem}

Interestingly, another characterization was obtained by Braun \emph{et al.}~\cite{braun:2009} who consider the advantage of guessing $X$ itself from $Y$, maximized over the input distribution $P_X$. That is,
\begin{theorem}[{\cite[Proposition 5.1] {braun:2009}}]  
    \label{thm:max-leakage-additive}
    \begin{align}
        \label{eq:thm-max-leakage-additive} 
        I_\infty(X,Y) = \sup_{P_X} \log \frac{ \sup_{P_{\hat{X}|Y}}  \Pr(X = \hat{X})}{\max_x P_X(x)}.
    \end{align}
\end{theorem}

\revised{Maximal Leakage has subsequently received considerable attention, whether in terms of applications such as hypothesis testing~\cite{LiaoSCT:2017}, generalization error~\cite{EspositoGI:2021}, biometric security~\cite{ShahrezaSM:23}, and others~\cite{EspositoG:22}; or in terms of extensions such as generalized gain functions~\cite{KurriSK:22}, pointwise maximal leakage~\cite{SaeidianCOS:23}, and Maximal $\alpha$-Leakage~\cite{LiaoKSC:19}, among others. }

\revised{More recently, Maximal Leakage has also been connected to Doeblin coefficients, a fundamental tool to study convergence of Markov chains~\cite{makurAS:24}. In particular, Makur and Singh~\cite{makurAS:24} introduce the ``max-Doeblin'' coefficient, which equivalent to $e^{\mathcal{L}(X  \to Y)}$. They show~\cite[Theorem 4]{makurAS:24} that max-Doeblin coefficient satisfies a coupling result, which generalizes the classical coupling result for total variation distance~\cite[Proposition 4.7]{levin2017markov}.} 

\subsubsection{Maximal $\alpha$-Leakage} \label{sec:alpha-leakage}

Liao \emph{et al.}~\cite{LiaoKSC:19} extend the Maximal Leakage framework by introducing a loss function parametrized by $\alpha \in (1,\infty)$:
\begin{align}
    \label{eq:loss-function}
    \ell_\alpha(x,y,P_{\hat{X}|Y}) =\frac{\alpha}{\alpha-1} \left( 1- P_{\hat{X}|Y}(x|y)^{\frac{\alpha-1}{\alpha}} \right).
\end{align}
\revised{The setting in which it was introduced assumes a Markov chain $X - Y - \hat{X}$, where $\hat{X}$ is an estimator of $X$, and $P_{\hat{X}|Y}$ denotes the strategy used to estimate $X$ from observations $Y$. In particular, $P_{\hat{X}|Y}(x|y)$ represents the probability of correctly estimating $X = x$ given $Y = y$. The family of losses $\ell_\alpha$ captures different ways of measuring the uncertainty associated with the strategy $P_{\hat{X}|Y}$.
In the limit as $\alpha \to 1$, the loss becomes
$$
\ell_1(x, y, P_{\hat{X}|Y}) = \log \frac{1}{P_{\hat{X}|Y}(x|y)},$$
while as $\alpha \to \infty$, it reduces to
$$
\ell_\infty(x, y, P_{\hat{X}|Y}) = 1 - P_{\hat{X}|Y}(x|y).$$
If $P_{\hat{X}|Y}(x|y) = 1$, both losses (and indeed all $\ell_\alpha$ for $\alpha \in [1, +\infty]$) are zero. However, if $P_{\hat{X}|Y}(x|y) = 0$, the two losses reflect uncertainty very differently: $\ell_1(x, y, 0) = +\infty$, while $\ell_\infty(x, y, 0) = 1$.
For intermediate values $\alpha \in (1, +\infty)$, the loss $\ell_\alpha$ interpolates between these two extremes: smaller values of $\alpha$ penalize incorrect estimates more heavily. Moreover, leveraging this tunable measure of loss}
 they define maximal $\alpha$-leakage as:
\begin{definition}[Maximal $\alpha$-Leakage~\cite{LiaoKSC:19}] 
    \label{def:max-alpha-leakage}
    \begin{align}
        &\mathcal{L}_\alpha^{\mathrm{max}} (X \!\! \to \!\! Y)= \sup_{U:U-X-Y} \frac{\alpha}{\alpha-1} \log \frac{\sup_{P_{\hat{X}|Y}} \E \left[ P_{\hat{X}|Y}(X|Y)^{\frac{\alpha-1}{\alpha}} \right] } { \sup_{P_{\hat{X}}} \E \left[ P_{\hat{X}}(X)^{\frac{\alpha-1}{\alpha}} \right] }.
        \label{eq:def-max-alpha-leakage}
    \end{align}
\end{definition}

It turns out that
\begin{theorem}[\cite{LiaoKSC:19}]
    \label{thm:max-alpha-leakage}
    Given a joint distribution $P_{XY}$ and $\alpha \in (1,\infty)$, maximal $\alpha$-leakage is given by
    \begin{align}
        \label{eq:max-alpha-leakage}
        \mathcal{L}_\alpha^{\mathrm{max}} (X \!\! \to \!\! Y) = \sup_{P_{\hat{X}}} I_\alpha(\hat{X},Y).
    \end{align}
    \revised{In particular,
    \begin{equation}
         \mathcal{L}_1^{\mathrm{max}} (X \!\! \to \!\! Y) = I(X;Y).
    \end{equation}}
    
\end{theorem}

\revised{Remarkably, Sibson $\alpha$-mutual information again appears as (part of) the answer. The proof of Liao \emph{et al.}~\cite{LiaoKSC:19}, in fact, uses and is stated in terms of Arimoto's mutual information~\cite{arimoto:77}, rather than Sibson's. However, the two notions are closely connected (they are equal when taking the supremum over $P_X$). We discuss this relationship and other generalized information measures in the next subsection.}

\subsection{Connection to Other Generalized Information Measures}\label{sec:otherRenyiInformations}

There have been several attempts to generalize mutual information along the same lines as R\'enyi's generalization of entropy and relative entropy. We summarize noteworthy attempts here and mention connections to Sibson $\alpha$-mutual information, where they exist.

\subsubsection{Arimoto mutual information}

Arimoto~\cite{arimoto:77} generalized mutual information based on the following characterization of Shannon mutual information for discrete random variables:
\begin{align}
    I(X;Y) = H(X) - H(X|Y).
\end{align}

\begin{definition}[Arimoto mutual information] \label{def:arimoto}
    Given discrete random variables $X$ and $Y$ and $\alpha \in (0,1) \cup (1,\infty)$, the Arimoto mutual information is defined as
    \begin{align} \label{eq:def-arimoto}
        I^A_\alpha (X,Y)  = H_\alpha(X) - H_\alpha(X|Y). 
    \end{align}
\end{definition}
It follows directly from~\Cref{eq:def-arimoto,eq:def-cond-renyi} that
\begin{align}
    \label{eq:arimotoMI-expanded}
    I^A_\alpha(X,Y) =  \frac{\alpha}{\alpha-1} \log \sum_{y \in \Y} \left( \sum_{x\in \X} \frac{P_{X}(x)^\alpha }{ \sum_{x' \in \X} P_X (x')^\alpha } P_{Y|X}(y|x)^\alpha \right)^{\frac{1}{\alpha}}.
\end{align}
By comparing with the expression given in~\Cref{eq:Ialpha-discrete-formula} for Sibson $\alpha$-mutual information for discrete alphabets, we get
\begin{proposition} \label{prop:arimoto-sibson-tilted}
    Given two discrete random variables $(X,Y) \in \X \times \Y$ with joint distribution $P_{XY} = P_X P_{Y|X}$ and $\alpha \in \range$,
    \begin{align} \label{eq:prop-arimoto-sibson-tilted}
        I_\alpha^A(X,Y) = I_\alpha( X_\alpha, Y),
    \end{align}
    where $ \ds P_{X_\alpha} (x) = \frac{P_X(x)^\alpha}{\sum_{x' \in \X} P_X(x')^\alpha} $, and $P_{X_\alpha Y} = P_{X_\alpha} P_{Y|X}$.
\end{proposition}
That is, Arimoto mutual information is equivalent to Sibson $\alpha$-mutual information evaluated at a ``tilted'' input distribution $P_{X_\alpha}$ which is proportional to $P_X^\alpha$ (and the same conditional $P_{Y|X}$). The tilting operation is reversible so that we may also write
\begin{align}
    I_\alpha(X,Y) = I^A_\alpha (X_{\frac{1}{\alpha}},Y),
\end{align}
where $P_{X_{\frac{1}{\alpha}}}$ is proportional to $P_X^\frac{1}{\alpha}$.
\begin{corollary} \label{corr:arimoto-sibson-equality}
    Given two discrete random variables $(X,Y) \in \X \times \Y$ and $\alpha \in \range$, if $P_X$ is uniform, then
    \begin{align}
        \label{eq:corr-arimoto-sibson-equality-unif}
        I_\alpha(X,Y) = I_\alpha^A(X,Y).
    \end{align}
    Moreover, for fixed $P_{Y|X}$,
    \begin{align}
        \sup_{P_X} I_\alpha (X,Y) = \sup_{P_X} I_\alpha^A(X,Y).
    \end{align}
\end{corollary}

One can derive several properties of Arimoto mutual information using the equality in~\Cref{prop:arimoto-sibson-tilted}, including: non-negativity (with $I_\alpha^A(X,Y)=0$ if and only if $X$ and $Y$ are independent), additivity, post-processing inequality (i.e., $I_\alpha^A(X,Z) \leq I_\alpha^A(X,Y)$ if $X-Y-Z$ is a Markov chain), invariance under injective transformations, as well as convexity/concavity properties with respect to $P_{Y|X}$ (for a fixed $P_X$). Moreover,
\begin{align}
    I_\alpha^A(X,Y) = I_\alpha(X_\alpha,Y) \leq H_\frac{1}{\alpha}(X_\alpha) = H_\alpha(X),
\end{align}
where the last equality follows from simply plugging in the relevant expressions. It is worth noting, however, that $I_\alpha^A(X,Y)$ does not satisfy the ``pre-processing'' inequality (also called linkage inequality). That is,
\begin{proposition}[{\cite[Proposition 1]{wuWIS:20}}]
    For any $\alpha \in \range$, there exists a triple of random variables $(X,Y,Z)$ satisfying the Markov chain $X-Y-Z$ and $I_\alpha^A(X,Z) > I_\alpha^A(Y,Z)$.
\end{proposition}

\subsubsection{Csisz{\'a}r mutual information}
Csisz{\'a}r~\cite{csiszar:95} considered the following characterization of Shannon mutual information
\begin{align}
    I(X;Y) = \min_{Q_Y} \E_{P_X} \left[ D \left( P_{Y|X}(.|X) \| Q_Y \right) \right]
\end{align}
and generalized it as follows:
\begin{definition}[Csisz\'ar mutual information] \label{def:csiszar-MI}
    Given a joint distribution $P_{XY}$ on $\X \times \Y$ and $\alpha \in \range$, the Csisz\'ar mutual information is defined as
    \begin{align}
        \label{eq:def-csiszar-MI}
        I_\alpha^C(X,Y) = \min_{Q_Y} \E_{P_X} \left[  D_\alpha \left( P_{Y|X}(.|X) \| Q_Y \right)\right].
    \end{align}
\end{definition}
As opposed to Sibson and Arimoto mutual information, there is no known closed-form formula for $I_\alpha^C(X,Y)$. However, maximizing $I_\alpha^C(X,Y)$ over $P_X$ again yields the same result:
\begin{proposition}
    \label{prop:csiszarMI-max}
    Fix $P_{Y|X}$. Then for all $\alpha \in \range$,
    \begin{align}
        \label{eq:prop-csiszarMI-max}
          \sup_{P_X} I_\alpha^C(X,Y) = \sup_{P_X} I_\alpha(X,Y) = \sup_{P_X} I^A_\alpha(X,Y). 
    \end{align}
\end{proposition}
Furthermore, Sibson and Csisz\'ar mutual information obey a strict ordering:
\begin{proposition}
    \label{prop:csiszar-sibson-ordering}
    For all $\alpha >1$ and all distributions $P_{XY}$,
    \begin{align}
        I_\alpha(X,Y) \geq I_\alpha^C(X,Y).
    \end{align}
    Moreover, for all $\alpha <1$ and all distributions $P_{XY}$,
    \begin{align}
        I_\alpha(X,Y) \leq I_\alpha^C(X,Y).
    \end{align}
\end{proposition}

\subsubsection{Lapidoth-Pfister mutual information}

Lapidoth and Pfister considered a minimization over product-of-marginals distributions $Q_XQ_Y$ instead of over $Q_Y$ only (as done in Sibson's definition), so that:

\begin{definition}[Lapidoth-Pfister mutual information~\cite{LapidothP:18}] Given a joint distribution $P_{XY}$ and $\alpha \in \range$, the Lapidoth-Pfister mutual information is defined as
\begin{align}
    I_\alpha^{LP} (X,Y) = \min_{Q_X Q_Y} D_\alpha (P_{XY} || Q_X Q_Y).
\end{align}
\end{definition}
It follows directly from the definitions that 
\begin{align} \label{eq:lapidoth-pfister-sibson}
I_\alpha^{LP} (X,Y) \leq I_\alpha (X,Y).
\end{align}
However, the capacities coincide for $\alpha > 1$:
\begin{proposition}[{\cite[Theorem V.1]{aishwarya:19}}]
    Given a conditional distribution $P_{Y|X}$ and $\alpha >1$,
    \begin{align}
        \label{eq:prop-lp-capacity}
        \sup_{P_X} I_\alpha^{LP}(X,Y) = \sup_{P_X} I_\alpha (X,Y).
    \end{align}
\end{proposition}
Moreover, Lapidoth-Pfister mutual information possesses several desirable properties, including: it is zero if and only if $X$ and $Y$ are independent, it satisfies the data processing inequality, it is bounded by $\log | \X |$, and it is additive over independent pairs. However, except for some special cases, there is no closed form for $I_\alpha^{LP}(X,Y).$

\section{Variational Representations}\label{sec:variational} 
Variational representations of information measures have many applications and implications. The most notable one is probably Donsker-Varadhan's representation of the Kullback-Leibler divergence~\cite{DonskerVaradhan:75,Varadhan:84}. This has been pivotal for linking the Kullback-Leibler divergence to hypothesis testing~\cite{sanov:58}, Wasserstein distances (and consequently, the concentration of measure phenomenon)~\cite{Marton:96,Marton:96b,Marton:98,RaginskyS:14}, generalization error of learning algorithms~\cite{RussoZ:16, XuR:17}, etc. In particular, these representations highlight the connection between information measures and spaces of functions and provide a powerful and elegant approach to linking information measures to various problems of applied nature (\textit{e.g.}, compression, guessing, testing, etc.~\cite{Esposito:22}).
We will now provide a variational representation of Sibson $\alpha$-mutual information.
In order to do so, one can undertake various paths:
\begin{itemize}
    \item leverage the connection between Sibson $\alpha$-mutual information and R\'enyi divergence (and, consequently, the connection between R\'enyi divergence and the Kullback-Leibler divergence~\cite[Theorem 30]{renyiDiv}). This is the approach undertaken to prove~\Cref{thm-Sibson-var-alternative};
    \item leverage the connection between Sibson $\alpha$-mutual information and norms, like we did in~\Cref{sec:normPerspective}. This approach leads to~\Cref{thm:varReprIalpha}.
\end{itemize}
In particular, one can see~\Cref{thm:varReprIalpha} as a corollary of~\Cref{thm-Sibson-var-alternative}. However, given the power of applicability of~\Cref{thm:varReprIalpha} and the insightful direct proof, it is stated as an independent result. Moreover, from~\Cref{thm:varReprIalpha} one can recover most of the results that employ Sibson's mutual information in applied settings: 
\begin{itemize}
    \item exponential concentration bounds for functions of random variables when the function is not independent with respect to the variables (see~\Cref{sec:depVsIndepPos}), with consequent applications in learning theory settings (see~\Cref{sec:concentrationLearning});
   
    \item relationships with the error exponent in hypothesis testing settings (see~\Cref{sec:hypothesisTesting});
     \item bounds on the expected generalization error of learning algorithms via transportation-cost like inequalities (see~\Cref{sec:tpcSibson} and~\Cref{thm:expGenErrSibson});
    \item applications in estimation theory via the construction of a generalized Fano's method (see~\Cref{sec:fanoMethodSibson}) and in Bayesian Risk Settings(see~\Cref{sec:bayesianFramework}).
\end{itemize}
This clearly shows the potential of these representations.
\subsection{Sibson $\alpha$-Mutual Information and Kullback-Leibler Divergence}

By analogy to~\Cref{prop:RenyiViaKL}, Sibson $\alpha$-mutual information can be connected to Kullback-Leibler divergence. This leads to the first variational representation we introduce:

\begin{theorem}\label{Lemma-SibsonViaKL}
For any $\alpha \in (0,1)\cup (1,\infty),$
\begin{align}
   &(1-\alpha)I_\alpha(X,Y)= \min_{R_{XY}} \left\{ \alpha D(R_{XY}\|P_{XY}) + (1-\alpha) D(R_{XY}\|P_{X}R_Y) \right\} \label{Lemma-SibsonViaKL-eq-main} 
\end{align}
Hence, $(1-\alpha)I_\alpha(X,Y)$ is the minimum of linear functions of $\alpha$ so it is concave in $\alpha$.
Equality is attained by selecting
$R_{XY} \propto P_{XY}^{\alpha} (P_XQ_Y^\star)^{1-\alpha},$
where $Q_Y^\star$ is given in~\Cref{eq:qstarIalpha}, and we note that for this joint distribution, the marginal distribution on $Y$ is given by $Q_Y^\star.$
\end{theorem}

The proof of this result is given in~\Cref{app:variational:Lemma-SibsonViaKL}.

In~\cite{shayevitz:2011} a variational representation for $I_\alpha$ in terms of Shannon information measures was also derived. In particular, the author introduces a variational representation for $I_\alpha(P,W)$ (therein denoted as $K_\alpha(P,W)$) in terms of maxima of sums of the Kullback-Leibler divergence and Csisz\'ar's $\alpha$-mutual information (which, in turn, can be expressed as a maximum of a sum between Shannon's mutual information and KL).
See~\cite[Theorem 1]{shayevitz:2011}.

\subsection{Variational Representations of Sibson $\alpha$-Mutual Information}

\begin{theorem}\label{thm-Sibson-var-alternative}
Assume that $P_{XY}\ll P_XP_Y$. For $\alpha>1,$ we have 
 \begin{align}
    I_\alpha(X,Y) &= \sup_{f:\X\times\mathcal{Y}\to\mathbb{R}} \frac{\alpha}{\alpha-1} \log \E_{P_{XY}}\left[e^{(\alpha-1)f(X,Y)}\right]- \E_{R_Y^*} \left[\log \E_{P_X}\left[e^{\alpha f(X,Y)}\right]\right] ,\label{thm-Sibson-var-alternative-eq-largealpha}
\end{align}
where
\begin{align}
    R_Y^*(y) &\propto \E_{P_X} \left[ P_{Y|X}(y|X) e^{(\alpha-1)f(X,y)} \right],\label{thm-Sibson-var-alternative-Eq-Rstar}\\
    \frac{dR_Y^*}{dP_Y}(y) &\propto \E_{P_{X|Y=y}} \left[e^{(\alpha-1)f(X,y)} \right].\label{thm-Sibson-var-alternative-Eq-Rstar2}
\end{align}
For $0<\alpha<1,$ we have
\begin{align}
  I_\alpha(X,Y) &= \sup_{f:\X\times\mathcal{Y}\to\mathbb{R}} \frac{\alpha}{\alpha-1} \log \E_{P_{XY}}\left[e^{(\alpha-1) f(X,Y)}\right] - \E_{Q_Y^{\star}} \left[\log \E_{P_X}\left[e^{\alpha f(X,Y)}\right]\right], \label{thm-Sibson-var-alternative-eq-smallalpha}
\end{align}
where $Q_Y^\star$ is given in~\Cref{eq:qstarIalpha}.
In both cases, equality is attained by selecting
\begin{align}
    e^{\alpha f(x,y)} & = \frac{\left(\frac{dP_{XY}}{dP_XP_Y}(x,y)\right)^{\alpha}}{\E_{P_{X}}\left[\left(\frac{dP_{XY}}{dP_XP_Y}(X,y)\right)^{\alpha}\right] }. \label{eq:equalityVarRepr}
\end{align}
\end{theorem}

The proof of this theorem can be found in~\Cref{app:variational:thm-Sibson-var-alternative}.

\begin{remark}
    Taking the limit of~\Cref{thm-Sibson-var-alternative-eq-largealpha} (or~\Cref{thm-Sibson-var-alternative-eq-smallalpha}) as $\alpha \rightarrow 1$ leads to the slightly strengthened Donsker-Varadhan representation of mutual information:
\begin{align}
    I(X;Y) & = \sup_f \E_{P_{XY}} [f(X,Y)] - \E_{P_Y} \left[\log \E_{P_X}\left[e^{f(X,Y)}\right]\right].
\end{align}
We note that this formula can be readily obtained by applying the Donsker-Varadhan representation separately for every $y,$ and then averaging over $Y.$
\end{remark}

The variational representation given in~\Cref{thm-Sibson-var-alternative} can be complemented by a second representation in a slightly different spirit. The inequality part of this second representation is, in fact, a weakening of~\Cref{thm-Sibson-var-alternative} through Jensen's inequality.
Specifically, the following variational representation is established.

\begin{theorem}\label{thm-Sibson-var-alternative-weaker}
Assume that $P_{XY}\ll P_XP_Y$. For $\alpha>1,$ we have 
 \begin{align}
    I_\alpha(X,Y) &= \sup_{f:\X\times\mathcal{Y}\to\mathbb{R}} \frac{\alpha}{\alpha-1} \log \E_{P_{XY}}\left[e^{(\alpha-1)f(X,Y)}\right] -  \log \E_{P_XR^*_Y}\left[e^{\alpha f(X,Y)}\right] ,\label{thm-Sibson-var-alternative-eq-largealpha-weaker}
\end{align}
where $R_Y^*$ is as in~\Cref{thm-Sibson-var-alternative}.
For $\alpha>0,$ we have
\begin{align}
  I_\alpha(X,Y) &= \sup_{f:\X\times\mathcal{Y}\to\mathbb{R}} \frac{\alpha}{\alpha-1} \log \E_{P_{XY}}\left[e^{(\alpha-1) f(X,Y)}\right]-  \log \E_{P_XQ_Y^{\star}}\left[e^{\alpha f(X,Y)}\right] , \label{thm-Sibson-var-alternative-eq-smallalpha-weaker}
\end{align}
where $Q_Y^\star$ is given in~\Cref{eq:qstarIalpha}.
In both cases, equality is attained by selecting the function $f(x,y)$ as in~\Cref{eq:equalityVarRepr}.
\end{theorem}

The proof of this theorem can be found in~\Cref{app:variational:thm-Sibson-var-alternative-weaker}.
\begin{remark}
    Notice that~\Cref{thm-Sibson-var-alternative-eq-smallalpha-weaker} can also be seen as a consequence of the variational representation of R\'enyi $\alpha$-divergence (\Cref{lemma-variationalRenyiDiv}). In fact, since $I_\alpha(X,Y)=D_\alpha(P_{XY}\|P_XQ_Y^\star)$, one can use the variational representation of $D_\alpha$ in order to recover~\Cref{thm-Sibson-var-alternative-eq-smallalpha-weaker}. To the best of our knowledge,~\Cref{thm-Sibson-var-alternative-eq-largealpha-weaker} instead cannot be recovered with this approach and thus we provide an explicit proof in Appendix~\ref{app:variational:thm-Sibson-var-alternative-weaker}.
\end{remark}

In extension to the considerations above, we now establish a third variational representation, which is, in many ways, more compact and likely more useful than the previous two theorems. So far, we have consistently refined the various variational representations in order to render them more applicable in settings of interest. In going from~\Cref{Lemma-SibsonViaKL} to~\Cref{thm-Sibson-var-alternative} we leveraged the variational representation of Sibson $\alpha$-MI in terms of measures and KL divergences to obtain one involving expected values of functions. In many applied settings, one is indeed typically interested in bounding expected values of functions (probabilities of events, expected errors, etc.). Thus,~\Cref{thm-Sibson-var-alternative} is arguably easier to use than~\Cref{Lemma-SibsonViaKL} in said settings. Then, we weakened~\Cref{thm-Sibson-var-alternative} in order to achieve~\Cref{thm-Sibson-var-alternative-weaker}, approaching a form that is closer to what one could retrieve when providing a variational representation for Shannon's mutual information via Donsker-Varadhan's result. In particular, taking the limit of $\alpha\to 1$ in~\Cref{thm-Sibson-var-alternative-weaker} one recovers exactly Donsker-Varadhan's representation of mutual information, \textit{i.e.}:
\begin{equation}
        I(X;Y) = \sup_{f} \E_{P_{XY}}\left[f(X,Y)\right] - \log \E_{P_XP_Y}\left[e^{f(X,Y)}\right]. \label{eq:varReprMI2}
\end{equation}
However,~\Cref{thm-Sibson-var-alternative-weaker} is still not immediate to employ in practical settings as one would have to compute $R_Y^\star$ or $Q_Y^\star$. We can thus weaken the result further in order to obtain yet another variational representation that still preserves the asymmetry between $X$ and $Y$ and can be easily and successfully employed in a variety of practical settings, as we will see in the following sections.
In particular, this representation can be seen as a corollary of~\Cref{thm-Sibson-var-alternative-weaker} (see~\Cref{app:variational:thm-Sibson-var-alternative-weaker-connection-to-VarReprSibs} for the corresponding derivation), but it can also be directly proved in an independent and insightful fashion via H\"older's inequality (see~\Cref{app:proofVarReprSibs}). Therefore, we prefer to state it as a separate theorem.

\begin{theorem}\label{thm:varReprIalpha}
Let $\alpha>1$ and assume that $P_{XY}\ll P_XP_Y$, then: 
\begin{equation}
\exp\left(\!\frac{\alpha-1}{\alpha}I_\alpha(X,Y)\!\right)\!\! =\!\!\! \!\!\sup_{g:\X\times\mathcal{Y}\to\mathbb{R}^+}\! \frac{\E_{P_{XY}}[g(X,Y)]}{\max_y \left(\E_{P_X}[g^\beta(X,y)]\right)^\frac1\beta },\label{eq:varReprIalpha1}
\end{equation}
where $\beta=\frac{\alpha}{\alpha-1}$.
Moreover, if $0<\alpha<1$ then:
\begin{align}
\exp\left(\!\frac{\alpha-1}{\alpha}I_\alpha(X,Y)\!\right) \!\! =\!\!\! \!\! 
\inf_{g:\X\times\mathcal{Y}\to\mathbb{R}^+}\! \frac{\E_{P_{XY}}[g(X,Y)]}{\min_y \left(\E_{P_X}[g^\beta(X,y)]\right)^\frac1\beta }.\label{eq:varReprIalpha1Neg}
\end{align}
\end{theorem}

Taking the limit of $\alpha\to\infty$ one recovers a novel variational representation for Maximal Leakage, stated below for ease of reference:
\begin{equation}
    \exp\left(\ml{X}{Y}\right) = \sup_{g:\X\times\Y\to\mathbb{R^+}} \frac{\E_{P_{XY}}[g(X,Y)]}{\max_y \E_{P_X}[g(X,y)]}. \label{eq:varReprLeakage}
\end{equation}

\revised{
As indicated by this equivalence,~\Cref{thm:varReprIalpha} is similar to formulations of privacy and security metrics. Indeed, it states that $I_\alpha$, for $\alpha >1$, is the multiplicative gain induced by observing $Y$, optimized over all gain functions (the case of $\alpha <1$ can be seen as a multiplicative decrease over all non-negative cost functions). Note that the optimization over $g$ subsumes the setup of guessing $X$ from $Y$ by choosing $g(X,Y) = \mathbf{1} \left\lbrace X = \arg \max_{x'} P_{X|Y}(x'|Y) \right\rbrace$
 (where ties should be broken, but can be broken arbitrarily). It also subsumes the setup of guessing any (possibly randomized) function of $X$ by noting that $U-X-Y-\hat{U}$ is a Markov chain implies $I_\alpha(X,Y) \geq I_\alpha (U,\hat{U})$ and choosing $g(U,\hat{U}) = \mathbf{1} \left\lbrace U = \hat{U} \right\rbrace$. This is in line with other recent variational approaches to privacy in general~\cite{lin2024graph,rodriguez2021variational}, and using Sibson $\alpha$-mutual information in particular~\cite{kamatsuka2025several,ding2024alpha}.
 
 It is worth noting here that the numerator in Theorem V.4 is an $L_1$-norm, whereas the denominator is an $L_\beta$-norm. This is akin to hypercontractivity bounds (where different norms are used at the input and output). As the $L_\beta$-norm is greater than the $L_1$-norm ($\beta >1$), this choice yields a smaller value of the optimization, $\frac{\alpha-1}{\alpha}I_\alpha(X,Y)$ instead of $\ml{X}{Y}$. As such, we get a trade-off between the choice of $\beta$ and the induced left-hand side.}

However, the function $g^\star$ achieving equality in~\Cref{eq:varReprLeakage} is not obvious from the proof of~\Cref{thm:varReprIalpha} and deserves an explicit treatise. One has that, in the case of~\Cref{eq:varReprLeakage}, equality is achieved by the following function:
\begin{equation}
g^\star(x,y) = \frac{\mathbbm{1}_{\left\{\substack{\text{argmax}\\ x}{\frac{dP_{XY}}{dP_XP_Y}(x,y)}\right\}}}{P_X\left(\left\{\substack{\text{argmax}\\ X}{\frac{dP_{XY}}{dP_XP_Y}(X ,y)}\right\}\right)}.~\label{eq:equalityML} 
\end{equation}
The argument to see why~\Cref{eq:equalityML} is true follows from essentially re-writing~\Cref{eq:equalityVarRepr} as follows:
\begin{equation}
    g^\star(x,y)^\beta = \frac{\left(\frac{dP_{XY}}{dP_XP_Y}(x,y)\right)^\alpha}{\left\lVert\frac{dP_{XY}}{dP_XP_Y}(X,y)\right\rVert^\alpha _{L^\alpha(P_X)} }.\label{eq:equalityIalpha2}
\end{equation}
One can then take the limit of $\alpha\to\infty$ and retrieve~\Cref{eq:equalityML}. Due to the presence of the power $\alpha$, the denominator in~\Cref{eq:equalityIalpha2} no longer converges to an $\esssup$. However, for a given $y$ the function $g^\star$ behaves (due to the presence of the norm in the denominator and the fact that the norm is taken with respect to a probability measure) \revised{as a normalized indicator function of $\argmax$}. 
Another interesting consideration comes from taking the limit of $\alpha\to 1$. Clearly, doing so naively in~\Cref{eq:varReprIalpha1} leads to a trivial bound of the form: $\E_{P_{XY}}[g(X,Y)]\leq \max_y \E_{P_X}[g(X,Y)]$. However, massaging the expressions in~\Cref{thm:varReprIalpha} and re-parametrising the set of functions one obtains the following:
\begin{equation}
    \begin{split}
        \frac{I_\alpha(X,Y)}{\alpha} &= \sup_{f} \frac{1}{\alpha-1}\log \E_{P_{XY}}\left[e^{(\alpha-1)f}\right]- \frac1\alpha \log \max_y \E_{P_X}\left[e^{\alpha f}\right]. \label{eq:varReprIalpha2}
    \end{split}
\end{equation}
Taking then the limit of $\alpha\to 1$ in~\Cref{eq:varReprIalpha2} leads to the following:
\begin{equation}
        I(X;Y) = \sup_{f} \E_{P_{XY}}\left[f(X,Y) \right] - \log \max_y \E_{P_X}\left[e^{f(X,Y)}\right]. \label{eq:varReprMI}
\end{equation}
\Cref{eq:varReprMI} is a point-wise worse variational representation for mutual information with respect to the one that can be obtained with Donsker-Varadhan (see~\Cref{eq:varReprMI2}).
Equality is achieved in both cases by $f^\star = 
\log \frac{dP_{XY}}{dP_XP_Y}$. Consequently, parametrised lower-bounds on the mutual information that one could achieve from~\Cref{eq:varReprIalpha1} do not represent a suitable parametrised generalisation of expressions that one could obtain via~\Cref{eq:varReprMI2}, as in the limit they would provide for a smaller lower-bound.
Also, the limit of $\alpha\to\infty$ in ~\Cref{eq:varReprIalpha2} leads to a trivial result: $\esssup_{P_{XY}(x,y)} f(x,y)\leq \esssup_{P_{X}P_Y}f(x,y)$ which follows from the absolute-continuity constraints necessary in order to define $I_\alpha(X,Y)$.
\revised{
\begin{remark}
    Recent research has shown growing interest in estimating information measures via neural networks~\cite{belghazi:2018,Birrell:21}. To leverage~\Cref{thm:varReprIalpha} for deriving an estimator of $I_\alpha$ using polynomials or neural networks, it is essential to understand when the function achieving equality—namely the one in~\Cref{eq:equalityML} or~\Cref{eq:equalityIalpha2}—is bounded or smooth.

The function in~\Cref{eq:equalityML} is generally not smooth. In fact, when the $\argmax$, $x^\star$, is unique, a discontinuity appears around $x^\star$. However, it remains bounded if $P_X$ is bounded away from zero (e.g., if the support of $X$ is compact).

For the function in~\Cref{eq:equalityIalpha2}, boundedness and smoothness depend on the Radon-Nikodym derivative. If $\left\lVert \frac{dP_{XY}}{dP_XP_Y} \right\rVert_{L^\infty(P_X)}$ is finite, the function is bounded. In particular, boundedness holds if there exists a constant $\kappa$ such that for every measurable set $E \subseteq \X \times \Y$, we have $P_{XY}(E) \leq \kappa P_XP_Y(E)$.

Regarding smoothness, if $P_X$, $P_Y$, and $P_{XY}$ admit smooth, bounded densities with respect to a common dominating measure (e.g., the Lebesgue measure), then the Radon-Nikodym derivative is smooth. This is the case, for instance, when $P_X$ and $P_Y$ have smooth and bounded densities and $Y = h(X) + N$, where $h$ is smooth and $N$ is independent of $X$ and also has a smooth density.

Empirically, it appears feasible to estimate $I_\alpha$ using~\Cref{eq:varReprIalpha1,eq:varReprIalpha1Neg}. Although the additional maximization in~\Cref{eq:varReprIalpha1} increases complexity compared to Shannon’s mutual information (see~\Cref{eq:varReprMI2}), this can be handled via a soft-max approximation, introducing a tunable parameter.

A simple architecture, similar to that used in~\cite{Birrell:21} for Rényi divergences, a single-hidden-layer network with 64 hidden units and about 257 parameters, effectively approximates $I_\alpha(X, Y)$ when $X$ and $Y$ are correlated Gaussians. The network used in the experiments was trained using Adam on $10^4$ samples with mini-batches of size 512.

\end{remark}

\begin{remark}
    Neural network estimators of~\Cref{eq:varReprIalpha1} (\Cref{thm:varReprIalpha}) can be leveraged for auditing differentially-private mechanisms~\cite{domingo2022auditing,nuradha2023pufferfish,koskela2024auditing,muthu2024nearly}: Since $\frac{\alpha-1}{\alpha} I_\alpha(X,Y)$ lower-bounds differential privacy, finding a function $g$ such that the ratio in~\Cref{eq:varReprIalpha1} exceeds $\varepsilon$ implies a differential privacy violation. This may yield more robust auditing compared to estimators using mutual information~\cite{nuradha2023pufferfish}, as $I_\alpha$ is larger. Developing and testing such auditing mechanisms is left as future work.
\end{remark}

}
We will now employ~\Cref{thm:varReprIalpha} in the settings described a the beginning of~\Cref{sec:variational} and show the versatility and potential it brings.



\section{Applications: Concentration of Measure}\label{sec:concentration}
Sibson $\alpha$-mutual information represents a fundamental object sitting at the intersection of Information Theory, Probability Theory, and Functional Analysis. As such, it is possible to provide fundamental results in concentration of measure and hypothesis testing settings. In particular, Sibson $\alpha$-mutual information allows us to link the expected values of a function of two random variables $X,Y$ when the random variables are dependent, to the expected value of the same function under independence. This connection allows us to employ the measure in:
\begin{itemize}
    \item providing concentration results extending McDiarmid/Hoeffding's inequality to functions of random variables in settings where the function depends on the random variables themselves. The concentration result can then in turn be used to provide bounds on the probability of having a large generalization error in Learning Theory settings;
    \item composite hypothesis testing settings in which one is testing for independence  but one of the two marginals is not fixed.
\end{itemize}
Similar results can be provided when one considers the conditional version of Sibson $\alpha$-mutual information (see~\Cref{sec:conditional}).
\subsection{Dependence vs Independence}\label{sec:depVsIndep}
\subsubsection{Case $\alpha>1$}\label{sec:depVsIndepPos}
An interesting feature of Sibson $\alpha$-mutual information comes from taking the norm-inspired perspective (\Cref{sec:normPerspective}) along with the Variational representations (\Cref{sec:variational}). In particular, one can see $I_\alpha$ as nested norms of the Radon-Nikodym derivative. For this section, consider~\Cref{eq:normRepresentationEveryAlpha}. Leveraging then~\Cref{thm:varReprIalpha}, given a function $f:\X\times\Y \to \mathbb{R}$ and two random variables $X,Y$ one can prove results relating: 
\begin{itemize}
    \item the expected value of $f$ under a joint measure $P_{XY}$;
    \item Sibson $\alpha$-mutual information between $X$ and $Y$;
    \item the nested norm of $f$ under the product of the marginals $P_X P_Y$;
\end{itemize} 
Part of the results presented in this section has already appeared in~\cite{EspositoGI:2021}.
One example of such a result is given in the following theorem:
\begin{theorem}\label{thm:upperBoundExpSibson}
    Let $P_{XY}$ be a joint measure. Let $f:\X\times\Y\to\mathbb{R}^+$ be a $P_{XY}$-measurable function, then one has that for every $\alpha>1$ and denoting with $\beta=\alpha/(\alpha-1)$,
    \begin{equation}
    \begin{split}
        \mathbb{E}_{P_{XY}}\left[f(X,Y)\right] &\leq \max_y \mathbb{E}^\frac1\beta_{P_{X}}\left[f^\beta(X,y)\right]\exp\left(\frac{\alpha-1}{\alpha}I_\alpha(X,Y)\right).
    \end{split}
    \end{equation}
    In particular, if one considers $\alpha\to\infty$ one recovers the following:
       \begin{equation}
        \mathbb{E}_{P_{XY}}\left[f(X,Y)\right] \leq \max_y \mathbb{E}_{P_{X}}\left[f(X,y)\right]\exp\left(\ml{X}{Y}\right).
    \end{equation}
\end{theorem}

This result is particularly interesting when one selects $f$ to be the indicator function of an event, which leads to the following corollary:
\begin{corollary}\label{thm:probBoundIalpha}
Let $E$ be a measurable event, then one has that for every $\alpha>1$ and denoting with $\beta=\alpha/(\alpha-1)$,
\begin{equation}
    P_{XY}(E) \leq \max_y \left(P_X(E_y)\right)^\frac1\beta\exp\left(\frac{\alpha-1}{\alpha}I_\alpha(X,Y)\right).
    \end{equation}
    In particular, if one considers $\alpha\to\infty$ one recovers the following:
       \begin{equation}
        P_{XY}(E) \leq \max_y P_X(E_y)\exp\left(\ml{X}{Y}\right),\label{eq:probBoundML}
    \end{equation}
    where, given $E\subseteq \X\times\Y $, one has that for a given $y\in\Y$ $E_y=\{x:(x,y)\in E\}$.
\end{corollary}
We will now provide three examples in which the bound expressed in~\Cref{eq:probBoundML}, is tight. The first setting represents an example in which the two random variables $X$ and $Y$ are independent and the bound is met with equality:
\begin{example}[independent case]
\label{tightness1}
   Suppose that $E$ is such that $P_X(E_y)=\zeta$ for all $y\in\Y$. In that case we have that, if $X$ and $Y$ are independent then $\ml{X}{Y}=0$ and:
    \begin{equation}
        \zeta=P_{XY}(E) \leq \max_y P_X(E_y)\exp(\ml{X}{Y}) =  \zeta.
    \end{equation}
\end{example}
The second example shows how even when the random variables are very dependent (\textit{e.g.}, $X=Y$), the bound can still be met with equality:
\begin{example}[strongly dependent case]
\label{tightness2}
   Consider the example presented in~\cite{BassilySNSY:18}: suppose $X=Y\sim\mathcal{U}([n])$ then we have that $\ml{X}{Y}=\log n$ and if $E=\{(x,y)\in [n]\times[n] | x=y\}$ then, 
    \begin{equation}
        1= P_{XY}(E) \leq \frac1n \cdot n = 1.
    \end{equation}
\end{example}
The last example is a classical information-theoretic setting in which the random variables are not independent nor one a deterministic function of the other: 
\begin{example}
Suppose $(X,Y)$ is a doubly-symmetric binary source (DSBS) with parameter $p$ for some $p<1/2$ \textit{i.e.}, $X\sim$Ber$(1/2)$, $P_{Y|X}$ is determined by a BSC$(p)$ and, consequently, $Y\sim$Ber$(1/2)$ . Let  $E = \{(x,y): x=y\}$. Then,
\begin{equation}
1 - p = P_{XY}(E) \leq \frac{1}{2}(2(1-p)) = 1 -p.
\end{equation}
\end{example}
The above examples essentially show that when the worst-case behavior (\textit{i.e.}, $\max_yP_X(E_y)$) matches with the average-case behavior (i.e., $\mathbb{E}_{P_Y}[P_X(E_y)]=P_XP_Y(E)$) then the bound involving maximal leakage can be tight regardless of the degree of dependence between the random variables.
Moreover, the following proposition shows that the bound is tight in the following strong sense: if we want to bound the ratio $P_{XY}(E)/(\esssup_{P_Y} P_X(E_y))$ as a function of $P_{Y|X}$ only (i.e., independently of $P_X$ and $E$), then $\exp \{ \ml{X}{Y} \}$ is the best bound we could get (proof in~\Cref{app:MLboundtightness}):

\begin{proposition} \label{prop:MLboundtightness}
Given finite alphabets $\X$ and $\Y$, and a  fixed conditional distribution $P_{Y|X}$, then there exist $P_X$ and $E$ such that~\Cref{eq:probBoundML} is met with equality. That is,
\begin{align}
\label{eq:MLboundtightness}
\revised{\max_{ E \subseteq \X \times \Y}  \max_{P_X}} \log \frac{P_{XY}(E)}{\esssup_{P_Y} P_X(E_Y) } = \ml{X}{Y}.
\end{align}
\end{proposition}

\revised{Notably, the given $(P_X,E)$ which achieve the maximum is such that $P_X(x) > 0 \Rightarrow x \in \argmax_y P_{Y|X}(y|x)$ for some $y \in \Y$, and $P_X(E_y)$ is constant.}

Unfortunately, a similar result does not seem to hold for $I_\alpha$ with $\alpha<\infty$.  
\subsubsection{Case $\alpha<1$}\label{sec:depVsIndepNeg}
Similarly to~\Cref{sec:depVsIndepPos} one can derive a series of bounds that relate expected values of functions (measures of events) under a joint with the expected value of the same function under independence and Sibson $\alpha$ with negative orders. The sign of inequality will be reversed with respect to when $\alpha>1$. 

\begin{theorem}\label{thm:lowerBoundExpSibson}
Let $P_{XY}$ be a joint measure. Let $f:\X\times\Y\to\mathbb{R}^+$ be a $P_{XY}$-measurable function, then one has that for every $\alpha\in(0,1)$ and $\beta=\frac{\alpha-1}{\alpha}$ one has that
\begin{equation}
 \mathbb{E}_{P_{XY}}[f(X,Y)] \min_y\mathbb{E}^{\frac{1}{\beta}}_{P_X}\left[f(X,y)^\beta\right] \exp\left(\frac{\alpha-1}{\alpha}I_\alpha(X,Y)\right).
\end{equation}
\end{theorem}
Recovering a bound that involves probabilities requires a bit more attention with respect to $\alpha>1$.
If $0<\alpha<1$ (which in turn implies $\beta<0$) and if there exists an $x$ with positive measure with respect to $P_X$, such that for every $y$ in the support of $Y$ $f(x,y)=0$, one recovers a trivial lower-bound on $\mathbb{E}_{P_{XY}}[f(X,Y)]$. This prevents us from setting $f = \mathbbm{1}_E$.
One can graphically compare~\Cref{thm:upperBoundExpSibson},~\Cref{thm:probBoundIalpha}, and ~\Cref{thm:lowerBoundExpSibson} via~\Cref{comparison}.

\begin{table*}[!hbpt]
\caption{Behaviour of the bounds expressed in~\Cref{thm:upperBoundExpSibson},~\Cref{thm:probBoundIalpha},~\Cref{thm:lowerBoundExpSibson}}
\label{comparison}
\centering
\begin{tabular}{ |p{2.5cm}||p{5cm}|p{5cm}|} 
 \hline \multicolumn{3}{|c|}{ } \\[-7pt]
 \multicolumn{3}{|c|}{Behaviour of the Bound $\mathbb{E}_{P_{XY}}[f(X,Y)]	\lesseqgtr h_\beta(f(X,Y))\cdot g(I_\alpha(X,Y))$,} \\
 \multicolumn{3}{|c|}{with $ g(I_\alpha(X,Y))=\exp\left((\alpha-1)/\alpha I_\alpha(X,Y)\right)$} \\[4pt]
 \hline
  Range of $\alpha$ &$0<\alpha<1 \implies \beta<0$ &$\alpha>1 \implies \beta>1$\\
 \hline $h_\beta(f(X,Y))$ 
 &$\min_{y}\mathbb{E}^\frac1\beta_{P_X}\left[f(X,y)^{\beta}\right]$ &$\max_{y}\mathbb{E}^\frac1\beta_{P_X}\left[f(X,y)^{\beta}\right]$  \\
 $h_\beta(\mathbbm{1}_E)$  & cannot be provided  & $\max_y (P_X(E_y))^\frac1\beta$\\  Inequality & $\mathbb{E}_{P_{XY}}[f]\geq h_\beta(f)\cdot g(I_\alpha(X,Y))$ & $\mathbb{E}_{P_{XY}}[f]\leq h_\beta(f)\cdot g(I_\alpha(X,Y))$ \\
 References & \Cref{thm:lowerBoundExpSibson}  & \Cref{thm:upperBoundExpSibson} and~\Cref{thm:probBoundIalpha} \\
 \hline
\end{tabular}
\end{table*}

\subsection{Concentration under Dependence and a Learning Setting}\label{sec:concentrationLearning}
The results presented in the previous section lend themselves to applications to concentration of measure and, consequently, to applications in Statistical Learning settings. This consideration represents the backbone of~\cite{EspositoGI:2021}. Consider a classical concentration of measure setting: let $X_1,\ldots,X_n$ be independent random variables over a metric space $(\mathcal{X},d)$ and $f:\mathcal{X}^n\to \mathbb{R}$ a $\kappa$-Lipschitz function with respect to $d$ \textit{i.e.}, 
\begin{equation}
    \sup_{x^n\neq \hat{x}^n} \frac{|f(x^n)-f(\hat{x}^n)|}{d(x^n,\hat{x}^n)} =  \kappa.
\end{equation}
One can typically prove that, under assumptions over $P_{X_i}$ that given a constant $\eta>0$
\begin{equation}
  \mathbb{P}(|f(X^n)-\mathbb{E}[f(X^n)]|\geq \eta) \leq 2\exp\left(-\frac{n\eta}{c \kappa^2} \right), \label{eq:concentrationIndependence}
\end{equation}
where $c$ is a constant typically depending on the approach utilized to prove the concentration result (for a survey, see~\cite{RaginskyS:14}). A natural question (arising from Statistical Learning settings) is: 
can one prove concentration even if $f$ is not independent of $X^n$? In particular, in Statistical Learning one typically assumes that one has access to a learning algorithm $\mathcal{A}:\mathcal{Z}^n\to \mathcal{H}$ with $\mathcal{Z}=\mathcal{X}\times\mathcal{Y}$ which takes as input a sequence of iid samples $Z_i\sim P_{Z}$ and produces as an output a \emph{hypothesis} $h\in\mathcal{H}$. The hypothesis is typically a function $h:\mathcal{X}\to\mathcal{Y}$ \textit{e.g.}, $h$ is a classifier that takes as input a picture and produces as an outcome the corresponding label (the species of the animal or the category of the object one is trying to classify). For an extensive introduction to the topic, the reader is referred to~\cite{shalevShwartzBD:14}.  One then typically wants to assess the performances of such a hypothesis $h$. A common way of doing so is considering the notion of \emph{generalization error}. Consider a loss function $\ell:\mathcal{Y}\times\mathcal{Y}\to\mathbb{R}^+$ one can define the empirical risk of $h$ on $Z^n=(X,Y)^n$ as follows : $L_{Z^n}(h) = \frac1n \sum_{i=1}^n \ell(h(X),Y)$. Similarly, one can define the expected risk $L_{P_Z}(h) = \mathbb{E}_{P_Z}[\ell(h(X),Y)]$. The generalization error of $\mathcal{A}$ is then defined as follows:
\begin{equation}
    \text{gen-err}(\mathcal{A},Z^n) = L_{Z^n}(\mathcal{A}(X^n))- L_{P_Z}(\mathcal{A}(X^n)). \label{eq:genErr}
\end{equation}
One can then try to bound the generalization error in expectation, as it is done in a large body of the literature (\cite{shalevShwartzBD:14,XuR:17,Yeung:09} etc.). Alternatively, one can try to bound~\Cref{eq:genErr} in probability. In this case, one has the following probability:
\begin{equation}
    \mathbb{P}(|\text{gen-err}(\mathcal{A},Z^n)|\geq \eta) \label{eq:concentrationDependence}
\end{equation} 
and one can see~\Cref{eq:concentrationDependence} as an instance of the left-hand side of~\Cref{eq:concentrationIndependence} with $f=\mathcal{A}(Z^n)$. Hence, in this case, one has that $f$ is \textbf{not} independent of $Z^n$. However, one can leverage the results presented in~\Cref{sec:depVsIndep} and, in particular,~\Cref{thm:probBoundIalpha} to provide a bound in this particular setting:
\begin{corollary}\label{thm:genErrBound}
    In a Learning setting as the one described above, if $\ell(h(x),y)=\mathbbm{1}_{h(x)\neq y}$ and for $\alpha>1$, one has the following:
    \begin{equation}
        \mathbb{P}(|\text{gen-err}(\mathcal{A},Z^n)|\geq \eta) \leq 2^\frac1\beta \exp\left(-\frac{n(\alpha-1)}{\alpha}\left(2\eta^2 - \frac{I_\alpha(Z^n,\mathcal{A}(Z^n))}{n}\right)\right).\label{eq:genErrSibs}
    \end{equation}
    In particular, if $\alpha\to\infty$ one recovers the following:
     \begin{equation}
    \mathbb{P}(|\text{gen-err}(\mathcal{A},Z^n)|\geq \eta) \leq  2 \exp\left(-n\left(2\eta^2 - \frac{\ml{Z^n}{\mathcal{A}(Z^n)})}{n}\right)\right).\label{eq:genErrML}
    \end{equation}
\end{corollary}

\begin{remark}
    ~\Cref{thm:genErrBound} holds in more general settings than the one described. In particular, the loss can be $\kappa$-Lipschitz with $\kappa<\infty$ however the so-called $0-1$ loss is a very popular choice in learning settings.
\end{remark}
\begin{remark}
    Re-considering the discussion at the beginning of the Section, one has that~\Cref{thm:genErrBound} and, in particular~\Cref{eq:genErrML},  represent a generalization of concentration inequalities when the function depends on the random variables themselves. Indeed, if $Z^n$ is independent of $\mathcal{A}(Z^n)$ then one has that $I_\alpha(Z^n,\mathcal{A}(Z^n))=0$ for every $\alpha$. In particular, $\ml{Z^n}{\mathcal{A}(Z^n)}=0$ and~\Cref{eq:genErrML} specialises in 
    \begin{equation}
       \mathbb{P}(|\text{gen-err}(Z^n,\mathcal{A}(Z^n))|\geq \eta)\leq 2\exp(-2n\eta^2) 
    \end{equation}
    \textit{i.e.}, it recovers McDiarmid's Inequality.
\end{remark}
The behaviour of~\Cref{eq:sibsMILowerBound} depends on the behaviour of $I_\alpha(Z^n,\mathcal{A}(Z^n))$ as a function of $n$. In general, one has that $I_\alpha(Z^n,\mathcal{A}(Z^n))$ is either finite and at most linear in $n$ or infinite. 
For instance, if $Z_i$ is supported over the entire real line (\textit{e.g.}, $Z_i$ follows a Gaussian distribution) and $\mathcal{A}$ is a deterministic mapping, then $I_\alpha(Z^n,\mathcal{A}(Z^n))$ is infinite. 
It is possible to see that in a variety of settings that are of interest in learning settings (differentially-privacy algorithms, variants of Stochastic Gradient Descent algorithms, etc.) that Maximal Leakage (and, consequently, Sibson's $I_\alpha$ for every $\alpha<\infty$) are finite and sub-linear in $n$ and thus provide exponential concentration in~\Cref{eq:genErrSibs}. 
\revised{Let us provide some concrete examples. Consider a noisy version of the popular Empirical Risk Minimization (ERM) algorithm (cf.~\cite[Corollary 5]{EspositoGI:2021}) \textit{i.e.}, an algorithm $\mathcal{A}$ such that
\begin{equation}
    \mathcal{A}(Z^n) = \argmin_{h\in\mathcal{H}}( L_{Z^n}(h)+N_h),
\end{equation}
where $N_h$ is independent exponential noise. Assume that $|\mathcal{H}|$ is finite (or countably infinite) and indexed by naturals. Let $b_j= j^{1.1}/n^{1/3}$ be the mean of noise corresponding the hypothesis $h_j$. Then $$\mathcal{L}(S\to\mathcal{A}(Z^n))=\sum_{j=1}^{|\mathcal{H}|} \log\left(1+\frac{1}{b_i}\right).$$ Consequently, in this case one has that:
$$\mathbb{P}(|\text{gen-err}(Z^n,\mathcal{A}(Z^n)\geq \eta) \leq 2\exp\left(-n\left(2\eta^2 - \frac{11}{n^{2/3}}\right)\right),$$
which will decay exponentially fast in $n$, for $n$ large enough. The idea of adding carefully calibrated exponential noise is inspired by differentially private algorithms. Many such mechanisms exhibit Maximal Leakage that grows sub-linearly with $n$; see~\cite[Lemmata 3,4,5, Corollary 12]{EspositoGI:2021} for formal statements. A broader set of examples is provided in~\cite{IssaEG:23}, where the Maximal Leakage of a class of learning algorithms—referred to as ``Noisy, Iterative Algorithms'' is shown to be bounded under various assumptions and noise models. This class includes noisy versions of SGD (Stochastic Gradient Descent), such as SGLD (Stochastic Gradient Langevin Dynamics).
} 
Notice that leveraging Sibson $\alpha$-mutual information (as opposed to Shannon's mutual information) allows to provide generalization error bound for over-parametrized models (\textit{e.g.}, deep neural networks) when trained using SGD-like algorithms even if the number of samples $n$ is finite~\cite{IssaEG:23}.

\subsection{Hypothesis Testing}~\label{sec:hypothesisTesting}
Hypothesis testing is another context where both~\Cref{thm:probBoundIalpha} and~\Cref{thm:probBoundConditionalIalpha} can be  utilized. In particular, composite hypothesis testing settings. Suppose one has access to a sequence of $n\geq 1$ iid samples $\{(X_i,Y_i)\}_{i=1}^n$ and suppose that one is interested in knowing, given a certain joint distribution $P_{XY}$, whether:
\begin{enumerate}
    \item $(X,Y)$ are jointly distributed according to $P_{XY}$ (\emph{null hypothesis});
    \item \revised{$X$ is distributed according to $P_X$ and independent of $Y$. \textit{I.e.}, the couple $(X,Y)$ is distributed according to $P_XQ_Y$ where $Q_Y$ is any probability distribution over the support of $Y$ (\emph{alternative hypothesis})}.
\end{enumerate}
\revised{This problem is connected to minimax converse results in channel coding theory and this connection has been formally stated in ~\cite[Section II]{Polyanskiy:13}.}
In these testing settings, $I_\alpha(X,Y)$ can be related to the error exponent in the following sense. Assume that, given $n\geq 1$, $T_n :\{ \mathcal{X}\times\mathcal{Y}\}^n\to\{0,1\}$ is a decision function that, upon observing the sequence of samples  $\{(X_i,Y_i)\}_{i=1}^n$ decides whether the null hypothesis is true ($T_n(\{(X_i,Y_i)\})=0$) or false ($T_n(\{(X_i,Y_i)\})=1$). One can commit two types of error in this case: the type-1 error, denoted with $p_n^1$, represent the probability of deciding the alternative hypothesis when the null hypothesis is true \textit{i.e.}, $p_n^1 = P_{XY}(T_n(\{(X_i,Y_i)\})=1)$ and the type-2 error, denoted with $p_n^2$. In the particular setting advanced here, with type-2 error we will denote the maximum probability (over all the possible choices of $Q_Y$) of choosing the null hypothesis when the alternative is true \textit{i.e.}, $p_n^2 = \sup_{Q_Y} P_XQ_Y(T_n(\{(X_i,Y_i)\})=0).$ The relevance of Sibson's $I_\alpha$ in this setting comes from asking the following question: if one assumes the type-2 error to decay exponentially fast with the number of samples $n$, what happens to the type-1 error? The reply to this question is encapsulated in the following result:
\begin{theorem}\label{thm:hypTesting}
    Let $n>0$ and $T_n : \{\X\times\Y\}^n \to \{0,1\}$ be a deterministic test, that upon observing the sequence $\{(X_i,Y_i)\}_{i=1}^n$ chooses either the null or the alternative hypothesis.
 Assume that there exists an $ R>0$ such that for all  $Q_Y$ one has that $ P_XQ_Y(T_n(\{(X_i,Y_i)\}_{i=1}^n) = 0) \leq \exp(-nR)$. If $\alpha\geq1$,
 \begin{align}
     1- p^1_n \leq \exp\left(-\frac{\alpha-1}{\alpha}n(R-I_\alpha(X,Y))\right).\label{eq:iAlphaHypothesisTesting}
 \end{align}
\end{theorem}
\revised{We believe \Cref{thm:hypTesting} to be known, however we could not find an appropriate reference in the literature.}
~\Cref{eq:iAlphaHypothesisTesting} essentially states that if one requires an exponential decay for the type-2 error with a rate $R$ larger than the Sibson $\alpha$-mutual information the probability of type-1 error will approach $1$. In a more technical sense, it is possible to define the so-called ``error-exponent pairs'' of the hypothesis testing problem (in the Hoeffding/Blahut sense~\cite{Hoeffding:65,Blahut:74}). These pairs can then be fully characterized via $I_\alpha$ however, such a characterization is outside the scope of this work. For more information, we refer the reader to~\cite{TomamichelH:18,LapidothP:18}.
\subsection{Generalized Transportation-Cost Inequalities}\label{sec:tpcSibson}
Leveraging the variational representations that have been provided in this work, one can provide novel transportation-cost-like inequalities. In particular, these results can be linked, much like the ones involving the Kullback-Leibler divergence, to concentration of measure. This link to concentration is different from the one established in~\Cref{sec:concentration}.
 For instance, it is known that one can prove that having a sub-Gaussian-like cumulant generating function $f$ (\textit{i.e.},  $
\log\E_{P_XP_Y}[\exp(\lambda f)]\leq \lambda^2\sigma^2/2$) implies a result akin to Pinsker's inequality:
\begin{equation}
    \E_{P_{XY}}[f]-\E_{P_XP_Y}[f] \leq \sqrt{2\sigma^2 I(X;Y)}.~\label{eq:tpcIneq}
\end{equation}
This property has been employed to relate the expected generalization error to information measures~\cite{XuR:17,EspositoG:22}. One can do something similar for $I_\alpha$ for $\alpha>0$, leveraging~\Cref{eq:varReprIalpha2}. 
Let us briefly introduce an object instrumental for the next result: $\varphi$-Entropies. Given a convex functional $\varphi$ and a random variable $X$ with measure $\xi$ then the $\varphi$-Entropy of $X$ under $\xi$ is defined as follows~\cite{Chafai:04}:
\begin{equation}
    \text{Ent}^\xi_\varphi(X)= \E_{\xi}[\varphi(X)] - \varphi(\E_{\xi}[X])).
\end{equation}
Notice that by Jensen's inequality $\text{Ent}_\varphi(X)\geq 0$. With a slight abuse of notation, we will utilize the same notation even for concave functionals $\varphi$. The positivity of $\text{Ent}_\varphi$ changes accordingly.
One can thus prove the following (see~\Cref{app:proof:thm:TpcIneqIalpha}):
\begin{theorem}\label{thm:TpcIneqIalpha}
    Let $0<\alpha<1$ and $\varphi(x)=\frac{1}{\alpha-1}\log(x)$.
    Let $f:\X\times\Y\to\mathbb{R}$ be a function such that there exists a constant $c$ such that for every $y\in\mathbb{R}$ and $\kappa\in\mathbb{R}$ one has that
    \begin{equation}
        \log\E_{P_X}[\exp(\kappa f(\cdot, y))] \leq \frac{\kappa^2 c}{2} - \text{Ent}^{P_{XY}}_\varphi(\exp((\alpha-1)\kappa f))\label{eq:subGaussGener}
    \end{equation}
    then
    \begin{equation}
         \E_{P_{XY}}[f]-\E_{P_XP_Y}[f] \leq \sqrt{\frac{2cI_\alpha(X,Y)}{\alpha}}.\label{eq:tcpIneqIalpha}
    \end{equation}
\end{theorem}
Notice that, since $0<\alpha<1$, $\varphi$ is a convex function and the corresponding Entropy is non-negative. Thus~\Cref{eq:subGaussGener} is more restrictive than sub-Gaussianity but~\Cref{eq:tcpIneqIalpha} can be tighter than~\Cref{eq:tpcIneq}.  Moreover, if $\alpha\to 1$ then $\text{Ent}_\varphi^{P_{XY}}(\exp((\alpha-1)\kappa f))\to 0$ as well as $I_\alpha(X,Y)/\alpha\to I(X;Y)$ and then,~\Cref{thm:TpcIneqIalpha} boils down to~\cite[Lemma 1]{XuR:17} \textit{i.e.}~\Cref{eq:tpcIneq}.
\begin{remark}
A different version of~\Cref{thm:TpcIneqIalpha} can be proved starting from~\Cref{thm-Sibson-var-alternative-eq-smallalpha}. In particular, instead of assuming that~\Cref{eq:subGaussGener} holds for every $y$ one can ask for it to hold in expectation with respect to $Q_Y^{\star}$ as defined in~\Cref{eq:qstarIalpha}. This would represent a relaxation of~\Cref{eq:subGaussGener}.
\end{remark}
\revised{
\begin{example}
    ~\Cref{eq:subGaussGener} is hard to verify in practice. However, if one has that $|f(X,Y)|\leq M$ a.e., $c = M^2(2-\alpha)$ satisfies~\Cref{eq:subGaussGener} for every $\kappa$. Indeed, assume that $f$ is zero mean with respect to $P_X$ for every $y$, as subtracting the mean of $f$ with respect to $P_X$ would not affect the argument we are about to present. The Hoeffding's Lemma yields:
    \begin{equation}
        \log\mathbb{E}_{P_X}[e^{\kappa f(X,y)}] \leq \frac{\kappa^2M^2}{2}, \label{eq:boundedCGF}
    \end{equation}
    along with 
    \begin{equation}
    \log\mathbb{E}_{P_{XY}}[e^{(\alpha-1)kf(X,Y)}] \leq (\alpha-1)\kappa\mathbb{E}_{P_{XY}}[f(X,Y)]+\frac{\kappa^2(\alpha-1)^2M^2}{2}.\label{eq:boundedCGF2}
    \end{equation}
    Given~\Cref{eq:boundedCGF2}, one has that 
       \begin{align}
           \text{Ent}^{P_{XY}}_{\varphi}((\exp(\alpha-1)\kappa f))  \nonumber &=\kappa \mathbb{E}_{P_{XY}}[f(X,Y)]-\frac{1}{\alpha-1}\log\mathbb{E}_{P_{XY}}[e^{(\alpha-1)\kappa f(X,Y)}] \\
           &\leq \frac{\kappa^2 (1-\alpha)M^2}{2}.
       \end{align}
       Consequently, finding a constant $c$ such that the following holds, suffices: \begin{align}
           \log\mathbb{E}[e^{kf(X,y)}]+\text{Ent}^{P_{XY}}_{\varphi}((\exp(\alpha-1)\kappa f))&\leq \frac{\kappa^2 M^2}{2} + \frac{\kappa^2 (1-\alpha)M^2}{2}  \\
 &\leq \frac{\kappa^2c}{2}.
       \end{align}
       Hence $c$ can be chosen such that
       \begin{equation}
           c \geq M^2(2-\alpha).
       \end{equation}
      and~\Cref{eq:subGaussGener} would hold.
\end{example}}
Clearly~\Cref{thm:TpcIneqIalpha} has been set so that one obtains a bound on the difference of expectation of some function $f$ with respect to the joint and product of the marginals (see~\Cref{eq:tcpIneqIalpha}). This result lends itself well to applications in learning theory. In particular, one can use it to provide a bound on the expected generalization error of a learning algorithm by setting $f$ to be the loss function and assuming it satisfies~\Cref{eq:subGaussGener}~\cite{XuR:17,EspositoG:22}.
\begin{corollary}\label{thm:expGenErrSibson}
    Let $0<\alpha<1$.
    Assume that $\frac1n\sum_i \ell (h,Z_i)$ satisfies~\Cref{eq:subGaussGener} under $P_{Z^n}$ for every $h$ in the hypothesis class, then \begin{equation}
        \mathbb{E}\left[\left|\text{gen-err}(\mathcal{A},Z^n)\right|\right]\leq \sqrt{\frac{2c I_\alpha(Z^n,\mathcal{A}(Z^n))}{\alpha}}
    \end{equation}
\end{corollary}
\begin{remark}

 An analogous result stemming from similar techniques to the ones employed in~\Cref{thm:TpcIneqIalpha} would connect the exponential integrability of a function $f$ to the exponential integrability of the same function under the joint. \textit{I.e.}, if one can bound the cumulant generating function of $f$ for every $y$ under $P_X$ then one can bound the cumulant generating function of $f$ under $P_{XY}$ with $I_\alpha(X,Y)/\alpha$. Said result would be reminiscent of~\Cref{thm:probBoundIalpha}, where an exponentially decaying bound on $P_X(E_y)$ for every $y$ and a bounded $I_\alpha/\alpha$ implies an exponentially decaying bound on $P_{XY}(E)$ for every event $E$.    
\end{remark}
\section{ Applications: Estimation Theory}\label{sec:estimationTheory}
Another stream of applications comes from estimation theory. In particular, one can employ Sibson $\alpha$-mutual information to derive a generalization of Fano's Inequality or in Bayesian estimation settings. In both cases, it is possible to establish impossibility results determining when estimation is not possible regardless of the number of observations one has access to. Moreover, although in a strict sense it does not represent an estimation setting, it is possible to link Sibson $\alpha$-MI to the problem of universal prediction. 

\subsection{Fano-Type Inequalities}\label{sec:fanoTypeIneq}

Recall that the binary R\'enyi divergence is defined for $(p,q) \in [0,1]^2$ as follows:
\begin{align}
    \label{eq:def-binary-renyi}
    d_\alpha(p||q) = \frac{1}{\alpha-1} \log \left( p^\alpha q^{1-\alpha} + (1-p)^\alpha (1-q)^{1-\alpha} \right).
\end{align}

Using the data processing inequality for R\'enyi divergence, one can derive the following:

\begin{theorem}[{~\cite{polyanskiyV:10,Rioul:21}}] \label{thm:fano-alpha}
    Suppose $\X$ is discrete, and let $P_{XY}$ be a joint distribution on $\X \times \Y$. Let $\eps_{X|Y}$ be the optimal probability of error of guessing $X$ from $Y$ (i.e., the error induced by the MAP rule). Then for any $\alpha \in [0,\infty]$,
    \begin{align} \label{eq:thm-fano-alpha}
        I_\alpha(X,Y) \geq d_\alpha \left( \eps_{X|Y} \| 1-p^\star \right),
    \end{align}
    where $p^\star = \max_x P_X(x)$. Moreover, $d_\alpha(\eps \| \delta)$ is a non-increasing function of $\eps$ on the interval $[0,\delta]$. Since $\eps_{X|Y} \leq 1-p^\star$, then
    \begin{align} \label{eq:thm-fano-alpha-2}
        \eps_{X|Y} \geq d_\alpha^{-1} \left( I_\alpha(X,Y) \| 1-p^\star\right),
    \end{align}
    where $d_\alpha^{-1}( \cdot \| 1-p^\star)$ is the inverse of $f: [0,1-p^\star] \rightarrow \mathbb{R}_+$ where $f(\eps) = d_\alpha(\eps \| 1-p^\star)$. 
\end{theorem}
\begin{remark}
    Note that, for a fixed $\delta \in [0,1)$, $f(\eps) = d_\alpha(\eps,\delta) \leq -  \log (1-\delta)$ since it is a non-increasing function of $\eps$ on the interval $[0,\delta]$. As such, if $I_\alpha(X,Y) \geq -\log p^\star$, the right-hand side of~\Cref{eq:thm-fano-alpha-2} will be considered 0 by convention.
\end{remark}
\begin{remark}
    \Cref{thm:fano-alpha} was derived for uniform $P_X$ by Polyanskiy and Verd{\'u}~\cite[Theorem 5.3]{polyanskiyV:10}. The more general version above is due to Rioul~\cite[Theorem 1]{Rioul:21}. For completeness, a proof is included in~\Cref{app:fano-1}.
\end{remark}

\begin{figure}
    \centering
    \includegraphics[scale = 0.6]{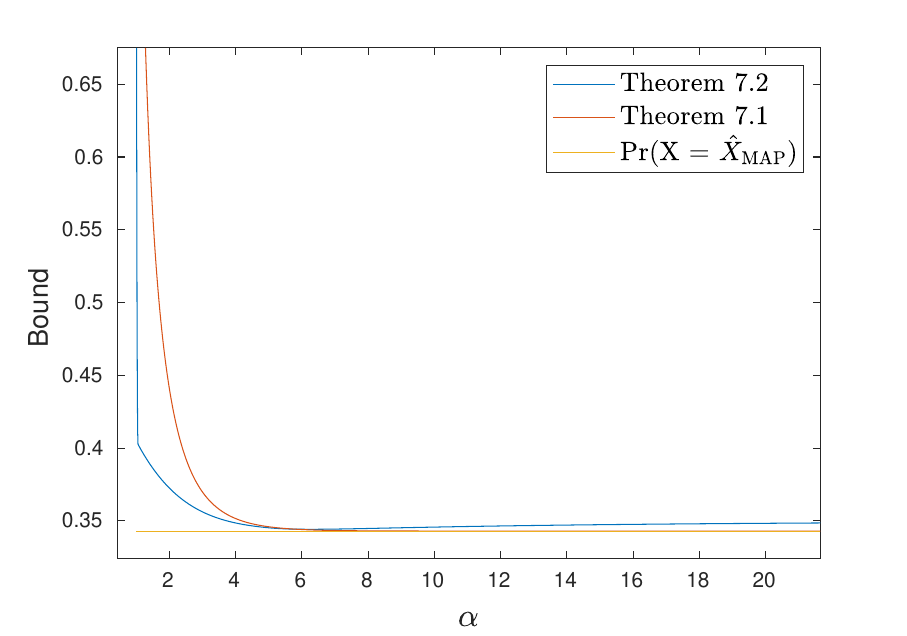}
    \caption{Comparison of Bounds of~\Cref{thm:fano-alpha,thm:fano-like} as a function of $\alpha$. Here, $n=3$, $X^n \stackrel{iid }{\sim} \mathrm{Ber}(1/2)$, and $P_{Y|X} = \mathrm{BSC}(0.3)$, so that $\Pr( X = \hat{X}_{MAP}(Y) ) = 0.7^3 = 0.343.$}
    \label{fig:comparison}
\end{figure}

 A downside of the above inequality is that it does not have a closed form, so it may be difficult to analyze. Alternatively, one could use the variational representation to obtain a parameterized family of bounds by choosing $f = \beta \mathbbm{1} \{ x = \hat{x}\}$ in~\Cref{thm-Sibson-var-alternative-weaker}.  Remarkably, optimizing the bound over $\beta$ may outperform the bound in~\Cref{thm:fano-alpha} (as seen in~\Cref{fig:comparison}). 
\begin{theorem} \label{thm:fano-like}
    Consider a joint distribution $P_{XY}$ where $X$ is a discrete random variable. Let $p^\star = \max_x P_X(x)$ and $\hat{X}_{\mathrm{MAP}}$ be the optimal estimator of $X$ given $Y$. Then for   all $\alpha >1$ and $\gamma >0$,
    \begin{align} 
         \Pr(X = \hat{X}_{\mathrm{MAP}}(Y))\leq  \frac{\left(p^\star(\gamma+1)^{\frac{\alpha}{\alpha-1}}+ \!\! 1-p^\star\right)^{\frac{\alpha-1}{\alpha}} \!\! \exp \! \left \lbrace \frac{\alpha-1}{\alpha} I_\alpha(X,Y) \right\rbrace -1 }{\gamma}.\label{eq:thm-fano-like}
        \end{align}
\end{theorem}
The proof is deferred to~\Cref{app:fano-like}. Taking the limit as $\gamma$ goes to infinity yields the following corollary:
\begin{corollary}  \label{corr:fano-like-gamma-inf}
Under the setting of~\Cref{thm:fano-like},
\begin{align} \label{eq:corr-fano-gamma-inf}
    \Pr(X = \hat{X}_{\mathrm{MAP}}(Y)) \leq  \left( p^\star e^{I_\alpha(X,Y)} \right)^{ \frac{\alpha-1}{\alpha}}.
\end{align}
\end{corollary}
Interestingly, whenever this bound is vacuous, so is the bound in~\Cref{thm:fano-alpha}, i.e., when $I_\alpha(X,Y) \geq - \log p^\star$. Otherwise, letting $\tilde{p} =$ right-hand side of~\Cref{eq:corr-fano-gamma-inf}, we get
\begin{align}
    d_\alpha(\tilde{p} \| p^\star) \! 
    = \! \frac{1}{\alpha-1} \! \log \! \left(  \! e^{(\alpha-1) I_\alpha(X,Y)} \! + \! (1 \!-p^\star) \! \left( \frac{1-\tilde{p}}{1-p^\star} \right)^{\alpha} \right). \notag
\end{align}
It is easy to check that $\tilde{p} > p^\star$, so that the second term inside the $\log$ is smaller than 1, and the first term dominates (especially for large $\alpha$). As such,~\Cref{thm:fano-alpha,eq:corr-fano-gamma-inf} are very close. 

As an illustration, consider $X^n \stackrel{iid }{\sim} \mathrm{Ber}(1/2)$, $n=3$, $P_{Y|X} = \mathrm{BSC}(0.3)$. We plot the various bounds in~\Cref{fig:comparison} as function of $\alpha$.
The bound in~\Cref{thm:fano-alpha} cannot be numerically distinguished from the bound in~\Cref{corr:fano-like-gamma-inf}, so we do not plot it here.

\subsubsection{Bound via Arimoto mutual information}
Moreover, using the connection to Arimoto mutual information (cf.~\Cref{prop:arimoto-sibson-tilted}) and the properties of conditional R\'enyi entropy, we get the following bound on the probability of error:
\begin{theorem}
    \label{thm:fano-alpha-Arimoto}
    Under the same setting of~\Cref{thm:fano-alpha}, for all $\alpha \in (0,1) \cup (1,\infty)$,
    \begin{align}
        \label{eq:thm-fano-alpha-Arimoto}
      \eps_{X|Y} 
      &\geq 1- \exp \left\lbrace -\frac{\alpha-1}{\alpha} \left( H_\alpha(X) - I^A_\alpha (X, Y) \right) \right\rbrace \\
      &= 1- \exp \left\lbrace -\frac{\alpha-1}{\alpha} \left( H_\alpha(X) - I_\alpha (X_\alpha, Y) \right) \right\rbrace,
    \end{align}
    where $ \ds P_{X_\alpha} (x) = \frac{P_X(x)^\alpha}{\sum_{x' \in \X} P_X(x')^\alpha} $.
\end{theorem}
It follows from~\Cref{corr:arimoto-sibson-equality} that, if $X$ is uniform, then we may rewrite $I_\alpha(X_\alpha, Y)$ as $I_\alpha(X,Y)$. The proof is deferred to~\Cref{app:fano-Arimoto}. In the context of the example in~\Cref{fig:comparison}, it is easy to check that the bound coincides with~\Cref{corr:fano-like-gamma-inf}.

\subsection{Generalized Fano Method}\label{sec:fanoMethodSibson}

Let $\cP$ be a family of probability measures on an alphabet $\Y$, and $\theta(P) \in \Theta$ be a parameter of interest. We are interested in estimating $\theta(P)$, with the quality of the estimate measured according to some given loss function $\ell: \Theta \times \Theta \rightarrow \mathbb{R}_+$ which induces a pseudo-metric space $(\Theta,\ell)$.  To that end, we observe $n$ i.i.d samples $Y_1, Y_2, \ldots, Y_n$ distributed according to (an unknown) $P \in \cP$ and generate an estimate $\hat{\theta}(X^n)$ to optimize a min-max objective:
\begin{align}
    \min_{ \hat{\theta(.)}} \max_{P \in \cP} \E_P \left[ \ell(\theta(P), \hat\theta(Y^n)) \right].
\end{align}

We are interested in the minimum number of samples necessary to guarantee the existence of an estimator with a small loss. A standard method to derive converse bounds in this setting is to reduce the estimation problem to a hypothesis testing problem on some subset of $\cP$~\cite{lecam:1973,assouad:1983,yu:1997}. As such, probability of guessing error bounds, and consequently mutual information bounds on the probability of guessing, have been used to derive converse bounds -- as in Fano's method~\cite{fano:1952,yu:1997}. In this section, we generalize Fano's method using results of~\Cref{sec:fanoTypeIneq}. First, we state a useful lemma:

\begin{lemma} \label{lem:Dalpha-bound-Ialpha}
Let $\alpha \in (0, \infty]$. For fixed $P_{Y|X}$, if there exists a $Q_Y$ such that for all $x$,
\begin{align}
    D_{\alpha} \left(P_{Y|X}(\cdot|x) \| Q_Y \right) &\le \beta,
\end{align}
then for all $P_X$,
\begin{align}
    I_{\alpha}(X,Y) & \le \beta.
\end{align}
\end{lemma}
This is a generalization of the ``standard'' lemma used in the Fano method, which only considers KL divergence (i.e., $\alpha=1$) and assumes $D(P_{Y|X}(.|x) || P_{Y|X}(.|x')) \leq \beta$ for all $x,x'$. That is, we slightly relax the condition and allow for any $\alpha \in (0,\infty]$. The proof is deferred to~\Cref{app:lem-fano-method}.

We are now ready to state the generalized Fano method:
\begin{theorem}\label{thm:genFano}[Generalized Fano method]
    Fix $\alpha >0$. Let $r\ge 2$ be an integer and set $\cJ=\{1,2,\ldots,r\}$. Consider  $S = \{P_1, P_2, \ldots, P_r\} \subseteq \cP$ such that there exists a distribution $P^\star$ (not necessarily in $\cP$) satisfying 
    \begin{align}
        D_{\alpha} (P_j || P^\star) & \leq \beta, \text{ for all } j = 1, 2, \ldots, r. \label{eq:thm-gen-fano-Dalpha-cond}
        \intertext{and}
        \ell \left(\theta(P_j), \theta(P_i) \right) & \geq \gamma, \text{ for all } j \neq i. \label{eq:thm-gen-fano-loss-cond}
    \end{align}
    Define $P_{Y|J}$ as $P_{Y|J=j}=P_j$ for all $j \in \cJ$. Then, for all distributions $Q$ on $\cJ$, 
    \begin{align}
        \max_{j \in \{1,2,\ldots, r\}} \E_{P_j} \left[ \ell ( \theta(P_j), \hat{\theta}(Y)) \right] \geq \frac{\gamma}{2} \eps_{J|Y},
    \end{align}
    where $\eps_{J|Y}$ is the optimal probability of error of guessing $J$ from $Y$. In particular, for all $\alpha >1$,
    \begin{align}
        &\max_{j \in \{1,2,\ldots, r\}} \E_{P_j} \left[ \ell ( \theta(P_j), \hat{\theta}(Y)) \right]\geq \frac{\gamma}{2} \left(1-\left(\max_j Q_J(j) e^{I_\alpha(J,Y)}\right)^{\frac{\alpha-1}{\alpha}} \right).
    \end{align}
\end{theorem}
In the last step, we used the bound in~\Cref{eq:corr-fano-gamma-inf} for illustrative purposes. Indeed, any bound on the probability of error, as given in~\Cref{thm:fano-like,thm:fano-alpha,thm:fano-alpha-Arimoto} would yield an analogous result. The proof follows along similar lines as~\cite{yu:1997} and is included in~\Cref{app:genFano}.

\subsection{Bayesian Risk}\label{sec:bayesianRisk}
\label{sec:bayesianFramework}
	Another example of application of~\Cref{thm:probBoundIalpha} comes from estimation procedures in a Bayesian setting. Let $\mathcal{W}$ denote the \revised{finite} parameter space and assume that we have access to a prior distribution over $\mathcal{W}$, $P_W$. Suppose that we observe $W\sim P_W$ through the family of distributions $\mathcal{P}= \{ P_{X|W=w}: w\in\mathcal{W} \}.$ Given a function $\phi:\mathcal{X}\to\hat{\mathcal{W}}$, one can then estimate $W$ from $X\sim P_{X|W}$ via $\phi(X)=\hat{W}$. Let us denote with $\ell:\mathcal{W}\times\hat{\mathcal{W}}\to \mathbb{R}^+$ a loss function, the Bayesian risk is defined as:
	\begin{equation}
		R_B= \inf_\phi \mathbb{E}_{P_{W\hat{W}}}[\ell(W,\phi(X))] = \inf_\phi\mathbb{E}_{P_{W\hat{W}}}[\ell(W,\hat{W})].\label{risk}
	\end{equation}
	Our purpose is to lower-bound $R_B$ using Sibson $\alpha$-mutual information. In particular, one can connect the expected value of $\ell$ under the joint $P_{W\hat{W}}$ to 
	\begin{itemize}
		\item the  expected value of the same function under the product of the marginals ($P_WP_{\hat{W}}$) or a ``small-ball probability'';
		\item the Sibson $\alpha$-mutual information, measuring how much information the observations retain about the parameter. 
	\end{itemize}
    \revised{Small-ball probabilities date back to~\cite{KuelbsLi:93} with more recent advances and applications in~\cite{Li:99,Bobkov:15, Xu:17}.}
	This particular technique allows us to retrieve a lower bound which is independent of the specific choice of the estimator $\phi$. 
	Let us denote the so-called small-ball probability  as follows
 \begin{equation}L_W(\rho)= \sup_{\hat{w}\in\mathcal{\hat{W}}} L_W(\hat{w},\rho) =  \sup_{\hat{w}\in\mathcal{\hat{W}}}P_W(\ell(W,\hat{w})< \rho).\label{smallBall}\end{equation}
 Then one can prove the following:
 	\begin{theorem}\label{thm:sibsMIResultBayesRisk}
		Consider the Bayesian framework just described. The following must hold for every $\alpha>1$ and $\rho>0$:
		\begin{equation}
			R_B\geq \rho\left(1- \exp\left(\frac{\alpha-1}{\alpha}\left(I_\alpha(W,X) + \log(L_W(\rho))\right) \right)\right). \label{eq:sibsMILowerBound}
    \end{equation}
    Taking the limit of $\alpha\to \infty$ one recovers the following:
    \begin{equation}
			R_B\geq \sup_{\rho>0}\rho\left(1- \exp\left(\ml{W}{X} + \log(L_W(\rho)) \right)\right). \label{eq:maximalLeakgeResultBayesRisk}
		\end{equation}
\end{theorem}

\subsubsection{Bernoulli Bias Estimation}
\begin{figure}
\label{fig:riskLowerBoundEx}
\centering
\includegraphics[scale=0.375]{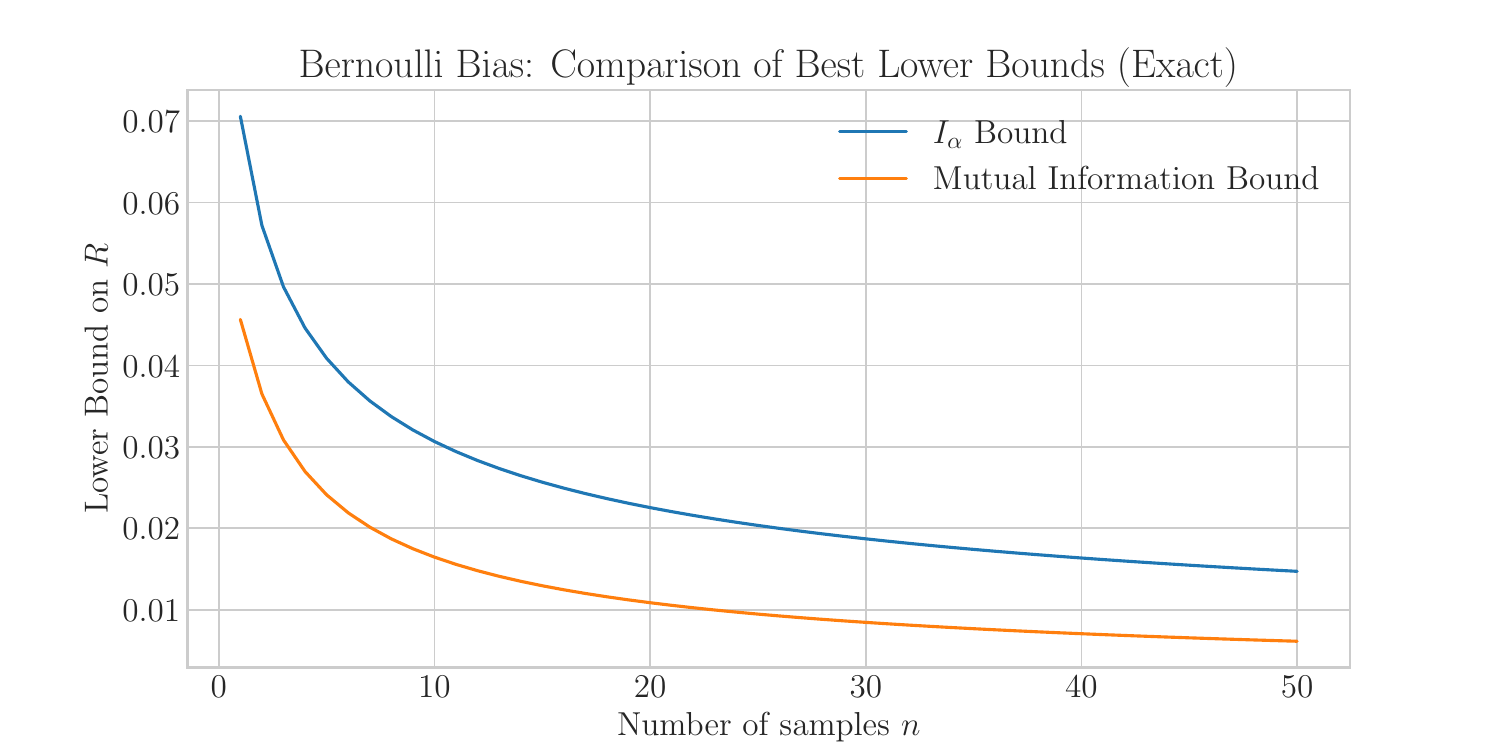}
			\label{fig:bernoulli_best_params}
   \caption{Comparison between the lower-bound provided by~\Cref{eq:iAlphaBernoulliBias} and {\cite[Equation~(19)]{XuR:17}}.
		The quantities are analytically maximized over $\rho$ and numerically optimized over $\alpha>1$.}
	\end{figure}

As an example of application of~\Cref{thm:sibsMIResultBayesRisk} consider the following setting. Assume that $W$ is uniformly distributed between $0$ and $1$ and that given $W=w$ our observations $X_i$ follow a Bernoulli distribution with parameter $w$. It is possible to see that \revised{using} the sample mean estimator \textit{i.e.}, $\hat{W} = \frac1n \sum_{i=1}^n X_i$: \begin{equation}R_B\leq \frac{1}{\sqrt{6n}}.\label{eq:bernoulliBiasUpperBound}\end{equation} Let us now provide a lower-bound on the problem leveraging~\Cref{thm:sibsMIResultBayesRisk}. In this case one has that \begin{equation}
		\begin{split}
		    &\exp\left(\frac{\alpha-1}{\alpha}I_\alpha(W,X^n)\right)= \sum_{k=0}^n \binom{n}{k}\left(\frac{\Gamma(k\alpha+1)\Gamma((n-k)\alpha+1)}{\Gamma(n\alpha+2)}\right)^\frac{1}{\alpha}.\label{eq:iAlphaBernoulliBiasValue}
		\end{split}
	\end{equation}
 Consider the loss function $\ell(w,\hat{w})=|w-\hat{w}|$, one can see that
    \begin{equation}
        L_W(\rho)=\sup_{\hat{w}}P_W(|W-\hat{w}|<\rho)=2\rho.
    \end{equation}
    Consequently,~\Cref{eq:sibsMILowerBound} specializes into: \begin{equation}
		R_B\geq \sup_{\rho>0}\sup_{\alpha>1}\rho\left(1-(2\rho)^\frac{\alpha-1}{\alpha}\exp\left(\frac{\alpha-1}{\alpha}I_\alpha(W,X^n)\right)\right) \label{eq:iAlphaBernoulliBias}.
	\end{equation}
For any given $\alpha>1$ the lower-bound in~\Cref{eq:iAlphaBernoulliBias} will be larger than similar lower-bounds involving Shannon's mutual information (cf.~\cite{XuR:17}). In~\Cref{fig:riskLowerBoundEx}, one can see the improvement that~\Cref{eq:iAlphaBernoulliBias} brings over~\cite[Equation 19)]{XuR:17} when numerically optimized over $\rho$ and $\alpha$. Given the difficulty in assessing the dependence on the various parameters in~\Cref{eq:iAlphaBernoulliBiasValue}, in order to better assess the power of~\Cref{thm:sibsMIResultBayesRisk} one can, for example, consider the limit of $\alpha\to\infty$. In that case, one can prove that \begin{equation}
        \ml{W}{X^n} \leq \log\left(2+\sqrt{\frac{\pi n}{2}}\right),\label{eq:upperBoundLeakageBernoulliBias}
    \end{equation}
leading us to the following lower-bound on the Bayesian risk:
\begin{align}
		R_B
		\geq \sup_{\rho>0} \rho\left(1-\left( 2+ \sqrt{\frac{\pi n}{2}}\right)2\rho\right). \label{eq:mlBernoulliBiasOverRho}
	\end{align}
 The quantity in~\Cref{eq:mlBernoulliBiasOverRho} does not depend on $\alpha$ anymore and can now be analytically optimized over $\rho$, leading to the following lower-bound:
\begin{equation}
		R_B\geq \frac{1}{8\left( 2+ \sqrt{\frac{\pi n}{2}}\right)}\label{eq:maximalLeakageBernoulliBias},\end{equation}
	which, for $n$ large enough (\textit{i.e.}, $n\geq 127/\pi \approx 41$), can be further lower-bounded as follows $$R_B\geq \frac{1}{5\sqrt{2\pi n}}.$$ Surprisingly, Maximal Leakage already offers a lower-bound that matches the upper-bound up to a constant (cf.~\Cref{eq:bernoulliBiasUpperBound}) and is simple to compute.~\Cref{eq:maximalLeakageBernoulliBias} provides a larger lower-bound than the one provided using mutual information (cf.~\cite[Corollary 2]{XuR:17}) for $n\geq 1$. Moreover, in order to achieve a lower-bound which behaves similarly to~\Cref{eq:bernoulliBiasUpperBound} in terms of the decay with respect to $n$, the authors in~\cite{XuR:17} need additional machinery and can only provide an \textit{asymptotic} lower bound on the risk.
    Another advantage of using the simple expression provided by Maximal Leakage is that Maximal Leakage depends on $P_W$ only through the support. This means that if one has access to an upper-bound on $L_W(\rho)$ that does not employ any additional knowledge of $P_W$, the resulting lower-bound on the risk would apply to any $W$ whose support is the interval $[0,1]$. This would render the approach even more general than it already is. For more details and examples of application, the reader is referred to~\cite{EspositoVG:23}.
    
\subsection{Universal Prediction}\label{sec:universalPrediction}

In~\cite{BondaschiG:22isit,bondaschi2023alphanml}, Sibson $\alpha$-mutual information is connected to the problem of universal prediction.
The quality of a universal predictor $\hat{p}$ is assessed by its regret with respect to a model class containing distributions $p_{\theta}(x^n),$ for all $\theta \in {\Theta},$ where $\Theta$ is the set of parameters.
To connect with Sibson $\alpha$-mutual information, we
consider the following generalized regret measure, defined for any $\alpha \geq 1$:
\begin{equation}
\begin{split}
R_{\alpha}(\hat{p}) &= \sup_{\theta\in\Theta} D_{\alpha}(p_{\theta}\|\hat{p})\\&= \sup_{\theta\in\Theta} \frac{1}{\alpha-1} \log \sum_{x^n\in\mathcal{X}^n} p_{\theta}(x^n) \left(\frac{p_{\theta}(x^n)}{\hat{p}(x^n)}\right)^{\alpha -1} \label{sec-universalprediction-eq-alpharegret}
\end{split}
\end{equation}
which we call $\alpha$-regret.
This regret measure itself is not standard, but interpolates between the two most important standard regret measures. Namely, the limit of $R_{\alpha}(\hat{p})$ as $\alpha \rightarrow 1$ is the standard average log-loss regret, while the limit of $R_{\alpha}(\hat{p})$ as $\alpha \rightarrow \infty$ is the standard worst-case log-loss regret (see~\cite{bondaschi2023alphanml} for an in-depth discussion).
The connection is then that the minimum regret is precisely given by the maximum value of Sibson $\alpha$-mutual information.
\begin{theorem}
\label{thm-alphaNML-MaxIAlpha}
For universal prediction with respect to $\alpha$-regret and a class of distributions $\{ p_\theta(x^n) : \theta \in \Theta\},$
assume that there exists a probability distribution $p^*(\theta)$ on $\Theta$ such that, with $W^*\sim p^*(\theta),$
\begin{equation}
I_{\alpha}(W^*,X^n) = \sup_{V} I_{\alpha}(V,X^n),
\end{equation}
where the supremum is over all random variables $V$ on $\Theta,$ and the conditional distribution of $X^n$ is given by $p_\theta(x^n).$
Then, the optimum predictor (minimizing the regret in~\Cref{sec-universalprediction-eq-alpharegret}) is
\begin{equation}
\hat{p}_{\alpha}(x^n) = \frac{\left\{\int_{\Theta} p^*(\theta)p_{\theta}^{\alpha}(x^n)\,d\theta\right\}^{1/\alpha}}{\sum_{\tilde{x}^n} \left\{\int_{\Theta} p^*(\tilde{\theta})p_{\tilde{\theta}}^{\alpha}(\tilde{x}^n)\,d\tilde{\theta}\right\}^{1/\alpha}},
\end{equation}
which is also referred to as the $\alpha$-NML with prior $p^*(\theta)$, and the corresponding minimum regret is
\begin{equation}
R_{\alpha}(\hat{p}_\alpha) = I_{\alpha}(W^*,X^n) .
\end{equation}
\end{theorem}

This theorem is proved in~\cite[Theorem 3]{bondaschi2023alphanml}.
The limiting cases of this theorem are well known.
For the limit $\alpha \rightarrow 1,$ the result is due to Gallager~\cite{gallager1} and often referred to as the capacity-redundancy theorem.
For the limit $\alpha\rightarrow\infty,$ the result is due to Shtarkov~\cite{shtarkov1}.
In this case, the optimum predictor is commonly referred to as {\it normalized maximum likelihood} (NML) predictor, and the corresponding regret as the {\it Shtarkov sum,} which in the language used here would be referred to as the maximal leakage from the parameter set $\Theta$ to the sequences $X^n,$ denoted as $\ml{\Theta}{X^n},$ as in~\Cref{eq-def-maxleakage} above.
We note that in this limiting case, the distribution $p^*(\theta)$ appearing in~\Cref{thm-alphaNML-MaxIAlpha} is immaterial since maximal leakage does not depend on the marginal distribution of the input.

\begin{remark}
    By the connection between universal prediction and universal compression, an analogous result can be obtained for universal compression if one studies the source coding setting advocated by Campbell~\cite{campbell1}. This is fully developed in~\cite{yagli1}, where it is shown that the minimal redundancy is equal to the maximum value of Sibson $\alpha$-mutual information is a sense similar to~\Cref{thm-alphaNML-MaxIAlpha}.
\end{remark}


\section{Extension to Conditional Sibson $\alpha$-Mutual Information}\label{sec:conditional}
\revised{There are multiple ways to define conditional mutual information, each highlighting different perspectives on dependence in a triplet of random variables.  Given that Sibson’s $\alpha$-mutual information generalizes mutual information, one can draw inspiration from these various approaches to formulate a conditional version. 
For example, one could define \begin{equation*}
    I(X;Y|Z)=I(X,Z;Y)-I(Z;Y).
\end{equation*}
Or as another example, one can also define 
\begin{equation*}
    I(X;Y|Z) = D(P_{XYZ}\|P_XP_{Z|X}P_{Y|Z}).
\end{equation*}
Or, as a third class of examples, one can define
\begin{align*}
    I(X;Y|Z) &= \min_{Q_{Z|X}} D(P_{XYZ}\|P_XQ_{Z|X}P_{Y|Z}) \text{ or }\\
    I(X;Y|Z) &= \min_{Q_{Y|Z}} D(P_{XYZ}\|P_XP_{Z|X}Q_{Y|Z}) \text{ or } \\
    I(X;Y|Z) &= \min_{Q_X} D(P_{XYZ}\|Q_XP_{Z|X}P_{Y|Z}).
\end{align*}
In the case of Kullback-Leibler divergence and Shannon's mutual information, all of these definitions (including those via minimization) coincide.
However, when dealing with R\'enyi divergences, each of these definitions leads to a different quantity, as we will see in the rest of the Section.}

Given $\alpha\neq 1$, it is somewhat unclear how one should define conditional information. Several candidate definitions exist and are all equally promising. In this section, we briefly discuss this context and point to some of the operational meanings that appear naturally as a consequence of the closed-form expressions that these information measures admit.
Mimicking the approach undertaken in~\Cref{def-SibsonalphafromDiv} along with the considerations made about Shannon's mutual information just above, we will define conditional Sibson $\alpha$-mutual information of $X$ and $Y$ given $Z$ as a minimization of R\'enyi divergences measuring how far the joint is from the Markov chain $X-Z-Y$. Differently from the two variables case, one can choose to minimize over different measures \textit{i.e.}, one can advance any of the following definitions
\begin{equation}
    I_\alpha(X,Y|Z) \stackrel{?}{=} \begin{cases} \min_{ Q_{Y|Z} } D_\alpha(P_{XYZ}\|P_{X|Z}Q_{Y|Z}P_Z) \text{ or } \\ \min_{ Q_{X|Z} } D_\alpha(P_{XYZ}\|Q_{X|Z}P_{Y|Z}P_Z) \text{ or }\\ \min_{ Q_{Z} } D_\alpha(P_{XYZ}\|P_{X|Z}P_{Y|Z}Q_Z).
    \end{cases}
\end{equation}
Moreover, one can consider different factorizations of the joint measure corresponding to the Markov chain $X-Z-Y$. 
Rather than making an explicit choice or analyzing every possible definition, we will briefly consider the similarities and the properties these objects possess. To do so, let us go through some examples and extrapolate said properties. We will distinguish the resulting information measure with a superscript indicating the family of measures over which perform the minimization. For the remainder of this section, consider a discrete setting. For instance, minimizing over conditional measures $Q_{Y|Z}$ leads to the following definition:
\begin{definition}
Let $X,Y,Z$ be three discrete random variables jointly distributed according to $P_{XYZ}$. Denote with $P_Z$ the corresponding marginal over the support of $Z$ and, for a given $z$, $P_{X|Z=z}$ the corresponding conditional distributions over the support of $X$. One can define the following conditional Sibson $\alpha$-mutual information: 
    \begin{equation}
    I^{Y|Z}_\alpha(X,Y|Z) = \min_{Q_{Y|Z}} D_\alpha(P_{XYZ}\|P_{X|Z}Q_{Y|Z}P_Z),\label{eq:conditionalY|Z}
\end{equation}
where the minimization is over all the conditional distributions $Q_{Y|Z=\cdot}$ over the support of $Y$.
\end{definition}

\Cref{eq:conditionalY|Z} admits a closed-form expression given in the following equation:
\begin{equation}
\begin{split}
    &I^{Y|Z}_\alpha(X,Y|Z) = \frac{\alpha}{\alpha-1}\log \E^\frac1\alpha_{P_Z}\left[\E^\alpha_{P_{Y|Z}}\left[\E^{\frac1\alpha}_{P_{X|Z}}\left[\left(\frac{P_{XYZ}}{P_ZP_{X|Z}P_{Y|Z}}\right)^\alpha\right]\right]\right]. \label{eq:closedFormCondY|Z}
\end{split}
\end{equation}
Similarly, if one chooses to minimize over $Q_Z$, the following definition arises:
\begin{definition}
Let $X,Y,Z$ be three discrete random variables jointly distributed according to $P_{XYZ}$. For a given $z$, denote with $P_{X|Z=z}$ the corresponding conditional distribution over the support of $X$ and with $P_{Y|Z=z}$ the corresponding conditional distribution over the support of $Y$. One can define the following conditional Sibson $\alpha$-mutual information: 
\begin{equation}
    I^{Z}_\alpha(X,Y|Z) = \min_{Q_{Z}} D_\alpha(P_{XYZ}\|P_{X|Z}P_{Y|Z}Q_Z),\label{eq:conditionalZ}
\end{equation}
where the minimization is over all the probability measures $Q_Z$ over the support of $Z$.
\end{definition}

\Cref{eq:conditionalZ} also admits a closed-form expression given in the following equation:
\begin{equation}
\begin{split}
    &I^{Z}_\alpha(X,Y|Z) =\frac{\alpha}{\alpha-1}\log \E_{P_Z}\left[\E^\frac1\alpha_{P_{Y|Z},P_{X|Z}}\left[\left(\frac{P_{XYZ}}{P_ZP_{X|Z}P_{Y|Z}}\right)^\alpha\right]\right].\label{eq:closedFormCondZ}
\end{split}
\end{equation}
The proofs of these statements follow from adaptations of the proof of~\Cref{thm:Ialpha}: every $Q_{\{\cdot\}}$ we are minimizing over in the second argument of $D_\alpha$ will have a corresponding expectation operator. Changing the measure of said expectation operator to the corresponding marginal $P_{\{\cdot\}}$ (or conditional) obtained from $P_{XYZ}$, along with an application of Jensen's inequality achieves the closed-form expression. Equality can then be proven by selecting $Q_{\{\cdot\}}^\star$ to be a carefully chosen tilting with respect to $P_{\{\cdot\}}$ (cf.~\Cref{eq:qstarIalpha}). For instance, if one minimizes over $Q_Z$, the corresponding $Q_Z^\star$ is the following:
\begin{equation}
\begin{split}
    &Q_Z^\star(z) =\frac{P_Z(z)\!\left(\!\sum_{x,y}  P_{XY|Z}(x,y,z)^{\alpha}(P_{X|Z}(x,z)P_{Y|Z}(y,z))^{1-\alpha}\!\right)^{\!\frac{1}{\alpha}}}{\mathbb{E}_{P_{Z}}\left[\!\left(\sum_{x,y} P_{XY|Z}(x,y,z)^{\alpha}(P_{X|Z}(x,z)P_{Y|Z}(y,z))^{1-\alpha}\! \right)^{\frac{1}{\alpha}}\!\right]}.
\end{split}
\end{equation}
Similarly, if one minimizes over $Q_{Y|Z}$ then the corresponding tilting, for a given $z$ is given in the following:
\begin{equation}
\begin{split}
    &Q_{Y|Z=z}^\star(y)=\frac{P_{Y|Z=z}(y)\left(\sum_{x} P_{XY|Z}(x,y,z)^{\alpha}(P_{X|Z}(x,z)P_Z(z))^{1-\alpha} \right)^{\frac{1}{\alpha}}}{\mathbb{E}_{P_{Y|Z=z}}\left[\left(\sum_{x} P_{XY|Z}(x,y,z)^{\alpha}(P_{X|Z}(x,z)P_Z(z))^{1-\alpha} \!\right)^{\frac{1}{\alpha}}\right]}.
\end{split}
\end{equation}
The expressions in~\Cref{eq:closedFormCondY|Z} and \Cref{eq:closedFormCondZ} are clearly distinct. 
In particular, one can see that $$\lim_{\alpha\to\infty} I_\alpha^{Y|Z}(X,Y|Z)=\ml{X}{Y|Z},$$ while the same does not hold for $I_\alpha^Z(X,Y|Z)$.
However, for both of them is true that $\lim_{\alpha\to 1} I_
\alpha^{\cdot}(X,Y|Z) = I(X;Y|Z)$.
Moreover, we do not know whether any of these satisfies a mutual information-like chain-rule \textit{i.e.},
\begin{equation}
    I_\alpha(X,(Y,Z))\leq I_\alpha(X,Z) + I^?_\alpha(X,Y|Z),
\end{equation}
even though, we do know that 
\begin{align}
 I(X;(Y,Z))&= I(X;Z) + I(X;Y|Z)  \\
 \ml{X}{(Y,Z)} &\leq \ml{X}{Z} + \ml{X}{Y|Z}.
\end{align}
Given the closed-form expressions and the technique leveraged in this work, we can, however, find applications for these quantities in hypothesis testing settings. Said links have been established in~\cite{TomamichelH:18,EspositoWG:21}. Moreover with a similar approach to~\Cref{sec:depVsIndep}, one can see $I_\alpha(X,Y|Z)$ as three nested norms of the Radon-Nikodym derivative of the joint $P_{XYZ}$ with respect to a measure corresponding to the Markov chain $X-Z-Y$. Similarly to the unconditional version, conditional $I_\alpha$ also involves a minimization and there is a variety of choices with respect to which measure to minimize over,~\Cref{sec:conditional}. Hence, as before, the norm corresponding to the measure we are minimizing over will be an $L^1$-norms, while the remaining two will be $L^\alpha$-norms. Similarly to the two variables case, given $P_{XYZ}$, one can relate via an inequality:
\begin{itemize}
    \item the expected value of a function (probability of event) under the joint $P_{XYZ}$;
    \item with the nested norms of the same function (probability of event) under the Markovity assumption $X-Z-Y$;
    \item nested norms of the Radon-Nikodym derivative of the joint with respect to the measure formalizing the Markov chain $X-Z-Y$ (\textit{e.g.}, conditional Sibson $\alpha$-mutual information).
    
\end{itemize}
For illustrative purposes, let us choose as a conditional version $I_\alpha^{Y|Z}$ (see~\Cref{eq:conditionalY|Z}). One can thus prove the following result:
\begin{theorem}\label{thm:functionBoundCondIalpha}
    Let $P_{XYZ}$ be a joint measure. Let $f:\X\times\Y\times Z \to\mathbb{R}^+$ be a $P_{XYZ}$-measurable function, then one has that for every $\alpha>1$ and denoting with $\beta=\alpha/(\alpha-1)$,
    \begin{equation}
    \begin{split}
        &\mathbb{E}_{P_{XYZ}}\left[f(X,Y,Z)\right]\leq\mathbb{E}^\frac{1}{\beta}_{P_Z}\left[\max_{y:P_{Y|Z}(y)>0} \mathbb{E}_{P_{X|Z}}\left[f^\beta(X,y,Z)\right]\right]\exp\left(\frac{\alpha-1}{\alpha}I^{Y|Z}_\alpha(X,Y|Z)\right).  
    \end{split}
    \end{equation}
    In particular, if one considers $\alpha\to\infty$ one recovers the following:
       \begin{equation}
       \begin{split}
      &\mathbb{E}_{P_{XYZ}}\left[f(X,Y,Z)\right]
            \leq\mathbb{E}_{P_Z}\left[\max_{y:P_{Y|Z}(y)>0} \mathbb{E}_{P_{X|Z}}\left[f(X,y,Z)\right]\right]\exp\left(\ml{X}{Y|Z}\right). 
       \end{split}
    \end{equation}
\end{theorem}

Consequently, similarly to before, one can relate probabilities of events under different distributions (joint and Markovian) selecting $f$ to be the indicator function of the event in~\Cref{thm:functionBoundCondIalpha}.
\begin{corollary}\label{thm:probBoundConditionalIalpha}
    Let $E$ be a measurable event, then one has that for every $\alpha>1$ and denoting with $\beta=\alpha/(\alpha-1)$,
    \begin{equation}
    \begin{split}
        &P_{XYZ}(E) \leq \mathbb{E}^\frac{1}{\beta}_{P_Z}\left[\max_{y:P_{Y|Z}(y)>0}P_{X|Z}(E_{y,Z})\right]\exp\left(\frac{\alpha-1}{\alpha}I_\alpha^{Y|Z}(X,Y|Z)\right).
    \end{split}
    \end{equation}
     where, given $E\subseteq \X\times\Y\times \mathcal{Z}$, one has that for a given $y\in\Y$ and $z\in\mathcal{Z}$, $E_{y,z}=\{x:(x,y,z)\in E\}$.
    In particular, if one considers $\alpha\to\infty$ one recovers the following:
    \begin{equation}
    \begin{split}
        &P_{XYZ}(E) \leq \mathbb{E}_{P_Z}\left[\max_{y:P_{Y|Z}(y)>0}P_{X|Z}(E_{y,Z})\right]\exp\left(\ml{X}{Y|Z}\right).
    \end{split}
    \end{equation}
\end{corollary}

\section{Discussion, Extensions, and Open Problems} \label{sec:discussion}

\subsection{Negative Values of $\alpha$}
\label{app:negativeAlpha}


In this paper, we have restricted attention to non-negative values of $\alpha.$ For this case, there is a rich and emerging literature. Needless to say, it is also tempting to consider negative values of $\alpha.$ Indeed, many of the techniques presented here can be extended to negative values of $\alpha,$ leading to interesting and useful results. However, providing a {\it meaningful definition} of Sibson $\alpha$-mutual information for negative values is non-trivial. 
 One can, for instance, define such an object following the classical approach, like in~\Cref{def-SibsonalphafromDiv}. The first step in this direction would be to extend the definition of R\'enyi divergence to negative $\alpha.$ Leaving the definition of $D_\alpha$ unaltered would violate the fundamental properties of a divergence, like non-negativity and the data-processing inequality, see e.g.~\cite[Section V]{renyiDiv}.
 Hence, in this case, one would need to either alter the definition of R\'enyi divergence itself or that of $I_\alpha$ as expressed in~\Cref{def-SibsonalphafromDiv}.  
 
Another possibility comes from undertaking the norm perspective considered in~\Cref{sec:normPerspective}. Indeed, one could extend the nested-norm functional in~\Cref{eq:normRepresentationEveryAlpha} to negative values of $\alpha$. This implies that the corresponding object will no longer be a norm (which is already the case when $0<\alpha<1$). 
Moreover, said object would immediately find applications in bounds \textit{e.g.}, 
 extending the result presented in~\Cref{sec:depVsIndepNeg} to negative values of $\alpha$ or in directly lower-bounding the Bayesian Risk (without employing Markov's inequality like we did in~\Cref{thm:sibsMIResultBayesRisk}). Some of these ideas were already advanced in~\cite{EspositoVG:23,EspositoVG:22}. However, one would incur into measure-theoretic subtleties and absolute continuity issues which are outside the scope of this work. Hence, these results will be formalised and presented in a separate manuscript.

\subsection{Probabilistic Variational Representation}
Drawing inspiration from~\cite{MaximalLeakage:20} one can also try to achieve a representation for $I_\alpha$ that mimics the one advanced in~\Cref{def:maxLeakage}. However, for finite $\alpha$ (as opposed to $\alpha\to\infty$ \textit{i.e.}, Maximal Leakage) one can only prove a lower bound:
\begin{theorem}\label{thm:markovChainRepresentation}
    Let $X,Y$ be two random variables jointly distributed according to $P_{XY}$ and let $\alpha\in (1,+\infty]$. The following holds true: \begin{equation}
    \exp\left(\frac{\alpha-1}{\alpha}I_\alpha(X,Y)\right) \geq \sup_{U:U-X-Y}  \frac{\mathbb{E}_{P_Y}\left[\left\lVert P_{U|Y}\right\rVert_{L^\alpha(P_U)}\right]}{\left\lVert\left\lVert P_{U|X}\right\rVert_{L^\beta(P_X)}\right\rVert_{L^\alpha(P_U)}}.\label{eq:varRepresProb}
\end{equation}
\end{theorem}
The following example shows the strictness of~\Cref{eq:varRepresProb} for every finite order $\alpha>1$.
\begin{example}[Strictness of~\Cref{eq:varRepresProb} for every finite order]
Equality in~\Cref{eq:varRepresProb} does not hold in general for any finite $\alpha>1$. Let
\[
X\sim\mathrm{Bernoulli}\left(\frac12\right),
\qquad
Y=X \quad\text{almost surely},
\]
and let $\beta=\alpha/(\alpha-1)$. For every auxiliary random variable $U$ satisfying $U-X-Y$, we have
\begin{align}
\mathbb{E}_{P_Y}\left[
\left\|
\frac{dP_{U|Y}}{dP_U}
\right\|_{L^\alpha(P_U)}
\right] &=
\frac12
\left\|
\frac{dP_{U|X=0}}{dP_U}
\right\|_{L^\alpha(P_U)}
+
\frac12
\left\|
\frac{dP_{U|X=1}}{dP_U}
\right\|_{L^\alpha(P_U)}
\nonumber\\
&\leq
\left[
\frac12
\left\|
\frac{dP_{U|X=0}}{dP_U}
\right\|_{L^\alpha(P_U)}^\alpha
+
\frac12
\left\|
\frac{dP_{U|X=1}}{dP_U}
\right\|_{L^\alpha(P_U)}^\alpha
\right]^{1/\alpha}
\nonumber\\
&=
\left\|
\left[
\frac12
\left(\frac{dP_{U|X=0}}{dP_U}\right)^\alpha
+
\frac12
\left(\frac{dP_{U|X=1}}{dP_U}\right)^\alpha
\right]^{1/\alpha}
\right\|_{L^\alpha(P_U)},
\label{eq:counterexample-power-mean}
\end{align}
where the inequality follows from the power-mean inequality \textit{i.e.}, given $a,b\geq 0$ and $\alpha>1$ one has that due to H\"older's inequality
\begin{equation}
    \frac{a+b}{2} \leq \left(\frac{a^\alpha+b^\alpha}{2}\right)^{1/\alpha}.
\end{equation}

Suppose first that $1<\alpha\leq2$. In this case $\alpha\leq\beta$, and hence, for $P_U$-almost every $u$,
\begin{align}
&\left[
\frac12
\left(\frac{dP_{U|X=0}}{dP_U}(u)\right)^\alpha
+
\frac12
\left(\frac{dP_{U|X=1}}{dP_U}(u)\right)^\alpha
\right]^{1/\alpha}\leq
\left[
\frac12
\left(\frac{dP_{U|X=0}}{dP_U}(u)\right)^\beta
+
\frac12
\left(\frac{dP_{U|X=1}}{dP_U}(u)\right)^\beta
\right]^{1/\beta}.
\end{align}
Combining this inequality with~\Cref{eq:counterexample-power-mean} gives
\begin{align}
&\frac{
\mathbb{E}_{P_Y}\left[
\left\|
\frac{dP_{U|Y}}{dP_U}
\right\|_{L^\alpha(P_U)}
\right]
}{
\left\|
\left[
\frac12
\left(\frac{dP_{U|X=0}}{dP_U}\right)^\beta
+
\frac12
\left(\frac{dP_{U|X=1}}{dP_U}\right)^\beta
\right]^{1/\beta}
\right\|_{L^\alpha(P_U)}
}
\leq1.
\end{align}
Taking $U$ to be constant gives
\[
\frac{dP_{U|X=0}}{dP_U}
=
\frac{dP_{U|X=1}}{dP_U}
=1,
\]
so both the numerator and denominator are equal to $1$. Therefore, the upper bound is attained.

Suppose now that $2\leq\alpha<\infty$. For the uniform distribution on the two-point alphabet $\{0,1\}$, one has the standard estimate 
$$
\lVert\cdot\rVert_{L^\alpha(P_U)}
\leq
2^{1/\beta-1/\alpha}
\lVert\cdot\rVert_{L^\beta(P_U)}
=
2^{1-2/\alpha}
\lVert\cdot\rVert_{L^\beta(P_U)}.
$$
Hence, for $P_U$-almost every $u$,
\begin{align}
\left[
\frac12
\left(\frac{dP_{U|X=0}}{dP_U}(u)\right)^\alpha
+
\frac12
\left(\frac{dP_{U|X=1}}{dP_U}(u)\right)^\alpha
\right]^{1/\alpha}
\nonumber\leq
2^{\frac1\beta-\frac1\alpha}
\left[
\frac12
\left(\frac{dP_{U|X=0}}{dP_U}(u)\right)^\beta
+
\frac12
\left(\frac{dP_{U|X=1}}{dP_U}(u)\right)^\beta
\right]^{1/\beta}.
\end{align}
Consequently,
\begin{align}
&\frac{
\mathbb{E}_{P_Y}\left[
\left\|
\frac{dP_{U|Y}}{dP_U}
\right\|_{L^\alpha(P_U)}
\right]
}{
\left\|
\left[
\frac12
\left(\frac{dP_{U|X=0}}{dP_U}\right)^\beta
+
\frac12
\left(\frac{dP_{U|X=1}}{dP_U}\right)^\beta
\right]^{1/\beta}
\right\|_{L^\alpha(P_U)}
}
\leq
2^{\frac1\beta-\frac1\alpha}
=
2^{1-\frac{2}{\alpha}}.
\end{align}
This upper bound is attained by taking $U=X$. Indeed, $P_U$ is then uniform on $\{0,1\}$ and
\[
\frac{dP_{U|X=0}}{dP_U}(u)
=
2\mathbbm{1}\{u=0\},
\qquad
\frac{dP_{U|X=1}}{dP_U}(u)
=
2\mathbbm{1}\{u=1\}.
\]
It follows that
\[
\left\|
\frac{dP_{U|X=0}}{dP_U}
\right\|_{L^\alpha(P_U)}
=
\left\|
\frac{dP_{U|X=1}}{dP_U}
\right\|_{L^\alpha(P_U)}
=
2^{1-\frac1\alpha},
\]
whereas
\begin{align}
&\left\|
\left[
\frac12
\left(\frac{dP_{U|X=0}}{dP_U}\right)^\beta
+
\frac12
\left(\frac{dP_{U|X=1}}{dP_U}\right)^\beta
\right]^{1/\beta}
\right\|_{L^\alpha(P_U)}
=
2^{\frac1\alpha}.
\end{align}
Thus, the resulting ratio is
\[
\frac{2^{1-\frac1\alpha}}{2^{\frac1\alpha}}
=
2^{1-\frac{2}{\alpha}}.
\]

We have therefore shown that
\[
\sup_{U:U-X-Y}
\frac{
\mathbb{E}_{P_Y}\left[
\left\|
\frac{dP_{U|Y}}{dP_U}
\right\|_{L^\alpha(P_U)}
\right]
}{
\left\|
\left[
\frac12
\left(\frac{dP_{U|X=0}}{dP_U}\right)^\beta
+
\frac12
\left(\frac{dP_{U|X=1}}{dP_U}\right)^\beta
\right]^{1/\beta}
\right\|_{L^\alpha(P_U)}
}
=
\begin{cases}
1, & 1<\alpha\leq2,\\[1mm]
2^{1-\frac{2}{\alpha}}, & 2\leq\alpha<\infty.
\end{cases}
\]
On the other hand, since $X$ is uniform and $Y=X$,
\[
I_\alpha(X,Y)
=
I_\alpha(X,X)
=
H_{1/\alpha}(X)
=
\log2,
\]
and therefore
\[
\exp\left(
\frac{\alpha-1}{\alpha}I_\alpha(X,Y)
\right)
=
2^{1-\frac1\alpha}.
\]
For every finite $\alpha>1$,
\[
2^{1-\frac1\alpha}
>
\begin{cases}
1, & 1<\alpha\leq2,\\[1mm]
2^{1-\frac{2}{\alpha}}, & 2\leq\alpha<\infty.
\end{cases}
\]
Hence, the inequality in~\Cref{eq:varRepresProb} is strict for every finite $\alpha>1$. As $\alpha\to\infty$, both sides converge to $2$, consistently with the equality known for maximal leakage.
\end{example} 
\begin{remark}
Taking the limit of $\alpha\to1^+$ in~\Cref{eq:varRepresProb} leads to a trivial result. Consider a discrete setting for simplicity. One has that taking said limit in~\Cref{eq:varRepresProb} leads to the following inequality:
\begin{equation}
\sum_y \sum_u p_{U|Y=y}(u) p_Y(y) p_{U}(u) \leq \sum_u p_U(u) \max_x p_{U|X=x}(u). \label{eq:trivialAlphaTo1}
\end{equation}
One can easily see that~\Cref{eq:trivialAlphaTo1} is always true as long as $U-X-Y$ form a Markov chain. Indeed:
\begin{align}
\sum_y p_Y(y) \sum_u p_{U|Y=y}(u)  p_{U}(u)    &= \sum_y p_Y(y) \sum_u p_{U}(u) \sum_x p_{U|X=x}(u) p_{X|Y=y}(x) \\
&\leq \sum_y p_Y(y) \sum_u p_{U}(u) \max_x p_{U|X=x}(u) \sum_x  p_{X|Y=y}(x) \\
&= \sum_u p_U(u) \max_x p_{U|X=x}(u).
\end{align}
\end{remark}
\section{Conclusions}
We provided a reference document for Sibson $\alpha$-mutual information.\\ In~\Cref{sec:propertiesknownresults} we presented its basic properties, along with fundamental and well-established results characterising the corresponding capacity, coding theorems as well as links to other generalized information measures.\\
In~\Cref{sec:variational}, we introduced a variety of variational representations for $I_\alpha$ that, to the best of our knowledge, are entirely novel: the first links Sibson $\alpha$-mutual information to the KL divergence and the other representations connect it to opportune functionals. Starting from~\Cref{Lemma-SibsonViaKL} we have consistently refined the various variational representations to render them more applicable in settings of interest. In going from~\Cref{Lemma-SibsonViaKL} to~\Cref{thm:varReprIalpha} we leveraged the variational representation of Sibson $\alpha$-MI in terms of measures to obtain one involving expected values of functions, removing the dependence of the variational characterization on the measures $R_Y^\star$ and $Q_Y^\star$. This allows us to more easily employ it in practical settings as is then shown in the subsequent sections.\\
In particular, in~\Cref{sec:concentration} we show how to employ said representations to re-derive results in concentration of measure when the random variables are not independent with some more concrete applications in bounding the generalization error of a learning algorithm. In~\Cref{sec:hypothesisTesting} we then highlight a link between composite hypothesis testing and $I_\alpha$. Moreover, leveraging once again the variational representations, we could produce a generalized transportation-cost-like inequality.\\In~\Cref{sec:estimationTheory} we present the link between the information measure and estimation problems. In particular, we produce Fano-type inequalities and a corresponding generalized Fano method (\Cref{sec:fanoTypeIneq,sec:fanoMethodSibson}) while in~\Cref{sec:bayesianRisk} we provide lower-bounds on the Bayesian risk via $I_\alpha$. To conclude, in~\Cref{sec:universalPrediction} we present a fundamental link between Sibson $\alpha$-mutual information and universal prediction.
In~\Cref{sec:conditional} we briefly mention a principled approach that can be undertaken to provide conditional versions of $I_\alpha$. The last section, ~\Cref{sec:discussion} considers potential extensions of $I_\alpha$ to negative values of $\alpha$ and takes a first step towards providing a representation similar in spirit to the one used to define Maximal Leakage.

\addtocontents{toc}{\setcounter{tocdepth}{1}} 

\appendix
\section{H\"older's Inequality} \label{app:Holder}

H\"older's inequality is a standard result, see e.g.~\cite[p.80]{Billingsley:95}. We include a statement and proof outline since the result is central to many of the proofs presented above and since the second case in the following theorem is not explicitly treated in most standard textbooks on the topic (but is of interest to us here).

\begin{theorem}[H\"older's Inequality]\label{thm-Holder}
Let $p$ and $q$ be H\"older conjugates, which means that they satisfy $\frac{1}{p} + \frac{1}{q} = 1.$
Two cases are possible:
\begin{enumerate}
\item $p,q \in [1,\infty].$ In this case,
\begin{align}
{\mathbb E}[|XY|] & \le  \left( {\mathbb E}[|X|^{p}] \right)^{\frac{1}{p}} \left( {\mathbb E}[|Y|^{q}] \right)^{\frac{1}{q}}.
\end{align}
In this case, the right hand side can be interpreted as the product of two norms.
\item $0<p<1,$ and thus, $q<0.$ In this case,
\begin{align}
{\mathbb E}[|XY|] & \ge  \left( {\mathbb E}[|X|^{p}] \right)^{\frac{1}{p}} \left( {\mathbb E}[|Y|^{q}] \right)^{\frac{1}{q}}.
\end{align}
\end{enumerate}
In both cases, equality is attained if and only if $|X|^p=\beta|Y|^q$ (almost surely) for some real number $\beta.$
\end{theorem}

\begin{remark}
Both bounds hold not just for probability measures, but for general measures. Specifically, we will sometimes use the case of Lebesgue measure.
\end{remark}

\begin{proof}
Assume, without loss of generality, $0<{\mathbb E}[|Y|^q]<\infty$ (otherwise the statement is trivial), and define
\begin{align}
    \tilde{P}_{XY}(x,y) &= P_{XY}(x,y)\frac{|y|^q}{{\mathbb E}[|Y|^q]}.
\end{align}
With this,
\begin{align}
    \frac{{\mathbb E}_{P_{XY}}[|XY|]}{{\mathbb E}[|Y|^q]} &= {\mathbb E}_{\tilde{P}_{XY}}[|X||Y|^{1-q}].
\end{align}
Now, if $p\ge 1$ and for $x\ge 0,$ $x^p$ is a convex function. Therefore, by Jensen's inequality,
\begin{align}
    \left( \frac{{\mathbb E}_{P_{XY}}[|XY|]}{{\mathbb E}[|Y|^q]} \right)^p &= \left({\mathbb E}_{\tilde{P}_{XY}}[|X||Y|^{1-q}]\right)^p \le {\mathbb E}_{\tilde{P}_{XY}}[|X|^p|Y|^{p(1-q)}].
\end{align}
Using the definition of $\tilde{P}_{XY},$
\begin{align}
    \left( \frac{{\mathbb E}_{P_{XY}}[|XY|]}{{\mathbb E}[|Y|^q]} \right)^p  \le {\mathbb E}_{P_{XY}}\left[|X|^p|Y|^{p(1-q)}\frac{|Y|^q}{{\mathbb E}[|Y|^q]}\right] = \frac{1}{{\mathbb E}[|Y|^q]} {\mathbb E}_{P_{XY}}\left[|X|^p|Y|^{p(1-q)+q}\right]
\end{align}
Now, note that $p+q-pq=0.$ Taking the $p$-th root on both sides gives the desired result, since $1-1/p=1/q.$

For $0\le p <1,$ the only change in the above is that $x^p$ is a concave function (for $x\ge 0$), hence Jensen's inequality is in the opposite direction.
\end{proof}

\section{Proofs for Section~\ref{sec:prelim}}

\subsection{Proof of~\Cref{prop:RenyiViaKL}}\label{app:prop:RenyiViaKL}


A detailed proof can be found in~\cite[Theorem 30]{renyiDiv}. Herein, we demonstrate the main technique. In particular, we focus on the case where $D_\alpha (P ||Q) < \infty$. Define a distribution $T$ such that 
\begin{align}
    \frac{dT}{dP} = \left(\frac{dP}{dQ}\right)^{\alpha-1} e^{-(\alpha-1)D_\alpha(P||Q)},
\end{align}
e.g., for discrete $\X$, we get $T(x) \propto P(x)^\alpha Q(x)^{1-\alpha}$. Then,
\begin{align}
   (1-\alpha) D_\alpha(P||Q) 
   & = - \log \E_{P}  \left[ \left( \frac{dP}{dQ} \right)^{\alpha-1} \right] \\
   &= - \log \E_R \left[ \left( \frac{dP}{dR} \right)^\alpha \left( \frac{dQ}{dR} \right)^{1-\alpha} \right]\\
    &\leq \E_R \left[- \log  \left( \frac{dP}{dR} \right)^\alpha - \log \left( \frac{dQ}{dR} \right)^{1-\alpha}\right]
    \\
    &= - \alpha \E_R \left[ \log \left( \frac{dP}{dR} \right)\right] - (1- \alpha)\E_R \left[ \log \left( \frac{dQ}{dR} \right) \right] \\
    &= \alpha D(R||P) + (1 - \alpha)D(R||Q),\label{Eq-RenyiViaKL-objective}
    \end{align}
    where the inequality is Jensen's inequality. 
    Therefore, the condition for equality is simply to select $R$ such that $ \left( \frac{dP}{dR} \right)^\alpha \left( \frac{dQ}{dR} \right)^{1-\alpha}$ is a constant, that is, $R = T$.

    Indeed, one can rewrite
         \begin{align}
      \alpha D(R||P) + (1 - \alpha)D(R||Q) &=  D(R\| T) + (1- \alpha)D_\alpha(P\|Q).
     \end{align}       
 Finally, the convexity and continuity of~\eqref{Eq-RenyiViaKL-objective} (as a function of $R$) follows from the convexity and continuity (in the first argument) of Kullback-Leibler divergence. 

\subsection{Proof of~\Cref{{lemma-variationalRenyiDiv}}}\label{app:lemma-variationalRenyiDivProof}
\begin{proof}
One has that
\begin{align} \mathbb{E}_P\left[e^{(\alpha-1)f}\right] &= \mathbb{E}_Q\left[e^{(\alpha-1)f}\frac{dP}{dQ}\right] \\ & \stackrel[\alpha>1]{\alpha<1}{\gtrless} \mathbb{E}^\frac1\beta_Q\left[e^{\beta(\alpha-1)f}\right]\mathbb{E}^\frac1\alpha_Q\left[\left(\frac{dP}{dQ}\right)^\alpha\right] ,\label{eq:holderVarRepr}
\end{align}
where the direction of the inequality in~\cref{eq:holderVarRepr} is determined by whether $\alpha<1$ (and one uses reverse H\"older's inequality) or $\alpha > 1$ (and one uses regular H\"older's inequality). Applying the logarithm on both sides and multiplying by $\frac{1}{\alpha-1}$ leads to:
\begin{align}
    \frac{1}{\alpha-1} \log \mathbb{E}_P\left[e^{(\alpha-1)f}\right] &\leq \frac{1}{\beta(\alpha-1)}\log\mathbb{E}_Q\left[e^{\beta(\alpha-1)f}\right] +\frac{D_\alpha(P\|Q)}{\alpha} \\
    &= \frac{1}{\alpha}\log\mathbb{E}_Q\left[e^{\alpha f}\right] +\frac{D_\alpha(P\|Q)}{\alpha}\label{eq:holderConjugates}
\end{align}
where~\cref{eq:holderConjugates} follows from the fact that $\beta(\alpha-1)=\alpha$ (indeed $\alpha$ and $\beta$ are H\"older's conjugates, i.e., $\frac1\beta = \frac{\alpha-1}{\alpha}$). Equality follows from selecting $f=\log\left(\frac{dP}{dQ}\right)$.
\end{proof}

\subsection{Proof of~\Cref{{thm:Ialpha}}}\label{app:proof:thm:Ialpha}

\begin{proof}
Suppose $\alpha \geq 1$ and $P_{XY}$ is not absolutely continuous with respect to $P_X P_Y$, then for any $Q_Y$ over $\Y$:
\begin{align}
    D_\alpha(P_{XY} || P_X Q_Y) \geq D (P_{XY} || P_X Q_Y) \geq D(P_{XY} || P_X P_Y) = + \infty.
\end{align}
Now consider the case in which $\alpha \in (0,1)$ or $P_{XY} \ll P_X P_Y$. Let $Q_Y$ be any probability measure over $\Y$:
    \begin{align}
        D_\alpha(P_{XY}\|P_XQ_Y) &= \frac{1}{\alpha-1} \log \E_{P_XQ_Y}\left[\left(\frac{dP_{XY}}{dP_XQ_Y}\right)^\alpha\right]\\
        &= \frac{1}{\alpha-1} \log \E_{Q_Y}\left[ \E_{P_X}\left[\left(\frac{dP_{XY}}{dP_XQ_Y}\right)^\alpha\right]\right]\\
        &= \frac{1}{\alpha-1} \log \E_{Q_Y}\left[ \left(\frac{dP_Y}{dQ_Y}\right)^\alpha\E_{P_X}\left[\left(\frac{dP_{X|Y}}{dP_X}\right)^\alpha\right]\right] \\
        &= \frac{1}{\alpha-1} \log \left(\E_{Q_Y}\left[ \left(\frac{dP_Y}{dQ_Y}\right)^\alpha\E_{P_X}\left[\left(\frac{dP_{X|Y}}{dP_X}\right)^\alpha\right]\right] \right)^\frac\alpha\alpha \\
        & \geq \frac{\alpha}{\alpha-1} \log\E_{Q_Y}\left[ \left(\frac{dP_Y}{dQ_Y}\right)\left(\E_{P_X}\left[\left(\frac{dP_{X|Y}}{dP_X}\right)^\alpha\right]\right)^\frac1\alpha\right] \label{eq:jensenProofIalphaClosedForm}\\
        &= \frac{\alpha}{\alpha-1} \log\E_{P_Y}\left[\left(\E_{P_X}\left[\left(\frac{dP_{X|Y}}{dP_X}\right)^\alpha\right]\right)^\frac1\alpha\right] \\
        &= I_\alpha(X,Y),
    \end{align}
    where~\Cref{eq:jensenProofIalphaClosedForm} follows from Jensen's inequality. Moreover, selecting $Q_Y=Q_Y^\star$ as defined in~\Cref{eq:qstarIalpha} one has that $D_\alpha(P_{XY}\|P_XQ_Y)=I_\alpha(X,Y)$.

    As for the limiting cases, we have
    \begin{align}
        I_\infty(X,Y) & = \min_{Q_Y} D_\infty(P_{XY} || P_X Q_Y) \\
        & = \min_{Q_Y} \lim_{\alpha \to \infty}  D_\alpha (P_{XY}|| P_X Q_Y) \\
        & \geq \lim_{\alpha \to \infty} \min_{Q_Y} D_\alpha (P_{XY}|| P_X Q_Y) \\
        & = \lim_{\alpha \to \infty} I_\alpha(X,Y) \\
        & = \lim_{\alpha \to \infty} \frac{\alpha}{\alpha-1} \log \E_{P_Y} \left[\left(\E_{P_X}\left[\left(\frac{dP_{XY}}{dP_XP_Y}\right)^\alpha\right]\right)^\frac{1}{\alpha}\right] \\
       & =  \log \E_{P_Y} \left[  \esssup_{P_X}  \frac{dP_{XY}}{dP_XP_Y} \right].
    \end{align}
    On the other hand, by choosing $Q^\star_Y$ such that
    \begin{align}
        \frac{d Q^\star_Y}{dP_Y} = \frac{  \esssup_{P_X}  \frac{dP_{XY}}{dP_XP_Y}} {\E_{P_Y} \left[  \esssup_{P_X}  \frac{dP_{XY}}{dP_XP_Y} \right]},
    \end{align}
    we get
    \begin{align}
        I_\infty(X,Y) \leq \log \E_{P_Y} \left[  \esssup_{P_X}  \frac{dP_{XY}}{dP_XP_Y} \right].
    \end{align}
    Finally, for $I_0(X,Y)$, note that
    \begin{align}
        I_0(X,Y) & = \min_{Q_Y} D_0(P_{XY} || P_XQ_Y) \\
        & = \min_{Q_Y} - \log P_XQ_Y \left\lbrace (x,y): \frac{dP_{XY}}{dP_XQ_Y}(x,y) > 0 \right\rbrace \\
        & = \min_{Q_Y} - \log P_XQ_Y \left\lbrace (x,y): \frac{dP_{X|Y}}{dP_X}(x,y) > 0, \frac{dP_Y}{dQ_Y}(y)>0 \right\rbrace \\
        & = \min_{ P_Y \ll Q_Y} - \log P_XQ_Y \left\lbrace (x,y): \frac{dP_{X|Y}}{dP_X}(x,y) > 0 \right\rbrace \\
        & = \essinf_{P_Y} -  \log P_X \left( \frac{dP_{XY}}{dP_XP_Y}(X,Y) > 0  \right).
    \end{align}
    \end{proof}

\subsection{Proofs of~\Cref{lem:pnorm-nondecreasing} and~\Cref{lem:l0norm}}\label{app:l0normProof}

\begin{proof}[Proof of~\Cref{lem:pnorm-nondecreasing}]
\begin{align}
    \| f \|_{L^p(\mu)} & = \left( \E_\mu [ |f(X)|^p ] \right)^{1/p} \\
    &= \left( \left( \E_\mu [ |f(X)|^p ] \right)^{q/p} \right)^{1/q} \\
    & \le \left(  \E_\mu \left[ \left(|f(X)|^p\right)^{q/p} \right]  \right)^{1/q}
\end{align}
where the last step is Jensen's inequality: observe that $q/p\ge 1,$ and thus, the function $f(y):=y^{q/p}$ is convex.
\end{proof}

\begin{proof}[Proof of~\Cref{lem:l0norm}]
    Consider $\varphi_p(X)= \frac1p \log \E_\mu \left[  |X|^p \right] - \E_\mu \left[ \log |X| \right]$ . One has that $\varphi_p(X)\geq 0$ by Jensen's inequality  \textit{i.e.}, $ \frac1p \log \E_\mu \left[  |X|^p \right] \geq \E_\mu \left[ \log |X| \right]$ . Moreover, one has that $\log(x)\leq x-1 $ for every $x>0$, thus
    \begin{align}
        \frac1p \log \E_\mu \left[  |X|^p \right] &\leq \frac{\left\lVert f \right\rVert_{L^\alpha(\mu)}^p-1}{p} \\
        &= \frac{\mathbb{E}_\mu\left[|X|^p-1\right]}{p} \xrightarrow{p\to 0^+} \mathbb{E}[\log|X|].
    \end{align}
    The last step follows as the function $(|x|^p-1)/p$ is monotone non-decreasing in $p>0$ and thus, by the monotone convergence theorem, one has that:
    \begin{equation}
        \lim_{p\to 0^+} \frac{\mathbb{E}[|X|^p-1]}{p} = \mathbb{E}\left[\lim_{p\to 0^+}\frac{|X|^p-1}{p}\right]=\mathbb{E}[\log|X|]
    \end{equation}
\end{proof}

\section{Proofs for~\Cref{sec:propertiesknownresults}}\label{app:proof:sec:propertiesknownresults}
\subsection{Proof of~\Cref{thm:properties}}\label{app:proof:thm:properties}
\begin{proof}
We provide short proofs for the individual properties given in~\Cref{thm:properties}.
    \begin{enumerate}[i)]
        \item The non-negativity of Sibson's $\alpha$-mutual information is a direct consequence of the non-negativity of R\'enyi divergence (see~\cite[Theorem 8]{renyiDiv}).
        \item To establish the data processing inequality, let $X-Y-Z$ be a Markov chain and let $K_{Z|Y}$ denote the corresponding Markov Kernel $K_{Z|Y}:(\mathcal{Y},\mathcal{F}) \to (\mathcal{Z},\mathcal{\hat{F}})$ induced by the transition probabilities $P_{Z|Y}$. Given any probability distribution over $Y,$ denoted by $Q_Y$, one can construct a corresponding a probability distribution over $Z$ via $K_{Z|Y}$ as follows, $Q_Z=Q_Y K_{Z|Y}$ where  $Q_Y K_{Z|Y}(z)=\E_{Q_Y}[P_{Z|Y}(z)]$. By the Data-Processing Inequality for $D_\alpha$ (\cite[Theorem 1]{renyiDiv}), one has that for every $Q_Y$:
\begin{equation}
    D_\alpha(P_{XZ}\|P_XQ_Z) \leq D_\alpha(P_{XY}\|P_XQ_Y)
\end{equation}
Taking the $\min$ with respect to $Q_Y$ leads to 
\begin{equation}
    D_\alpha(P_{XZ}\|P_XQ_Z) \leq I_\alpha(X,Y).
\end{equation}
Moreover, since $I_\alpha(X,Z) \leq D_\alpha(P_{XZ}\| P_X \hat{Q}_Z)$ for every $\hat{Q}_Z$ the statement follows by selecting $\hat{Q}_Z=Q_YK_{Z|Y}$. 
The other inequality follows by observing that considering the Kernel determined by $P_{XZ|XY}$ and denoted by $K_{XZ|YZ}$ one has that $P_{XZ}=P_{YZ}K_{XZ|YZ}$ while $P_XQ_Z = (P_XQ_Y)K_{XZ|YZ}$. Consequently by the Data-Processing Inequality for $D_\alpha$ one has that 
\begin{equation}
    D_\alpha(P_{XZ}\|P_XQ_Z) \leq D_\alpha(P_{YZ}\|P_YQ_Z).
\end{equation}
The statement follows from a similar argument as above.
        \item The invariance of Sibson's $\alpha$-mutual information to injective transforms is a consequence of the data processing inequality. Indeed, one has that $X-Y-g(Y)$ form a Markov Chain, thus $I_\alpha(X,Y)\geq I_\alpha(X,g(Y))$. Moreover, since $g$ is injective and invertible, one also has $X-g(Y)-Y$ and, consequently $I_\alpha(X,g(Y))\geq I_\alpha(X,Y)$. We thus have that $I_\alpha(X,Y)=I_\alpha(X,g(Y))$. Applying the same reasoning to $X$ and $f(X)$ yields the result.
        
        \item The additivity of Sibson's $\alpha$-mutual information for independent pairs of random variables can be established by direct evaluation, as follows:
    \begin{align}
        \frac{\alpha-1}{\alpha }I_\alpha (X^n,Y^n) & =  \log \E_{P_{Y^n}} \left[\left(\E_{P_{X^n} }\left[\left(\frac{dP_{X^nY^n}}{dP_{X^n}P_{Y^n}}\right)^\alpha\right]\right)^\frac{1}{\alpha}\right] \\
        & \stackrel {\text{(a)}} = \log   \E_{P_{Y^n}} \left[\left(\E_{P_{X^n} }\left[\left( \prod_{i=1}^n \frac{dP_{X_iY_i}}{dP_{X_i}P_{Y_i}}\right)^\alpha\right]\right)^\frac{1}{\alpha}\right] \\
        & \stackrel {\text{(b)}} = \log   \prod_{i=1}^n \E_{P_{Y^n}} \left[   \left(\E_{P_{X^n} }\left[\left( \frac{dP_{X_iY_i}}{dP_{X_i}P_{Y_i}}\right)^\alpha\right]\right)^\frac{1}{\alpha}\right] \\
        & = \sum_{i=1}^n \log   \E_{P_{Y^n}} \left[  \left(\E_{P_{X^n}} \left[\left( \frac{dP_{X_iY_i}}{dP_{X_i}P_{Y_i}}\right)^\alpha\right]\right)^\frac{1}{\alpha}\right] \\
        & = \sum_{i=1}^n \log \E_{P_{Y_i}} \left[  \left(\E_{P_{X_i} }\left[\left( \frac{dP_{X_iY_i}}{dP_{X_i}P_{Y_i}}\right)^\alpha\right]\right)^\frac{1}{\alpha}\right].
    \end{align}
    where (a) and (b) follow from the independence assumption.        
\end{enumerate}

\end{proof}

\subsection{Proof of~\Cref{thm:boundsviaRenyi}}\label{app:proof:boundsviaRenyi}
\begin{proof}
    By the Data-Processing Inequality of $I_\alpha$ (see~\Cref{thm:dpiIalpha}) one has that:
    \begin{align}
        I_\alpha(X,Y) \leq I_\alpha(X,X) &= \frac{\alpha}{\alpha-1}\log\sum_{x}P_X(x)\left(\sum_{\tilde{x}}\left(\frac{P_{XX}(x,\tilde{x})}{P_X(X)P_X(\tilde{x})}\right)^\alpha P_X(\tilde{x})\right)^\frac1\alpha\\
        &= \frac{\alpha}{\alpha-1}\log\sum_{x}P_X(x)\left(\sum_{\tilde{x}}\left(\frac{\delta_{x=\hat{x}}}{P_X(\tilde{x})}\right)^\alpha P_X(\tilde{x})\right)^\frac1\alpha \\
        &= \frac{\alpha}{\alpha-1}\log\sum_{x}P_X(x) P_X(x)^{\frac{1-\alpha}{\alpha}}\\
        &= \frac{\alpha}{\alpha-1}\log\sum_{x}P_X^\frac1\alpha(x) \\
        &= H_{\frac1\alpha}(X). 
    \end{align}
Moreover,
\begin{align}
    I_\alpha(X,Y) & = \frac{\alpha}{\alpha-1} \log \sum_y \left( \sum_x P_X(x) P_{Y|X}(y|x)^\alpha \right)^{\frac{1}{\alpha}} \\
    & \stackrel{\text{(a)}} \leq \frac{\alpha}{\alpha-1} \log \sum_y \left( \sum_x P_X(x) P_{Y|X}(y|x) \right)^{\frac{1}{\alpha}} \\
    & =  \frac{\alpha}{\alpha-1} \log \sum_y P_{Y}(y)^{\frac{1}{\alpha}} \\
    & = H_{\frac{1}{\alpha}} (Y).
\end{align}
    where (a) follows from the fact that $\sign(\alpha-1) P_{Y|X}(y|x)^{\alpha} \leq \sign(\alpha-1) P_{Y|X}(y|x)$. Equality holds if and only if for all $(x,y)$, $P_{Y|X}(y|x)^\alpha = P_{Y|X}(y|x)$, i.e., $P_{Y|X} \in 
    \{0,1\}$.
\end{proof}

\subsection{Proof of~\Cref{thm:properties-alpha}}\label{app:proof:thm:properties-alpha}

\begin{enumerate}[i)]    \item  The asymmetry of Sibson's $\alpha$-mutual information follows from the fact that $\min_{Q_Y} D_\alpha(P_{XY}\|P_XQ_Y)$ is, in general, different from $\min_{Q_X}D_\alpha(P_{XY}\|Q_XP_Y)$. An example is shown in~\Cref{ex:bec}.
    \item The continuity in $\alpha$ if $I_\alpha$ follows from the continuity in $\alpha$ of $D_\alpha$~\cite[Theorem 7]{renyiDiv}.
    \item The non-decreasability in $\alpha$ of Sibson's $\alpha$-mutual information follows from the following chain of inequalities:
    \begin{align}
        I_{\alpha_1}(X,Y) &= \min_{Q_Y} D_{\alpha_1}(P_{XY}\|P_XQ_Y) \\
        &\leq D_{\alpha_1}(P_{XY}\|P_XQ^{\star,\alpha_2}_Y) \\
        &\leq D_{\alpha_2}(P_{XY}\|P_XQ^{\star,\alpha_2}_Y) \label{eq:nonDecr} \\
        &= I_{\alpha_2}(X,Y),
    \end{align}
    where $Q_Y^{\star,\alpha_2}=\argmin_{Q_Y}D_{\alpha_2}(P_{XY}\|P_XQ_Y)$ and~\Cref{eq:nonDecr} follows from the non-decreasability of $D_\alpha$ with respect to $\alpha$ (see~\cite[Theorem 3]{renyiDiv}).
\end{enumerate}

\subsection{Proof of~\Cref{thm:convexityproperties}}\label{proof:convexityproperties}
\begin{proof} The convexity properties follow from relating $I_\alpha$ to norms \textit{i.e.}, from~\Cref{eq:normRepresentationAlphaLargerThanOne}. Indeed, for a given $P_X$, convexity in $P_{Y|X}$, whenever the conditional exists, follows from Minkowski's inequality. To see this, take $\lambda\in (0,1)$ and two conditional distributions on $\Y$, $P_{Y_1|X},P_{Y_2|X}$, then one has that:
    \begin{align}
    &\mathbb{E}_{P_Y}\left[\left\lVert \frac{\lambda P_{Y_1|X}+(1-\lambda)P_{Y_2|X}}{P_Y}\right\rVert_{L^\alpha(P_X)}\right] \\&\hspace{1em} \leq  \lambda\mathbb{E}_{P_Y}\left[\left\lVert \frac{ P_{Y_1|X}}{P_Y}\right\rVert_{L^\alpha(P_X)}\right] + (1-\lambda)\mathbb{E}_{P_Y}\left[\left\lVert \frac{ P_{Y_2|X}}{P_Y}\right\rVert_{L^\alpha(P_X)}\right].
    \end{align}
    Similarly, if $\alpha\in (0,1)$ concavity with respect to $P_{Y|X}$ follows from reverse Minkowski's inequality. Moreover, if $\alpha\in(0,1)$, since $\frac{\alpha}{\alpha-1}\log(x)$ is a convex non-increasing function and $\exp\left(\frac{\alpha-1}{\alpha}I_\alpha(X,Y)\right)$ is concave in $P_{Y|X}$, then convexity of $I_\alpha(X,Y)$ in $P_{Y|X}$ follows from the composition of the two functions (see~\cite[Eq. (3.10)]{boyd}), thus proving Property~\labelcref{prop:convexConditional}.
    If $\alpha \in (1,\infty)$, then concavity of $\exp\left(\frac{\alpha-1}{\alpha}I_\alpha(X,Y)\right)$ in $P_X$ follows from linearity of $\left\lVert \frac{P_{Y|X}}{P_Y}\right\rVert_{L^\alpha(P_X)}^\alpha$ with respect to $P_X$ and concavity of $x^\frac1\alpha$. Moreover, since $\frac{\alpha}{\alpha-1}\log(x)$ is concave non-decreasing, concavity of $I_\alpha(X,Y)$ in $P_X$ follows from the composition of the two functions (see again~\cite[Eq. (3.10)]{boyd}). The convexity of $\exp\left(\frac{\alpha-1}{\alpha}I_\alpha(X,Y)\right)$ with respect to $P_X$ for $\alpha\in(0,1)$ follows from a similar argument. 
    \end{proof}

\subsection{Proof of~\Cref{thm:Riouletal}}\label{app:proof:thm:Riouletal}

\begin{proof}
We write out the proof for the case of discrete and finite random variables, as follows:
    \begin{align}
        I_\alpha(X, (Y,Z)) &= \frac{\alpha}{\alpha-1} \log \sum_{y,z} \left( \sum_{x}P_X(x) P_{Y,Z|X}(y,z|x)^\alpha \right)^{1/\alpha} \\
          &= \frac{\alpha}{\alpha-1} \log \sum_{y,z} \left( \sum_{x} P_X(x) P_Z(z)^\alpha P_{Y|X,Z}(y|x,z)^\alpha \right)^{1/\alpha} \label{app:proof:thm:Riouletal:eq:ind} \\
          &= \frac{\alpha}{\alpha-1} \log \sum_{y} \sum_z P_Z(z) \left( \sum_{x} P_X(x)  P_{Y|X,Z}(y|x,z)^\alpha \right)^{1/\alpha} \\
          & \le \frac{\alpha}{\alpha-1} \log \sum_{y} \left( \sum_{x,z} P_X(x) P_Z(z) P_{Y|X,Z}(y|x,z)^\alpha \right)^{1/\alpha} \label{app:proof:thm:Riouletal:eq:jensen}\\
          & = \revised{I_\alpha((X,Z), Y)},
    \end{align}
    where~\Cref{app:proof:thm:Riouletal:eq:ind} holds since $X$ and $Z$ are independent, and~\Cref{app:proof:thm:Riouletal:eq:jensen} is due to Jensen's inequality applied to the expectation taken with respect to $Z$ and distinguishing the two cases of $\alpha\ge 1$ and $0 \le \alpha < 1$.

    Can this be somehow generalized?

\end{proof}

\subsection{Proof of~\Cref{thm:tensorization}}\label{app:proof:thm:tensorization}

\begin{proof}
This theorem can be established along the following lines:
    \begin{align}
      \frac{\alpha-1}{\alpha}  I_\alpha (X, Y^n) & =  \log  \sum_{y^n} \left( \sum_x P(x) P(y^n|x)^{\alpha} \right)^{ \frac{1}{\alpha}}  \\
      & = \log  \sum_{y^n} \left( \sum_x P(x) \prod_{i=1}^n P(y_i|x)^{\alpha} \right)^{ \frac{1}{\alpha}}  \\
      & = \log  \sum_{y^n} \left( \sum_x  \prod_{i=1}^n P(x)^{\frac{1}{\beta_i}} P(y_i|x)^\alpha \right)^{ \frac{1}{\alpha}}  \\
      & \stackrel{\text{(a)}} \leq \log  \sum_{y^n} \left(  \prod_{i=1}^n   \left( \sum_x  P(x) P(y_i|x)^{\beta_i \alpha} \right)^{\frac{1}{\beta_i}} \right)^{ \frac{1}{\alpha}}  \\
      & = \log  \sum_{y^n}   \prod_{i=1}^n   \left( \sum_x  P(x) P(y_i|x)^{\beta_i \alpha} \right)^{\frac{1}{\beta_i\alpha }}  \\
      & = \log \prod_{i=1}^n \left( \sum_{y_i} \left( \sum_x  P(x) P(y_i|x)^{\beta_i \alpha} \right)^{\frac{1}{\beta_i\alpha }} \right) \\
      & = \sum_{i=1}^n \frac{\beta_i \alpha-1}{\beta_i \alpha} I_{\beta_i \alpha}(X,Y_i),
    \end{align}
    where (a) follows from H\"older's inequality (for more than two functions).
\end{proof}
\subsection{Proof of~\Cref{thm:sibson-capacity}} \label{app:sibson-capacity-proof}

For completeness, we reproduce Csisz\'ar's proof~\cite{csiszar:95}:
\begin{align}
    \sup_{P_X} I_\alpha(X,Y) & 
    = \sup_{P_X} \inf_{Q_Y} D_\alpha(P_{XY} || P_X Q_Y) \\
    & = \sup_{P_X} \inf_{Q_Y} \frac{1}{\alpha-1} \log \E_{P_{XY}}  \left[   \left( \frac{dP_{Y|X}}{d Q_Y} \right)^{\alpha-1}     \right] \\
    & = \sup_{P_X} \inf_{Q_Y} \frac{1}{\alpha-1} \log \left( \sign(\alpha-1) \cdot \sign (\alpha-1) \cdot  \E_{P_X} \left[ \E_{P_{Y|X}} \left[   \left( \frac{dP_{Y|X}}{d Q_Y} \right)^{\alpha-1}  \right]   \right]   \right) \\
    & = \frac{1}{\alpha-1} \log \left( \sign(\alpha-1) \cdot \sup_{P_X} \inf_{Q_Y} \sign (\alpha-1)  \E_{P_X} \left[ \E_{P_{Y|X}} \left[   \left( \frac{dP_{Y|X}}{d Q_Y} \right)^{\alpha-1}  \right]   \right] \right).
\end{align}
It is easy to verify that
\begin{align}
    \sign (\alpha-1)  \E_{P_X} \left[ \E_{P_{Y|X}} \left[   \left( \frac{dP_{Y|X}}{d Q_Y} \right)^{\alpha-1}  \right]   \right] 
\end{align}
is convex and continuous in $Q_Y$ and linear (hence, concave) and continuous in $P_X$. As such, $\sup$ and $\inf$ can be swapped to yield:
\begin{align}
    \sup_{P_X} I_\alpha(X,Y) & 
    = \frac{1}{\alpha-1} \log \left( \sign(\alpha-1) \cdot  \inf_{Q_Y}  \sup_{P_X} \sign (\alpha-1)  \E_{P_X} \left[ \E_{P_{Y|X}} \left[   \left( \frac{dP_{Y|X}}{d Q_Y} \right)^{\alpha-1}  \right]   \right] \right). \\
    & = \frac{1}{\alpha-1} \log \left( \sign(\alpha-1) \cdot \inf_{Q_Y} \sup_{x \in \X} \sign (\alpha-1) e^{(\alpha-1) D_\alpha (P_{Y|X=x} || Q_Y  ) } \right) \\
    & = \inf_{Q_Y} \sup_{x \in \X} D_\alpha \left( P_{Y|X=x} || Q_Y \right).
\end{align}

\subsection{Proof of~\Cref{thm-symmetric-channel-cap}}\label{app:proof:thm:symmetric-channel-cap}
\begin{proof}
    First, we note that for any input distribution $P_X,$
    \begin{align}
        I_\alpha(X,Y) & \le D_\alpha ( P_X P_{Y|X} \| P_X U_Y),\label{eq-thm-symmetric-channel-cap-upperbound}
    \end{align}
    where $U_Y$ denotes the uniform distribution on $Y.$ Second, we note that the right-hand side of the last equation does not actually depend on $P_X$ due to the symmetry of $P_{Y|X},$ which can be seen by writing out
     \begin{align}
        D_\alpha ( P_X P_{Y|X} \| P_X U_Y) &= \frac{1}{\alpha-1} \log \sum_{(x,y) \in \X\times\Y} P(x)^\alpha P(y|x)^\alpha P(x)^{1-\alpha}U(y)^{1-\alpha} \\
        &= \frac{1}{\alpha-1} \log \frac{1}{|\Y|^{1-\alpha}}\sum_{(x,y) \in \X\times\Y} P(x) P(y|x)^\alpha \\
        &= \frac{1}{\alpha-1} \log \frac{1}{|\Y|^{1-\alpha}}\sum_{x \in \X} P(x) \sum_{y \in \Y}P(y|x)^\alpha.
    \end{align}
    By the symmetry assumption on $P_{Y|X},$ the last sum over $y$ is the same for all values of $x.$
    Therefore,
     \begin{align}
        D_\alpha ( P_X P_{Y|X} \| P_X U_Y) &=  \frac{1}{\alpha-1} \log \frac{1}{|\Y|^{1-\alpha}} \sum_{y \in \Y}P(y|x)^\alpha,
    \end{align}
    which does not depend on $P(x).$
    Finally, we observe that if $P_X=U_X,$ due to the symmetry of $P_{Y|X}$ and using the formula from~\Cref{eq:qstarIalpha}, the resulting $Q^\star_Y(\{y\})$ is also uniform, thus attaining the upper bound in Equation~\eqref{eq-thm-symmetric-channel-cap-upperbound} with equality.
\end{proof}
\subsection{Proof of~\Cref{prop:csiszarMI-max}}\label{app:proof:prof:csiszarMI-max}
\begin{proof}
    The second equality was shown in~\Cref{corr:arimoto-sibson-equality}. The first equality follows from noting that:
    \begin{align}
        \sup_{P_X} I_\alpha^C(X,Y)  & = \sup_{P_X} \min_{Q_Y} \E_{P_X} \left[  D_\alpha \left( P_{Y|X}(.|X) \| Q_Y \right)\right] \\
        & \stackrel{\text{(a)}} = \min_{Q_Y} \sup_{P_X} \E_{P_X} \left[  D_\alpha \left( P_{Y|X}(.|X) \| Q_Y \right)\right] \\
        & = \min_{Q_Y} \max_{x \in \X}  D_\alpha \left( P_{Y|X=x} \| Q_Y \right) \\
        & \stackrel{\text{(b)}} = \sup_{P_X} I_\alpha(X,Y),
    \end{align}
    where (a) follows from the fact that $D_\alpha(P ||Q)$ is convex in $Q$~\cite[Theorem 12]{renyiDiv}, and the expectation is linear in $P_X$, and (b) follows from~\Cref{thm:sibson-capacity}.
\end{proof}
\subsection{Proof of~\Cref{prop:csiszar-sibson-ordering}}\label{app:proof:prop:csiszar-sibson-ordering}
\begin{proof}
    Consider $\alpha >1$. Then,
    \begin{align}
        I_\alpha(X,Y) & = \min_{Q_Y} D_\alpha( P_{XY} || P_X Q_Y)  \\
        & = \min_{Q_Y} \frac{1}{\alpha-1} \log e^{(\alpha-1) D_\alpha( P_{XY} || P_X Q_Y) } \\
        & =  \min_{Q_Y} \frac{1}{\alpha-1} \log \E_{P_X} \left[ e^{(\alpha-1) D_\alpha(P_{Y|X}(.|X) || Q_Y) }       \right] \\
        & \geq \min_{Q_Y}\frac{1}{\alpha-1} \E_{P_X} \left[ \log  e^{(\alpha-1) D_\alpha(P_{Y|X}(.|X) || Q_Y) }       \right] \\
        & = \min_{Q_Y} \E_{P_X} \left[ D_\alpha (P_{Y|X}(.|X) || Q_Y) \right],
    \end{align}
    where the inequality follows from Jensen's inequality. For $\alpha <1$, the proof follows along the same steps, except that the inequality is flipped because $\alpha-1 < 0$.
\end{proof}

\section{Proofs for~\Cref{sec:variational,sec:concentration}} 

\subsection{Proof of~\Cref{Lemma-SibsonViaKL}}\label{app:variational:Lemma-SibsonViaKL}

\begin{proof}
We start by observing
\begin{align}
   (1-\alpha)I_\alpha(X,Y) &= (1-\alpha) \min_{Q_Y} D_\alpha(P_{XY}\|P_XQ_Y)
\end{align}   
Consider first $0<\alpha<1.$ Then, using~\Cref{prop:RenyiViaKL},
\begin{align}
   (1-\alpha)I_\alpha(X,Y) &=  \min_{Q_Y} (1-\alpha) D_\alpha(P_{XY}\|P_XQ_Y) \\
   &= \min_{Q_Y} \min_{R_{XY}} \{\alpha D(R_{XY}\|P_{XY}) + (1-\alpha) D(R_{XY}\|P_XQ_Y) \}\\
   &= \min_{R_{XY}} \{\alpha D(R_{XY}\|P_{XY}) + (1-\alpha) \min_{Q_Y} D(R_{XY}\|P_XQ_Y) \} \\
   &= \min_{R_{XY}} \{\alpha D(R_{XY}\|P_{XY}) + (1-\alpha)  D(R_{XY}\|P_XR_Y) \} 
\end{align}
and the minimum is attained for
\begin{align}
 R_{XY} &\propto P_{XY}^{\alpha} (P_XQ_Y^*)^{1-\alpha}, \label{Lemma-SibsonViaKL-eq-RXYforequality}
\end{align}
where $Q_Y^*$ is given in~\Cref{eq:qstarIalpha}.
Note that for this choice of $R_{XY},$ the corresponding marginal on $Y$ is exactly $Q_Y^*.$

Likewise, for $\alpha>1,$ we observe (again using~\Cref{prop:RenyiViaKL})
\begin{align}   
(1-\alpha)I_\alpha(X,Y) &= (1-\alpha) \min_{Q_Y} D_\alpha(P_{XY}\|P_XQ_Y) \\
&=  \max_{Q_Y}(1-\alpha) D_\alpha(P_{XY}\|P_XQ_Y) \\
&=  \max_{Q_Y} \min_{R_{XY}} \{\alpha D(R_{XY}\|P_{XY}) + (1-\alpha) D(R_{XY}\|P_XQ_Y) \} \\
&\le  \min_{R_{XY}} \max_{Q_Y} \{\alpha D(R_{XY}\|P_{XY}) + (1-\alpha) D(R_{XY}\|P_XQ_Y) \} \\
&\le  \min_{R_{XY}}  \{\alpha D(R_{XY}\|P_{XY}) + (1-\alpha) \min_{Q_Y} D(R_{XY}\|P_XQ_Y) \},
\end{align}
where the last step holds since $(1-\alpha)$ is negative.
The inner minimum is attained for $Q_Y=R_Y,$ proving the inequality part of~\Cref{Lemma-SibsonViaKL-eq-main}.
Again, selecting $R_{XY}$ as in~\Cref{Lemma-SibsonViaKL-eq-RXYforequality} attains equality throughout.
\end{proof}

\subsection{Proof of~\Cref{thm-Sibson-var-alternative}}\label{app:variational:thm-Sibson-var-alternative}

\begin{proof}
Observe that
\begin{align}
    D(R_{XY}\|P_XR_Y) &= \E_{R_Y}\left[ D(R_{X|Y}(\cdot|Y)\|P_X(\cdot))  \right].
\end{align}
From the Donsker-Varadhan variational representation of Kullback-Leibler divergence, we have that for any distribution $R_{XY}$ and any (marginal) $P_X,$
\begin{align}
    \E_{R_Y} [\log \E_{P_X}[e^{\alpha f(X,Y)}]] & \ge \E_{R_Y} [\E_{R_{X|Y}}[\alpha f(X,Y)]] - D(R_{XY}\|P_XR_Y).\label{eq:proof-thm-Sibson-var-alternative-eq1}
\end{align}
Let $X,Y$ be distributed according to $P_{XY}.$
From Lemma~\ref{Lemma-SibsonViaKL}, for $\alpha > 1,$ we have
\begin{align}
    I_\alpha(X,Y) & \ge -\frac{\alpha}{\alpha-1} D(R_{XY}\|P_{XY}) + D(R_{XY}\|P_XR_Y).
\end{align}
Combining, we get
\begin{align} &\E_{R_Y} [\log \E_{P_X}[e^{\alpha f(X,Y)}]] \nonumber \\
   &\hspace{5em} \ge - I_\alpha(X,Y) +\frac{\alpha}{\alpha-1} \left( 
     \E_{R_Y} [\E_{R_{X|Y}}[(\alpha-1) f(X,Y)]] - D(R_{XY}\|P_{XY}) 
     \right).
\end{align}
This bound holds for all choices of $R_{XY}.$ Let us select
\begin{align}
R_{XY}^*(x,y) &\propto P_{XY}(x,y) e^{(\alpha-1)f(x,y)}
\end{align}
and observe that the corresponding marginal $R^*_Y(y)$ is precisely as in~\Cref{thm-Sibson-var-alternative-Eq-Rstar}. For this distribution, one can readily verify (by plugging in and evaluating) that
\begin{align}
     \E_{R_Y^*} [\E_{R_{X|Y}^*}[(\alpha-1) f(X,Y)]] - D(R_{XY}^*\|P_{XY}) &= \log \E_{P_{XY}}[e^{(\alpha-1)f(X,Y)}],
\end{align}
which proves the inequality part of Theorem~\ref{thm-Sibson-var-alternative} for the case $\alpha>1.$
Equality is achieved by selecting the function $f(x,y)$ to satisfy
\begin{align}
    e^{\alpha f(x,y)} &=  \frac{\left(\frac{dP_{XY}}{dP_XP_Y}(x,y)\right)^\alpha}{\left\lVert\frac{dP_{XY}}{dP_XP_Y}(X,y)\right\rVert^\alpha _{L^\alpha(P_X)} } = \frac{\left(\frac{dP_{XY}}{dP_XP_Y}(x,y)\right)^\alpha}{\E_{P_{\tilde{X}}}\left[\left(\frac{dP_{XY}}{dP_XP_Y}(\tilde{X},y)\right)^{\alpha}\right] },
\end{align}
where $\tilde{X}$ denotes a random variable such that $P_{\tilde{X}} = P_X.$
To see that with this choice, we indeed have equality, we first observe that the second summand in~\Cref{thm-Sibson-var-alternative-eq-largealpha} vanishes. For the first summand, we observe
\begin{align}
    \frac{\alpha}{\alpha-1} \log \E_{P_{XY}}[e^{(\alpha-1)f(X,Y)}] &=  \frac{\alpha}{\alpha-1} \log \E_{P_{XY}}\left[\left(\frac{\left(\frac{dP_{XY}}{dP_XP_Y}(X,Y)\right)^\alpha}{\E_{P_{\tilde{X}}}\left[\left(\frac{dP_{XY}}{dP_XP_Y}(\tilde{X},Y)\right)^{\alpha}\right] }\right)^{\frac{\alpha-1}{\alpha}}\right] \\
    &=  \frac{\alpha}{\alpha-1} \log \E_{P_{Y}} \E_{P_X}\left[\left(\frac{dP_{XY}}{dP_XP_Y}(X,Y)\right)\left(\frac{\left(\frac{dP_{XY}}{dP_XP_Y}(X,Y)\right)^\alpha}{\E_{P_{\tilde{X}}}\left[\left(\frac{dP_{XY}}{dP_XP_Y}(\tilde{X},Y)\right)^{\alpha}\right] }\right)^{\frac{\alpha-1}{\alpha}}\right] \\
    &= \frac{\alpha}{\alpha-1} \log \E_{P_{Y}} \left[ \frac{\E_{P_X}\left[\left(\frac{dP_{XY}}{dP_XP_Y}(X,Y)\right)^{\alpha}\right]}{\left(\E_{P_{\tilde{X}}}\left[\left(\frac{dP_{XY}}{dP_XP_Y}(\tilde{X},Y)\right)^{\alpha}\right] \right)^{\frac{\alpha-1}{\alpha}}}\right]\\
    &= \frac{\alpha}{\alpha-1} \log \E_{P_{Y}} \left[ \left( \E_{P_X}\left[\left(\frac{dP_{XY}}{dP_XP_Y}(X,Y)\right)^{\alpha}\right] \right)^{1/\alpha}\right], \label{thm-Sibson-var-alternative-eq-equalityproof}
\end{align}
which, from~\Cref{thm:Ialpha-Eqn-explicitdef}, is exactly equal to $I_\alpha(X,Y).$

To cover the case $0<\alpha<1,$ we again start from~\Cref{eq:proof-thm-Sibson-var-alternative-eq1}.
Additionally, writing the Donsker-Varadhan variational representation with the function $-(1-\alpha)f(x,y),$
\begin{align}
     \log \E_{P_{XY}}\left[e^{(\alpha-1) f(X,Y)}\right] & \ge -(1-\alpha) \E_{R_{XY}}\left[ f(X,Y)\right] - D(R_{XY}\|P_{XY}).
\end{align}
Let us now divide this by $(1-\alpha)$ (which is positive).
Let us also divide~\Cref{eq:proof-thm-Sibson-var-alternative-eq1} by $\alpha$
and add up the two inequalities to obtain
\begin{align}
 \frac{1}{\alpha}\E_{R_Y} \left[\log \E_{P_X}\left[e^{\alpha f(X,Y)}\right]\right] + \frac{1}{1-\alpha} \log \E_{P_{XY}}\left[e^{(\alpha-1) f(X,Y)}\right] \nonumber &\ge\\  &\hspace{-6em}- \frac{1}{\alpha} D(R_{XY}\|P_XR_Y) - \frac{1}{1-\alpha} D(R_{XY}\|P_{XY}).
\end{align}
Multiplying both sides by $\alpha(1-\alpha)$ (which is positive) gives
\begin{align}
 (1-\alpha)\E_{R_Y} \left[\log \E_{P_X}\left[e^{\alpha f(X,Y)}\right]\right] + \alpha \log \E_{P_{XY}}\left[e^{(\alpha-1) f(X,Y)}\right]  &\ge 
 \nonumber\\  &\hspace{-6em}- \left\{(1-\alpha) D(R_{XY}\|P_XR_Y) + \alpha D(R_{XY}\|P_{XY}) \right\}.
\end{align}
This bound holds for all $R_{XY}.$ By~\Cref{Lemma-SibsonViaKL}, let us select
\begin{align}
 R_{XY}^{**} &\propto P_{XY}^{\alpha} (P_XQ_Y^*)^{1-\alpha}, 
\end{align}
where $Q_Y^*$ is given in~\Cref{eq:qstarIalpha}.
Recall that for this distribution, the corresponding marginal distribution $R_Y$ for $Y$ is exactly $Q_Y^*.$
Then, we obtain
\begin{align}
  (1-\alpha)\E_{Q_Y^*} \left[\log \E_{P_X}\left[e^{\alpha f(X,Y)}\right]\right] + \alpha \log \E_{P_{XY}}\left[e^{(\alpha-1) f(X,Y)}\right]  & \ge  - (1-\alpha)I_\alpha(X,Y).
\end{align}
Dividing on both sides by $(\alpha-1),$ which is negative, leads to
\begin{align}
  I_\alpha(X,Y) &\ge \frac{\alpha}{\alpha-1} \log \E_{P_{XY}}\left[e^{(\alpha-1) f(X,Y)}\right] - \E_{Q_Y^*} \left[\log \E_{P_X}\left[e^{\alpha f(X,Y)}\right]\right].
\end{align}
Equality is achieved by selecting $f(x,y)$ in the same way as in the case $\alpha>1,$ the derivation applies without any changes.
\end{proof}

\subsection{Proof of~\Cref{thm-Sibson-var-alternative-weaker}}\label{app:variational:thm-Sibson-var-alternative-weaker}

\begin{proof}
For the inequality part, we have that~\Cref{thm-Sibson-var-alternative-eq-largealpha-weaker} and~\Cref{thm-Sibson-var-alternative-eq-smallalpha-weaker} follow directly from~\Cref{thm-Sibson-var-alternative-eq-largealpha} and~\Cref{thm-Sibson-var-alternative-eq-smallalpha}, respectively, by a simple application of Jensen's inequality.
What is initially less clear is that the representation is not only a lower bound, but is, in fact, tight, as claimed in the theorem.
To see this for $\alpha>1,$ it suffices to observe that if we select $f(x,y)$ as in~\Cref{eq:equalityVarRepr}, then a simple calculation reveals that for the second summand in~\Cref{thm-Sibson-var-alternative-eq-largealpha-weaker}, we have
\begin{align}
  \E_{P_XR^*_Y}\left[e^{\alpha f(X,Y)}\right] &=1,
\end{align}
which actually holds irrespective of the choice of the distribution for $Y$ (as long as $X$ and $Y$ are independent). For the first summand, the calculation is identical to that leading up to~\Cref{thm-Sibson-var-alternative-eq-equalityproof}, which completes the proof for $\alpha>1.$
The proof for $0<\alpha<1$ is identical.
\end{proof}

\subsubsection{Derivation of~\Cref{thm:varReprIalpha} from~\Cref{thm-Sibson-var-alternative-weaker}}\label{app:variational:thm-Sibson-var-alternative-weaker-connection-to-VarReprSibs}

\begin{proof}
Here, we show how the inequality part of~\Cref{thm:varReprIalpha} can be obtained from~\Cref{thm-Sibson-var-alternative-weaker}.

For $\alpha>1,$ we start from~\Cref{thm-Sibson-var-alternative-eq-largealpha-weaker} and further lower bound by
 \begin{align}
    I_\alpha(X,Y) &\ge  \frac{\alpha}{\alpha-1} \log \E_{P_{XY}}\left[e^{(\alpha-1)f(X,Y)}\right] -  \log \max_y \E_{P_X}\left[e^{\alpha f(X,y)}\right] .
\end{align}
Multiplying on both sides by $(\alpha-1)/\alpha$ (which is positive) and exponentiating leads to
 \begin{align}
    \exp\left( \frac{\alpha-1}{\alpha} I_\alpha(X,Y) \right) &\ge  \frac{ \E_{P_{XY}}\left[e^{(\alpha-1)f(X,Y)}\right] }{  \left( \max_y \E_{P_X}\left[e^{\alpha f(X,y)}\right] \right)^{\frac{\alpha-1}{\alpha}}}.
\end{align}
To complete the proof, we observe that $\beta = \frac{\alpha}{\alpha-1},$ and we define $g(x,y) = e^{(\alpha-1)f(x,y)},$ which implies that $g^\beta(x,y) = e^{\alpha f(x,y)}.$ Since $(\alpha-1)/\alpha$ is positive, we can move the maximum outside of the exponentiation in the denominator of the last equation, which completes the proof.

For $0 < \alpha < 1,$ we start from~\Cref{thm-Sibson-var-alternative-eq-smallalpha-weaker} and further lower bound by
\begin{align}
  I_\alpha(X,Y) &\ge \frac{\alpha}{\alpha-1} \log \E_{P_{XY}}\left[e^{(\alpha-1) f(X,Y)}\right] -  \log \max_y \E_{P_X}\left[e^{\alpha f(X,y)}\right] .
\end{align}
Multiplying on both sides by $(\alpha-1)/\alpha$ (which is negative) and exponentiating leads to
\begin{align}
  \exp\left( \frac{\alpha-1}{\alpha} I_\alpha(X,Y) \right) &\le \frac{ \E_{P_{XY}}\left[e^{(\alpha-1) f(X,Y)}\right]  }{ \left( \max_y \E_{P_X}\left[e^{\alpha f(X,y)}\right] \right)^{\frac{\alpha-1}{\alpha}} }.
\end{align}
To complete the proof, we observe that $\beta = \frac{\alpha}{\alpha-1},$ and we define $g(x,y) = e^{(\alpha-1)f(x,y)},$ which implies that $g^\beta(x,y) = e^{\alpha f(x,y)}.$ Since $(\alpha-1)/\alpha$ is negative, when we move the maximum outside of the exponentiation in the denominator of the last equation, it becomes a minimum, which completes the proof.
\end{proof}

\subsection{Proof of~\Cref{thm:varReprIalpha}}\label{app:proofVarReprSibs}
Given a positive-valued function $g:\X\times \Y \to \mathbb{R}^+$ one has that if $P_{XY}\ll P_XP_Y$:
\begin{align} \mathbb{E}_{P_{XY}}\left[g\right] &= \mathbb{E}_{P_XP_Y}\left[g\frac{dP_{XY}}{dP_XP_Y}\right] \\ 
&= \mathbb{E}_{P_Y}\left[\mathbb{E}_{P_X}\left[g\frac{dP_{XY}}{dP_XP_Y}\right]\right]\\ & \stackrel[\alpha>1]{\alpha<1}{\gtrless} \mathbb{E}_{P_Y}\left[\mathbb{E}^\frac1\beta_{P_X}\left[g^{\beta}\right]\mathbb{E}^\frac1\alpha_{P_X}\left[\left(\frac{dP_{XY}}{dP_XP_Y}\right)^\alpha\right]\right]\label{eq:holderVarReprSibsMI1} \\
&\stackrel[\alpha'>1]{\alpha'<1}{\gtrless} \mathbb{E}_{P_Y}^\frac{1}{\beta'}\left[\mathbb{E}^\frac{\beta'}{\beta}_{P_X}\left[g^{\beta}\right]\right]\mathbb{E}^\frac{1}{\alpha'}_{P_Y}\left[\mathbb{E}^\frac{\alpha'}\alpha_{P_X}\left[\left(\frac{dP_{XY}}{dP_XP_Y}\right)^\alpha\right]\right]\label{eq:holderVarReprSibsMI2},
\end{align}
where $\frac1\alpha+\frac1\beta =\frac{1}{\alpha'}+\frac{1}{\beta'} = 1$ and, moreover, if $0<\alpha<1$ then so is $\alpha'$. If $\alpha,\alpha'>1$ then $\beta,\beta'>1$, taking the limit of $\alpha'\to 1$ which implies that $\beta'\to\infty$ one retrieves that:
\begin{equation}
    \mathbb{E}_{P_{XY}}[g] \leq \exp\left(\frac{\alpha-1}{\alpha}I_\alpha(X,Y)\right) \esssup_{P_Y}\left[\mathbb{E}_{P_X}^\frac1\beta\left[g^\beta\right]\right].
\end{equation}
Similarly, if $\alpha,\alpha'<1$ then $\beta,\beta'<1$, taking the limit of $\alpha'\to 1$ which implies that $\beta'\to\infty$ one retrieves that:
\begin{equation}
    \mathbb{E}_{P_{XY}}[g] \geq \exp\left(\frac{\alpha-1}{\alpha}I_\alpha(X,Y)\right) \essinf_{P_Y}\left[\mathbb{E}_{P_X}^\frac1\beta\left[g^\beta\right]\right].
\end{equation}
Equality follows by selecting $g$ as in~\Cref{eq:equalityIalpha2} and simple algebraic manipulations similar to the ones used to prove~\Cref{thm-Sibson-var-alternative}.

\subsection{Proof of~\Cref{prop:MLboundtightness}} \label{app:MLboundtightness}
\begin{proof}
 Define a function $g: \Y \rightarrow \X$ such that $g(y) \in \argmax_{x \in \X} \Pm_{Y|X} (y|x)$, and let $\X_g \subseteq \X$ be the image of $g$. Now, let $\Pm_X$ be the uniform distribution over $\X_g $, and $E = \{ (x,y): x=g(y) \}$. Then, for any $y \in \Y$, 
\begin{align} \label{eq:pxey}
 E_y = \{ g(y) \} \Rightarrow \Pm_X(E_y)  = \frac{1}{|\X_g|}. 
\end{align}
So we get 
\begin{align}
P_{XY} (E) & = \sum_{(x,y) \in E} \Pm_{XY}(x,y) \\
& = \sum_{y \in \Y} \sum_{x \in E_y} P_{X} (x) P_{Y|X} (y|x) \\
& = \sum_{y \in \Y} P_X(g(y)) P_{Y|X} (y| g(y) ) \\
& = \frac{1}{|\X_g|} \sum_{y \in \Y} \max_x P_{Y|X} (y|x),
\end{align}
where the last equality follows from~\eqref{eq:pxey} and the definition of $g$.
\end{proof}

\subsection{Proof of~\Cref{thm:TpcIneqIalpha}}\label{app:proof:thm:TpcIneqIalpha}
\begin{proof}
    Assume that $\mathbb{E}_{P_XP_Y}[f]=0$. If this is not true the following argument follows by selecting $f=f-\mathbb{E}_{P_XP_Y}[f]$.
     By~\Cref{thm:varReprIalpha} one has that for every function $f$ and every $\kappa$
    \begin{align}
    \frac{I_\alpha(X,Y)}{\alpha} &\ge \frac{1}{\alpha-1} \log \E_{P_{XY}}\left[e^{(\alpha-1)\kappa f}\right] -  \frac1\alpha \log \max_y \E_{P_X}\left[e^{\alpha \kappa f(X,y)}\right] \\
    &\geq \frac{1}{\alpha-1} \log \E_{P_{XY}}\left[e^{(\alpha-1)\kappa f}\right] -   \log \max_y \E_{P_X}\left[e^{\kappa f(X,y)}\right]\\
    &\geq \frac{1}{\alpha-1} \log \E_{P_{XY}}\left[e^{(\alpha-1)\kappa f}\right] - \frac{\kappa^2 c}{2} + \text{Ent}_\varphi^{P_{XY}}(\exp((\alpha-1) \kappa f) .
\end{align}
This implies that:
\begin{align}
    \frac{I_\alpha(X,Y)}{\alpha} + \frac{c\kappa^2 }{2} &\geq \frac{1}{\alpha-1} \log \E_{P_{XY}}\left[e^{(\alpha-1)\kappa f}\right]+  \text{Ent}_\varphi^{P_{XY}}(\exp((\alpha-1)\kappa f)\label{eq:jensenTpc}\\
    &= \frac{1}{\alpha-1} \log \E_{P_{XY}}\left[e^{(\alpha-1)\kappa f}\right]+\mathbb{E}_{P_{XY}}[\kappa f]-\frac{1}{(\alpha-1)}\log\mathbb{E}_{P_{XY}}\left[e^{(\alpha-1) \kappa f}\right]\label{eq:assumptionUseTpc} \\
    &= \kappa \mathbb{E}_{P_{XY}}[f].
\end{align}
Where~\Cref{eq:jensenTpc} follows from Jensen's inequality and the convexity of $x^\frac1\alpha$ with $0<\alpha<1$, while~\Cref{eq:assumptionUseTpc} follows from~\Cref{eq:subGaussGener}.
Thus one has the following:
\begin{equation}
    \mathbb{E}_{P_{XY}}[f] \leq \inf_{\kappa>0} \frac{\frac{I_\alpha(X,Y)}{\alpha}+\frac{c\kappa^2}{2}}{\kappa} = \sqrt{2c \frac{I_\alpha(X,Y)}{\alpha}}.
\end{equation}
\end{proof}

\section{Proofs for~\Cref{sec:estimationTheory}}

\subsection{Proof of~\Cref{thm:fano-alpha}} \label{app:fano-1}
\begin{proof}
    Let $\hat{X}$ be the optimal estimator of $X$ from $Y$. Then the Markov chain $X-Y-\hat{X}$ holds. Then,
    \begin{align}
        I_\alpha(X,Y) & \geq I_\alpha \left(X, \hat{X} \right) \\
        & \stackrel{\text{(a)}} = D_\alpha \left(P_{X \hat{X}} || P_X Q^\star_{\hat{X}} \right) \\
        & \stackrel{\text{(b)}} \geq d_\alpha \left( {P_{X \hat{X}}}( X =\hat{X}) \| 
        {P_X Q^\star_{\hat{X}}} ( X =\hat{X}) \right), 
    \end{align}
    where in (a) $Q^\star_{\hat{X}}$ is defined as in~\Cref{thm:Ialpha}, and (b) follows from the data processing inequality. Now note that,
    \begin{align} \label{eq:bound-guess-prob}
        {P_X Q^\star_{\hat{X}}} ( X =\hat{X})  = \sum_{x \in \X} P_X(x) Q^\star_{\hat{X}}(x) \leq \max_x P_X(x) \leq {P_{X \hat{X}}}( X =\hat{X}),
    \end{align}
    where the last inequality follows from the optimality of the MAP rule. It then follows from Lemma 11 of~\cite{Rioul:21} that
    \begin{align}
        I_\alpha(X,Y) \geq d_\alpha \left( \Pr_{P_{X \hat{X}}}( X =\hat{X}) \| \max_x P_X(x) \right) = d_\alpha \left (\eps_{X|Y} \| 1- \max_x P_X(x) \right).
    \end{align}
    The fact that $d_\alpha(\eps,\delta)$ is a non-increasing function of $\eps$ on the interval $[0,\delta]$ follows from simple differentiation:
    \begin{align}
        \frac{\partial }{\partial \eps} d_\alpha(\eps, \delta) & = \frac{\partial }{\partial \eps} \frac{1}{\alpha-1} \log \left( \eps^\alpha \delta^{1-\alpha} + (1-\eps)^\alpha (1-\delta)^{1-\alpha} \right) \\
        & = \frac{1}{\alpha -1} \frac{ \alpha  \left(   \left( \frac{\eps}{\delta} \right)^{\alpha-1} - \left( \frac{1-\eps}{1-\delta} \right)^{\alpha-1}  \right)  }{ \eps^\alpha \delta^{1-\alpha} + (1-\eps)^\alpha (1-\delta)^{1-\alpha}} \\
        & \leq 0,
    \end{align}
    where the last inequality follows from the fact that, for $(\eps,\delta) \in [0,1]^2$, $\ds \frac{\eps}{\delta} \leq \frac{1-\eps}{1-\delta}$ if and only if $\eps \leq \delta$.
\end{proof} Note that the bound can be tightened if one can derive a tighter bound on $ \Pr_{P_X Q^\star_{\hat{X}}} ( X =\hat{X})$ (compared to $\max_x P_X(x)$). This is hindered by the complicated expression for $Q^\star_{\hat{X}}$.

\subsection{Proof of~\Cref{thm:fano-like}} \label{app:fano-like}

Consider $\alpha >1$, $\beta > 0$, and let $f(X,\hat{X}) = \frac{\beta}{\alpha-1}\mathbbm{1} \{X = \hat{X}\}$. For short, let $\hat{p} = \Pr_{P_{XY}}( X =\hat{X})$ and $\hat{q} =  \Pr_{P_{X}Q^\star_{Y}}( X =\hat{X})$. Then, by~\Cref{thm-Sibson-var-alternative-weaker},
\begin{align}
    I_\alpha(X,Y) 
    & \geq I_\alpha(X,\hat{X}) \notag \\
    & \geq \frac{\alpha}{\alpha-1} \log \left( \hat{p} e^\beta + 1- \hat{p} \right) - \log \left( \hat{q} e^{ \frac{\alpha}{\alpha-1}\beta} + 1- \hat{q} \right) \notag \\
    & \geq  \frac{\alpha}{\alpha-1} \log \left( \hat{p} (e^\beta-1) + 1 \right)- \log \left( p^\star e^{ \frac{\alpha}{\alpha-1}\beta} + 1- p^\star \right), 
\end{align}
where the second inequality follows from the fact that $e^\beta -1 > 0$ (for $\beta >0$), and $\hat{q} \leq \max_x P_X(x)$ since $Y$ is independent of $X$ under $P_X Q^\star_Y$. Setting $\gamma = e^\beta -1 >0$ and rearranging terms yields~\cref{eq:thm-fano-like}.

\subsection{Proof of~\Cref{thm:fano-alpha-Arimoto}} \label{app:fano-Arimoto}
\begin{proof}
      $\frac{\alpha-1}{\alpha} H_\alpha(X,Y)$ is monotonically increasing in $\alpha$~\cite[Proposition 2]{SasonV:17}. As such,
    \begin{align}
        H_\infty(X|Y) \geq \frac{\alpha-1}{\alpha} H_\alpha(X,Y) = \frac{\alpha-1}{\alpha} (H_\alpha(X) - I_\alpha^A(X,Y)) =  \frac{\alpha-1}{\alpha} (H_\alpha(X) - I_\alpha(X_\alpha,Y)), 
    \end{align}
    where the first equality follows from the definition of Arimoto mutual information (cf.~\Cref{def:arimoto}) and the second equality follows from~\Cref{prop:arimoto-sibson-tilted}. The theorem follows by noting that
    \begin{align}
        H_\infty (X|Y) =  - \log \E_{P_Y} \left[  \max_{x \in \X} P_{X|Y}(x|Y) \right] = - \log \left( 1- \eps_{X|Y} \right).
    \end{align}
\end{proof}

\subsection{Proof of~\Cref{lem:Dalpha-bound-Ialpha}} \label{app:lem-fano-method}

\begin{proof}
For $\alpha = 1$, the statement is straightforward since
\begin{align}
    I(X;Y) = \min_{Q_Y} D(P_{XY} || P_X Q_Y ) = \min_{Q_Y} \E_{P_X} \left[ D(P_{Y|X}(.|X) || Q_Y) \right] \leq \beta.
\end{align}
For $\alpha < \infty$, an elementary proof can be given by observing
    \begin{align}
        I_\alpha(X,Y) & = \min_{Q_Y} D_\alpha (P_{XY} || P_X Q_Y) \\
        & = \min_{Q_Y} \frac{1}{\alpha-1} \log  \exp \left\lbrace   (\alpha-1)D_\alpha(P_{XY} || P_X Q_Y)  \right\rbrace \\
        & = \min_{Q_Y} \frac{1}{\alpha-1} \log \E_{P_X} \left[  \exp \left\lbrace   (\alpha-1)D_\alpha(P_{Y|X}(.|X) || Q_Y)  \right\rbrace \right] \\
        &  =  \frac{1}{\alpha-1} \log \E_{P_X} \left[  \exp \left\lbrace   (\alpha-1) \min_{Q_Y} D_\alpha(P_{Y|X}(.|X) || Q_Y)  \right\rbrace \right] \\
        & \leq \beta,
    \end{align}
    where the last inequality follows from the hypothesis of the lemma. \\
    Finally, for $\alpha = \infty$, then for all $x$
    \begin{align}
         D_\infty(P_{Y|X}(.|x) || Q_Y) \leq \beta \Rightarrow D_\alpha(P_{Y|X}(.|x) ||  Q_Y) \leq \beta \text{ for all } \alpha < \infty.
    \end{align}
    Hence, by the proof for $\alpha < \infty$, we get $I_\alpha(X,Y) \leq \beta$ so that
    \begin{align}
        I_\infty(X,Y) = \lim_{\alpha \rightarrow \infty} I_\alpha(X,Y) \leq \beta.
    \end{align}
\end{proof}

\subsection{Proof of~\Cref{thm:genFano}} \label{app:genFano}

\begin{proof}
    Consider a random variable $J \in \cJ$ with distribution $Q$ and set $P_{Y|J}$ as in the statement of the theorem. Let $\eps_{J|Y}$ be the optimal probability of error of guessing $J$ from $Y$. Fix any estimator $\theta$, and consider a sub-optimal rule $\hat{J}(Y)$ as follows:
    \begin{align}
        \hat{J}(Y) \in \argmin_{j \in \{1,2,\ldots,r\}} \ell \left( \hat\theta(Y), \theta(P_j) \right), 
    \end{align}
    where ties are broken arbitrarily.
Then,
\begin{align}
    \eps_{J|Y}  & \stackrel{\text{(a)}} \leq \Pr \left( \hat{J}(Y) \neq J \right) 
    \stackrel{\text{(b)}} \leq \Pr \left(  \ell \left( \hat{\theta}(Y), \theta(P_J)  \right) \geq \frac{\gamma}{2} \right) 
     \stackrel{\text{(c)}} \leq \frac{2}{\gamma} \E \left[   \ell \left( \hat{\theta}(Y), \theta(P_J) \right) \right],
\end{align}
where (a) follows from the optimality of $\eps_{J|Y}$, (b) follows from the fact that $ \ell \left( \hat{\theta}(Y), \theta(P_J) \right) < \gamma/2 $ implies $J = \argmin_{i \in \{1,2,\ldots,r\}} \ell \left( \hat\theta(Y), \theta(P_i) \right)$ by~\cref{eq:thm-gen-fano-loss-cond} (and the fact that pseudo-metrics satisfy the triangle inequality), and (c) follows from Markov's inequality. Now,
\begin{align}
    \max_{j \in \{1,2,\ldots, r\}} \E_{P_j} \left[ \ell ( \theta(P_j), \hat{\theta}(Y)) \right] 
    & \geq \E_{Q P_{Y|J}} \left[ \ell ( \theta(P_J), \hat{\theta}(Y)) \right] \\
    & \geq \frac{\gamma}{2} \eps_{J|Y} \\
    & \geq \frac{\gamma}{2} \left(1-\left(\max_j Q_J(j) e^{I_\alpha(J,Y)}\right)^{\frac{\alpha-1}{\alpha}} \right),
\end{align}
where the last inequality follows from~\Cref{eq:corr-fano-gamma-inf}.
\end{proof}

\newpage

\bibliographystyle{IEEEtran}
\bibliography{SibsonAlphaBib}

\end{document}